\RequirePackage{ifpdf}
\documentclass[hyper,letterpaper,11pt]{JHEP3}
\usepackage{amsmath,amssymb,multirow,array,graphicx,mathrsfs}
\usepackage[normalem]{ulem}
\renewcommand{\nc}{\newcommand}
\nc{\beq}{\begin{equation}}
\nc{\eeq}{\end{equation}}

\nc{\MFa}{\mathfrak{a}} % define the KK gauge field
\nc{\MFs}{\mathfrak{s}} % define the g_00 component 
\nc{\JV}{\mathrm{J}_{\scriptscriptstyle\mathrm{V}}} % define the vector current operator
\nc{\Seff}{S_{\rm eff}} % define the symbol for the Wilsonian effective action
\nc{\C}{\mathscr{C}} % define the calagraphic C which gives the trivially conserved current
\nc{\Wstatic}{W_{\rm hydrostatic}}
\nc{\be}{\begin{equation}}
\nc{\ee}{\end{equation}}
\nc{\ba}{\begin{eqnarray}}
\nc{\ea}{\end{eqnarray}}

\nc{\al}{\alpha}
\nc{\ga}{\gamma}
\nc{\de}{\delta}
\nc{\ep}{\epsilon}
\nc{\ze}{\zeta}
\nc{\et}{\eta}

\nc{\ka}{\kappa}
\nc{\la}{\lambda}
\nc{\rh}{\rho}
\nc{\si}{\sigma}
\nc{\ta}{\tau}
\nc{\up}{\upsilon}
\nc{\ph}{\phi}
\nc{\ch}{\chi}
\nc{\ps}{\psi}
\nc{\om}{\omega}
\nc{\Ga}{\Gamma}
\nc{\De}{\Delta}
\nc{\La}{\Lambda}
\nc{\Si}{\Sigma}
\nc{\Up}{\Upsilon}
\nc{\Ph}{\Phi}
\nc{\Ps}{\Psi}
\nc{\Om}{\Omega}
\nc{\ptl}{\partial}
\nc{\del}{\nabla}
\nc{\bi}{\bibitem}
\nc{\Lie}{\mathfrak{L}} % define the lie derivative
\nc{\ab}[2]{^{#1}_{\phantom{#1}#2}}
\def\coeff#1#2{{\textstyle {\frac {#1}{#2}}}}

\nc{\tf}{\tilde{f}} % define the tilde'd f's - Wilsonian
\nc{\f}{f} % define the Latin f's - Wilsonian
\nc{\p}{p} % define a Latin p - Wilsonian
\renewcommand{\a}{a} % define a Latin a - Wilsonian
\renewcommand{\ta}{\tilde{a}} % define a tilde'd Latin a - Wilsonian
\nc{\tF}{\tilde{\varphi}} % define the tilde'd Greek f - 1PI
\nc{\F}{\varphi} % define the Greek f - 1PI
\renewcommand{\P}{P} % define a Greek P - 1PI
\nc{\A}{\alpha} % define a Greek a - 1PI
\nc{\tA}{\tilde{\alpha}} % define a Greek tilde'd a - 1PI
\renewcommand{\j}{j} % define the j's that sit in the identically conserved current $\mathscr{C}$ - Wilsonian
\nc{\tj}{\tilde{j}} % define the tilde'd j's similarly - Wilsonian
\nc{\J}{\chi} % define the equivalent of j that sits in $\mathscr{C}$ - 1PI
\nc{\tJ}{\tilde{\chi}} % define the equivalent of tilde'd j that sits in $\mathscr{C}$ - 1PI

\title{Chiral conductivities and effective field theory}
\author{Kristan Jensen, Pavel Kovtun, and Adam Ritz \\
Department of Physics and Astronomy, University of Victoria, \\ 
Victoria, BC V8W 3P6, Canada \\
Email: kristanj@uvic.ca, pkovtun@uvic.ca, aritz@uvic.ca
}
\abstract{We construct the three-dimensional effective field theory which reproduces low-momentum static correlation functions in four-dimensional quantum field theories with $U(1)$ axial anomalies and a dynamical vector gauge field, in thermal equilibrium.  We compute radiative corrections to parity-violating chiral conductivities, to leading order in the effective theory. All of the anomaly-induced transport is susceptible to radiative corrections, except for certain two-point functions which are required by symmetry to vanish.}
\date{\today}

\begin{document}

\section{Introduction and summary}
\label{S:introduction}

\subsection{Introduction}
Anomalies are a ubiquitous feature of quantum field theories. Their usefulness is tied to the fact that they are exact and so may be determined even at strong coupling. This exactness is a consequence of certain non-renormalization properties, and allows non-perturbative insight via tools such as anomaly matching~\cite{Weinberg2}. There has been a recent resurgence of interest in anomalies in the context of field theory at nonzero temperature and chemical potential. In particular, it is now known that in the hydrodynamic limit there are certain first-order transport coefficients, on the same footing as viscosity or conductivity, which are fixed in terms of anomaly coefficients and thermodynamic quantities~\cite{Son:2009tf,Neiman:2010zi,Jensen:2012kj,Son:2004tq}.

Such anomaly-induced transport is dissipationless, as may be seen by an entropy production analysis within hydrodynamics. In addition, the corresponding transport coefficients are determined by Euclidean Kubo formulae~\cite{Amado:2011zx}, meaning that they are thermodynamic parameters, and may be measured in equilibrium. Thanks to their thermodynamic nature, the anomalous transport coefficients can also be understood from Euclidean field theory~\cite{Banerjee:2012iz, Jensen:2012jy}.
This anomaly-induced hydrodynamic transport provides a macroscopic manifestation of microscopic anomaly-induced physics, such as the chiral magnetic effect (CME)~\cite{Vilenkin:1980fu,Alekseev:1998ds, Fukushima:2008xe, Kharzeev:2009p}, chiral separation effect (CSE)~\cite{Vilenkin:1980ft, Metlitski:2005pr, Newman:2005as}, and the chiral vortical effect (CVE)~\cite{Vilenkin:1979ui}. Specifically, for a theory with a vector current $J^\mu_{V, cov}$ and an axial current $J^\mu_{A, cov}$, coupled to the corresponding background non-dynamical sources $V_\mu$ and $A_\mu$, anomalies give rise to the following additional terms in the hydrodynamic constitutive relations:
\begin{align}
\begin{split}
  J^\mu_{V,cov} & = \dots + \xi_V w^\mu + \xi_{VV} B^\mu + \xi_{VA} B^\mu_A\,, \\
  J^\mu_{A,cov} & = \dots + \xi_A w^\mu + \xi_{AV} B^\mu + \xi_{AA} B^\mu_A \,.
\end{split}
\end{align}
Here $u^\mu$ is the fluid velocity, $B^\mu = \frac12 \epsilon^{\mu\nu\rho\sigma}u_\nu F_{V,\rho\sigma}$, and $B^\mu_A = \frac12 \epsilon^{\mu\nu\rho\sigma}u_\nu F_{A,\rho\sigma}$ are the corresponding vector and axial magnetic fields, and $w^\mu = \epsilon^{\mu\nu\rho\sigma}u_\nu \nabla_\rho u_\sigma$ is the vorticity. The subscript denotes covariant currents, as we explain later. The CME is usually associated with the $\xi_{VV}$ term, the CSE with the $\xi_{AV}$ term and the CVE with the $\xi_A$ term. See~\cite{Zakharov:2012vv} for a recent review. 

Most of the results on anomalous transport are derived for theories whose anomalies involve only global symmetries. In this case, if we schematically denote the anomalous current divergence as $\nabla J_A = c F \tilde{F}$, the background field strength on the right-hand side is non-dynamical, the symmetry currents are non-perturbatively defined objects, and the corresponding anomalies are exact: under an anomalous background gauge transformation, the partition function picks up a definite, anomalous phase which is built out of the background fields that couple to the global currents. The anomalous current non-conservation should be considered on the same footing as the non-conservation of the energy-momentum tensor in an external gauge field.

The notion of anomalous transport becomes more involved when the theory in question has anomalies involving a gauge symmetry (see e.g. \cite{Neiman:2010zi}).  For example, if the field strength $F$ is dynamical, the Ward identity $\nabla J _A= c F \tilde{F}$ (where $J_A$ is a global current)  needs to be interpreted with the appropriate definitions of the renormalized field operators $J_A$ and $F$. While the Adler-Bardeen theorem \cite{ab} still allows for exact statements to be made given specific kinematics, taking the expectation value of the Ward identity may involve loop corrections with the anomalous vertex entering as a sub-diagram. These so-called `rescattering' corrections can modify the effective anomaly coefficient when the Ward identity is evaluated in a given state, see e.g.~\cite{adler, Adler:2004qt}. In the hydrodynamic regime, these corrections affect the values of the anomalous transport coefficients, since analogous loop diagrams involving the anomaly triangle contribute to the Kubo formulae. For the CVE coefficient, these corrections were studied recently in~\cite{Golkar:2012kb,Hou:2012xg}.

When considering hydrodynamic transport, two additional points need to be kept in mind. First, if the field strength $F$ is dynamical, 
the current $J_A$ is no longer conserved, which means that $J_A$ is only relevant to hydrodynamics on time scales which are short compared to the time scale of current non-conservation. Second, even in the absence of anomalies, dynamical $U(1)$ gauge fields become extra hydrodynamic degrees of freedom, changing the equations of hydrodynamics to those of magneto-hydrodynamics (MHD). Ignoring loops of $U(1)$ gauge fields, the transport coefficients are to be thought of as transport coefficients in MHD. Quantum loops of $U(1)$ gauge fields invalidate MHD as a classical description, meaning that some care should be taken when interpreting loop corrections such as those computed in~\cite{Golkar:2012kb,Hou:2012xg} as corrections to hydrodynamic transport coefficients. This is analogous to loop corrections to transport coefficients due to the normal hydrodynamic collective modes~\cite{Kovtun:2011np} (which do not contribute in the static limit).

In this paper, we undertake a systematic study of theories with $U(1)$ anomalies and a dynamical $U(1)$ gauge field, from the effective field theory point of view. As we discuss below, we find that all of the anomaly-induced transport is susceptible to quantum corrections, except for certain two-point functions which are required by symmetry to vanish.

\subsection{Summary}
\label{S:result}
We consider four-dimensional field theories at nonzero temperature $T$, which have general covariance  and a gauged $U(1)_V$ symmetry corresponding to a vector current $J_V^{\mu}$. In the zero-gauge coupling limit, where $U(1)_V$ becomes a global symmetry, we assume the theory has a $U(1)_A$ axial current $J_A^\mu$ with a $U(1)^3$ $AVV$ anomaly, a $U(1)^3$ $AAA$ anomaly, and an $ATT$ mixed axial-gravitational anomaly. We consider theories which have negligible explicit $U(1)_A$-breaking at the cutoff scale. A simple example is QED with massless Dirac fermions.

As reviewed in Section~\ref{S:review}, the Ward identity for the {\it consistent} current takes the form,
\be
\label{E:axialWI}
\nabla_{\mu} J_A^{\mu} = \frac{1}{4}\epsilon^{\mu\nu\rho\sigma}\left[ c_g F_{V,\mu\nu}F_{V,\rho\sigma}+ \frac{\bar{c}_g}{3}F_{A,\mu\nu}F_{A,\rho\sigma}+ c_m R\ab{\alpha}{\beta\mu\nu}R\ab{\beta}{\alpha\rho\sigma}\right],
\ee 
in terms of the corresponding field strengths. The gauge fields are normalized so that the anomaly coefficients $c_g$, $\bar{c}_g$ and $c_m$ are pure numbers, with no factors of the gauge coupling. Gauging $U(1)_V$ explicitly breaks $U(1)_A$, so that $J_A^{\mu}$ becomes merely one of the many non-conserved pseudovector operators of the theory. 
We will be interested in the regime where the $U(1)_V$ gauge interactions are weak, as in electromagnetism, and thus the corrections can be studied within perturbation theory. 

In thermal equilibrium, the static response of these theories may be computed in three-dimensional Euclidean effective theory: for the low-momentum degrees of freedom, one may dimensionally reduce on the Euclidean time circle and obtain a three-dimensional effective action $\Seff$ on the spatial slice. In dimensionally reducing on the thermal circle, we implicitly integrate out all nonzero Matsubara modes. We then consider the effective description for sufficiently low momenta $p\leq \Lambda$ so that the only remaining degrees of freedom of the Wilsonian effective theory are the bosonic Matsubara zero modes of the photon. The cutoff may at most be $\Lambda\sim T$, the scale of the first nonzero Matsubara mode. Such effective theories are well known in the context of hot QCD and QED~\cite{Braaten:1995cm, Braaten:1995jr, Andersen:1995ej}.

When $U(1)_V$ is a global symmetry, 
we can couple the theory to external time-independent sources $V_\mu$, $A_\mu$, and $g_{\mu\nu}$ for the vector current, axial current, and the energy-momentum tensor. We parametrize the Lorentzian-signature fields as in~\cite{Banerjee:2012iz},
\begin{align}
\begin{split}
\label{E:sources}
V & = V_0(x)(dt+\MFa_i(x)dx^i) + \hat{V}_i(x)dx^i\,,
\\
A & = A_0(x)(dt+\MFa_i(x)dx^i) + \hat{A}_i(x)dx^i\,,
\\
g & = -e^{2\MFs(x)}(dt+\MFa_i(x)dx^i)^2 + \hat{g}_{ij}(x)dx^idx^j\,.
\end{split}
\end{align}
After $U(1)_V$ is gauged, the effective description of the static equilibrium may be given in terms of a three-dimensional effective action $\Seff[V,A,g]$ for $V_\mu$, the Matsubara zero modes of the $U(1)_V$ gauge field continued to real time. We are assuming that $V_{\mu}$, which we call the ``photon'', is the lightest field on the spatial slice.
The effective action must be invariant under coordinate reparametrization and $U(1)_V$ gauge transformations. Due to the $AVV$ anomaly, $U(1)_A$ is no longer a symmetry of the theory when $V_\mu$ is dynamical, hence $\Seff$ is not invariant under $U(1)_A$ gauge transformations. When expressed in terms of the fields in \eqref{E:sources}, $\Seff$ should be invariant under spatial diffeomorphisms, spatial reparametrizations of time and stationary $U(1)_V$ transformations. Under spatial reparametrizations of time (which we will call $U(1)_{KK}$), $\MFa_i$ transforms as a connection, with all other fields neutral. Under time-independent $U(1)_V$ transformations, $\hat{V}_i$ transforms as a $U(1)$ connection with all other fields neutral. We identify the local temperature $T =  e^{-\MFs}\beta^{-1}$ and chemical potentials $\mu_V = e^{-\MFs}V_0$, $\mu_A = e^{-\MFs}A_0$, where $\beta$ is the coordinate periodicity of Euclidean time.%
\footnote{
	Note that $\mu_A$ as defined is conjugate to the {\it consistent} (but non-conserved)
	axial charge; to be distinguished from the genuine chemical potential $\mu_5$ 
	conjugate to the conserved axial charge, as discussed in Section~\ref{S:discuss}.
	}

It will prove useful to consider both the Wilsonian effective action $\Seff[V,A,g]$, and the corresponding three-dimensional 1PI effective action $\Gamma[\langle V \rangle, A,g]$. The latter is generically a nonlocal functional, but since we consider static background fields at finite temperature, only the static magnetic photon remains as a massless degree of freedom and, due to gauge invariance, its derivative couplings soften the infrared behaviour.
We will analyze $\Seff$ and $\Gamma$ in a derivative expansion when the fields are slowly varying, which will in turn fix the small-momentum structure of static correlation functions.
We treat the $U(1)$ gauge fields and the metric as ${\cal O}(1)$, so that the background field strengths are ${\cal O}(\ptl)$. This is appropriate for studying the response of the system to infinitesimal background fields but does not capture the effect of having a finite background magnetic field or a curved geometry. The three-dimensional effective action has the form
\begin{equation}
\label{E:effSresult}
\Seff = \Seff^{(0)} + \Seff^{(1)} + \Seff^{(2)} + \dots,
\end{equation}
where the superscripts denote the order in the derivative expansion. To lowest order,
\begin{equation}
\label{E:Seff00}
\Seff^{(0)} = \int\! (dt)\, d^3x \,e^{\MFs}\sqrt{\hat{g}} \,\p(T,\mu_V,\mu_A, \hat{A}^2)\,, 
\end{equation}
where $\hat{A}^2\equiv \hat{A}_i\hat{A}_j \hat{g}^{ij}$. The notation $(dt)$ in (\ref{E:Seff00}) stands for $(-i\beta)$. We chose to write the three-dimensional action in a form that mimics the four-dimensional notation in order to facilitate the comparison with the four-dimensional generating functional of Ref.~\cite{Jensen:2012jh}, and to easily access the four-dimensional correlation functions. The arguments of the effective action are the time-independent fields, which can be viewed as the Matsubara zero modes continued to real time. In particular, the chemical potentials $\mu_V$, $\mu_A$ are real. The geometric factor $e^{\MFs} \sqrt{\hat{g}}$ is just $\sqrt{-g}$. The function $p$ is the equilibrium pressure of the theory subject to spatially uniform external sources.
To first order in the derivative expansion, we have
\begin{align}
\label{E:effSexp}
\begin{split}
\Seff^{(1)} &= \int\! (dt)\, d^3x\, \tilde{\epsilon}^{\, ijk} \left[\frac{\tilde{c}_{VV}}{2\beta} \hat{V}_i\, \ptl_j \hat{V}_k -\frac{\tilde{c}_V}{\beta^2} \hat{V}_i\,\ptl_j \MFa_k + \frac{\tilde{c}}{2\beta^3} \MFa_i \ptl_j \MFa_k \right. \\
& \qquad\qquad\qquad+  \tf_{AA} \hat{A}_i\,\ptl_j \hat{A}_k + \tf_{AV} \hat{A}_i \ptl_j \hat{V}_k + \tf_{Aa} \hat{A}_i \ptl_j \MFa_k 
\bigg] + \delta S^{(1)}_{\rm axial}\,,
\end{split}
\end{align}
where the epsilon-symbol satisfies $\tilde{\epsilon}^{\,123}=1$, the Chern-Simons terms do not depend on the metric, and $\delta S^{(1)}_{\rm axial}$ (shown explicitly in~\eqref{S1axial}) contains further $U(1)_A$-violating terms that do not contribute to 
the chiral response at low order in perturbation theory.

The coefficients in Eq.~\eqref{E:effSexp} fall into two categories. Invariance under $U(1)_{KK}$ and $U(1)_V$ implies that $\tilde{c}_{VV}$, $\tilde{c}_V$ and $\tilde{c}$ must be constant, and dimensional analysis dictates the inverse powers of $\beta$ present in Eq.~\eqref{E:effSexp}. In contrast, the $\tf$'s are unconstrained and may be nontrivial functions of $T$, $\mu_V$, $\mu_A$, and $\hat{A}^2$. In particular, they may receive corrections beyond the tree-level contributions associated with anomalies. 
We find
\begin{align}
\begin{split}
\label{corr}
 \tilde{f}_{AA} &= -\frac{2\bar{c}_g A_0}{3}+\frac{\tilde{c}_{AA}}{2\beta}+ \de \tf_{AA}, \\
 \tilde{f}_{AV} &= -2c_gV_0 + \frac{\tilde{c}_{AV}}{\beta}+\de \tf_{AV},\\
 \tilde{f}_{Aa} &= -c_g V_0^2 -\frac{\bar{c}_g A_0^2}{3} - \frac{\tilde{c}_A}{\beta^2}+ \de \tf_{Aa},
 \end{split}
\end{align}
where $c_g$, $\bar{c}_g$ and $c_m$ are the chiral anomaly coefficients for consistent currents shown in \eqref{E:axialWI} and described in more detail in Section~\ref{S:review}. The corrections $\delta \tilde f$, which may be nontrivial functions of $T$, $\mu_V$, $\mu_A$, and $\hat{A}^2$, arise from integrating out non-zero Matsubara modes of all the fields in the theory, plus the high momentum component of the photon zero modes. Finally, the coefficients $\tilde{c}_{AA},\tilde{c}_{AV},\tilde{c}_A$ are constants, which are allowed by the symmetries and will be discussed further below. The corrections $\delta \tilde f$ may be computed through matching to the microscopic theory. On the other hand, the effective theory (\ref{E:effSresult}) can be used to systematically compute the infrared loop corrections due to the photon zero modes.

We may employ the Wilsonian effective action~\eqref{E:effSresult}, and the associated 1PI effective action,  to compute correlation functions of currents in equilibrium. Defining the currents conjugate to the axial vector and metric sources, in Section~\ref{S:effectiveS} we compute the equilibrium one-point functions, and the zero-frequency, low-momentum two-point functions, in a flat background. To first order in derivatives and to linear order in the metric perturbation, the expressions for the axial and momentum currents then take the following form in equilibrium with constant temperature and chemical potentials,
\begin{align}
\label{E:1pt}
\begin{split}
\langle J_A^i\rangle_{\rm equil.}   =&  
  +2\tF_{AA} B_A^i + \frac{1}{2}\left(2\mu_A \tF_{AA} +\mu_V \tF_{AV} - \tF_{Aa}\right) \Om^i +\tF_{AV} {B}^i  + \cdots \\
\langle T^{0i} \rangle_{\rm equil.} = & 
 -2\tF_{Aa} B_A^i + \frac{1}{2}\left(\mu_A \tF_{Aa}+ \mu_V\tilde{c}_{V}T^2 +\tilde{c}\,T^3 \right)\Om^i +\tilde{c}_{V}T^2 {B}^i + \cdots 
\end{split}
\end{align}
where $B_A^i = \tilde{\ep}^{\,ijk} \ptl_j A_k$, $\Om^i = -2\tilde{\ep}^{\,ijk} \ptl_j \MFa_{k}$, and ${B}^i = \tilde{\ep}^{\,ijk} \ptl_j \langle {V}_k\rangle$. The dots denote terms proportional to $A_i$, $g_{0i}$, and $\ptl_i\langle V_0\rangle$. As a consequence of the anomalies, the currents are not invariant under $U(1)_A$ gauge transformations.
The coefficients $\tF_i$ are the 1PI analogues of the $\tf_i$ coefficients appearing in $\Seff$, and have a similar structure in the flat background,
\begin{align}
\label{corr2}
 \tF_{AA,AV,Aa} &= \tf_{AA,AV,Aa} +  {\cal O}(V\textrm{-loops}).
\end{align}
In addition to the nonzero Matsuabara mode corrections in (\ref{corr}), these coefficients also receive corrections from infrared-sensitive loops of the photon Matsubara zero modes. We present computations of the leading one-loop corrections to these parameters in Section~\ref{S:pert}.

The equilibrium expressions (\ref{E:1pt}) for the consistent currents can be viewed as definitions of various chiral conductivities, however one should be careful in interpreting (\ref{E:1pt}) and (\ref{corr2}) as hydrodynamic constitutive relations.
From~\eqref{E:1pt} one can see that the only symmetry-protected parts of the chiral conductivities are those which depend on the $\tilde{c}$'s alone.
Under {\em four-dimensional} CPT transformations, the coefficients $\tilde{f}_{AA}, \tilde{f}_{AV}, \tilde{c}_{AA}, \tilde{c}_{AV}, \tilde{c}_{VV},$ and $\tilde{c}$ are CPT-violating, while $\tilde{f}_{Aa}$ and $\tilde{c}_{V}$ are CPT-preserving. It follows that $\tilde{c}_{AA}, \tilde{c}_{AV}, \tilde{c}_{VV}$ and $\tilde{c}$ vanish in the absence of CPT-violating sources. In our case, both $\mu_A$ and $\mu_V$ act as CPT-violating sources, and so $\tilde{c}_{AA}, \tilde{c}_{AV}, \tilde{c}_{VV}$ and $\tilde{c}$ could in principle be odd functions of $A_0/|A_0|$ and $V_0/|V_0|$.%
\footnote{
	When the Chern-Simons term arises by integrating out a massive charged Dirac fermion 
	in 2+1 dimensions,
	there is an analogous non-analytic dependence of the Chern-Simons coefficient
	on the fermion mass~\cite{Redlich:1983dv}.
}
In particular, a constant axial chemical potential $\mu_5$ (as distinct from $\mu_A$) will induce the CPT-violating parameter $\tilde{c}_{VV}$. We will retain the CPT-violating constants to facilitate contact with physics at nonzero $\mu_5$.

In practice, the Chern-Simons couplings $\tilde{c}_{AA}, \tilde{c}_{AV}, \tilde{c}_{VV}, \tilde{c}_A, \tilde{c}_V$, and $\tilde{c}$ may be computed in the microscopic theory by integrating out massive matter fields at one loop~\cite{Coleman:1985zi, Golkar:2012kb}. Remarkably, the coefficients $\tilde{c}_V$ and $\tilde{c}_A$ have been shown to be proportional to the mixed $VTT$ and $ATT$ gauge-gravitational anomaly coefficients~\cite{Jensen:2012kj,Golkar:2012kb}, giving
\begin{equation*}
\tilde{c}_V = 0\,, \qquad \tilde{c}_A = - 8\pi^2 c_m\,,
\end{equation*}
where $\tilde{c}_V$ vanishes because there is no $VTT$ anomaly in our theories.

In writing the one-point functions in (\ref{E:1pt}), we did not quote the vector current for the dynamical $U(1)_V$. The vector current is more subtle, since it is constrained even at the classical level by the equation of motion for $V_\mu$, as we discuss in Sect.~\ref{S:effectiveS}.  In the effective theory, at scales below $T$ the charged degrees of freedom decouple, and the vector current $\JV^\mu$ is identically conserved. This is a manifestation of a more general feature that the consistent current can differ from the generator of $U(1)_V$ transformations by identically conserved currents, whose precise form needs to be fixed by matching to the microscopic theory. In particular, this is true in QED, where the conventional dimension-3 vector current can mix with the identically conserved current $\frac{1}{\sqrt{-g}}\partial_{\nu}(\sqrt{-g}F^{\mu\nu}_V)$~\cite{Collins:2005nj}. 

In Section~\ref{S:effectiveS}, we show how the $\tilde{f}$'s and $\tilde{c}$'s appear in hydrodynamics, which normally makes use of {\it covariant} axial and vector currents, which differ from consistent currents. We explicitly compute the zero-frequency, low momentum two-point functions, shown in Eq.~(\ref{E:Kubo1}), (\ref{E:Kubo2}). The general structure of the ${\cal O}(k)$ terms is represented schematically as follows,

\begin{minipage}{0.4\textwidth}
\begin{align*}
 &\lim_{k\to 0}\frac{\epsilon_{ijk}k^k}{k^2}\langle j^i_{1}(k)j^j_{2}(-k)\rangle=
 \end{align*}
 \end{minipage}
 \begin{minipage}{0.6\textwidth}
 \hspace{-.5cm}\includegraphics[trim={3cm 22.2cm 0cm 1.2cm},clip,width=0.9\textwidth]{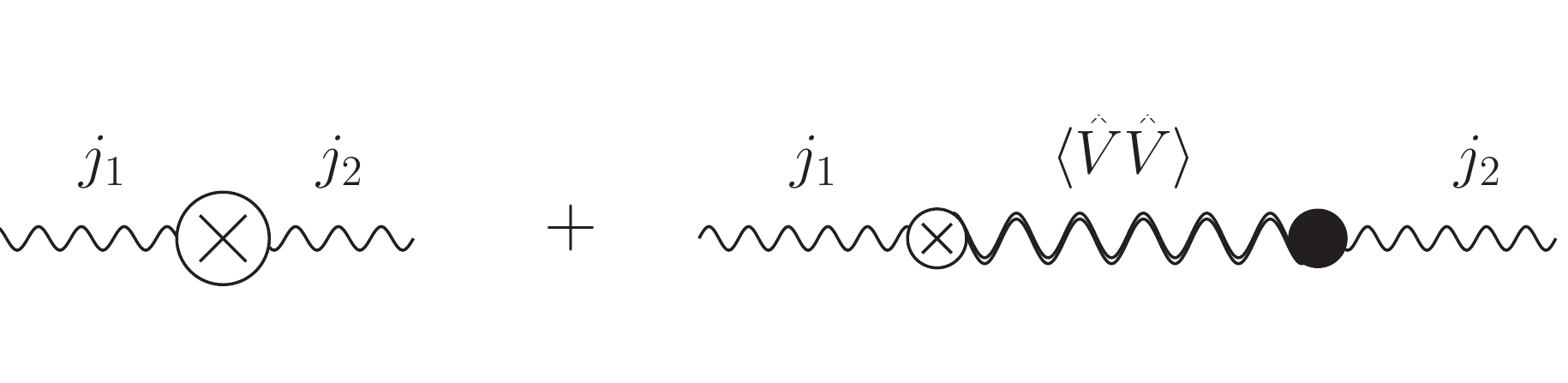}
\end{minipage}
where $j_{1,2}$ denote any of the currents under consideration. The first diagram represents the direct variation 
of the 1PI effective action, while the second diagram reflects the need to build up the full connected
correlator by accounting for contributions that combine other 1PI vertices with exact propagators for the massless spatial photon. However, we note that if $\tilde{c}_{VV}\neq0$, the spatial photon is (anti-)screened through a topological mass and the second class of contributions above changes. We present the full structure of these low-momentum two-point functions in Section~\ref{S:Gamma}.

The applicability of anomalous transport seemingly relies on the corrections in~\eqref{corr},~\eqref{corr2} being sufficiently small, which means that exact statements about the chiral conductivities are rather limited. However, we note that certain combinations of two-point functions do lead to interesting vanishing theorems which persist in the presence of these perturbative corrections. In particular, setting all constants which violate four-dimensional CPT to zero we have 
\begin{align}
\label{E:const}
\begin{split}
&\lim_{k\to 0}\frac{\epsilon_{ijk}k^k}{k^2}\langle \JV^i(k) \JV^j(-k)\rangle= \lim_{k\to 0}\frac{\epsilon_{ijk}k^k}{k^2}\langle \JV^i(k)Q^j(-k)\rangle= 0,
\end{split}
\end{align}
where $Q^i = T^{0i}-\mu_V \JV^i - \mu_A J_A^i$ is the heat current. The corresponding low-momentum behavior of $\langle Q^i(k)Q^j(-k)\rangle$ gives a result which is nonzero, but perturbatively suppressed at weak $U(1)_V$ gauge coupling. The one-point functions (\ref{E:1pt}) clearly exhibit terms identifiable with the chiral-vortical and chiral-separation conductivities. The result (\ref{E:const}) implies zero chiral magnetic conductivity, in agreement with the result in~\cite{Rubakov:2010qi} for the consistent vector current at non-zero $\mu_A$ (as distinct from $\mu_5$; see Section~\ref{S:discuss} for further discussion), when $U(1)_V$ is a global symmetry. 

The effective action~\eqref{E:effSresult}, currents~\eqref{E:1pt}, and the Kubo formulae in~(\ref{E:Kubo1}), (\ref{E:Kubo2})  summarize the main results of the present work. For theories where the anomaly is shared between a global symmetry current and a weakly coupled gauge sector, the $\tF$'s will be perturbatively close to the results obtained when the gauge sector is non-dynamical. In weakly coupled QED, for instance, the leading correction to $\tF_{Aa}$ is of order $e^2$ in the electromagnetic coupling~\cite{Golkar:2012kb,Hou:2012xg}. For the QCD sector, we anticipate similar corrections to the conductivities associated with non-singlet $SU(N_f)_A$ chiral currents in the quark-gluon plasma,
which only have electromagnetic anomalies. In contrast, the singlet $U(1)_A$ axial current in QCD has a gluonic $U(1)_A\times SU(3)^2$ anomaly, which will lead to large corrections to the $\tilde{\varphi}$ coefficients, as observed in lattice calculations with $SU(3)$ replaced with $SU(2)$~\cite{Braguta:2013loa}. The singlet $U(1)_A$ chiral conductivity in QCD is not fixed by anomaly coefficients (and is not  well-defined in hydrodynamics), except at asymptotically high temperatures where the gluonic sector becomes weakly coupled. We will comment on the application of our results in different physical regimes in Section~\ref{S:discuss}.

In the rest of this work we expand on the observations of this section. We briefly review anomalies and Ward identities in Section~\ref{S:review1}, their manifestation in hydrodynamics in Section~\ref{S:review2}, and anomalous hydrostatics in Section~\ref{S:review3}. In Section~\ref{S:effectiveS} we discuss theories with a dynamical $U(1)_V$ gauge sector, formulating MHD, the prescription to compute correlation functions, and the three-dimensional effective theory (\ref{E:effSresult}) suitable for hydrostatics. We use this effective theory to compute the one-loop corrections to the chiral conductivities in Section~\ref{S:pert}. We conclude with a brief discussion in Section~\ref{S:discuss}.

\section{Non-dynamical background fields}
\label{S:review}

\subsection{Anomalous conservation laws}
\label{S:review1}
Let us begin by considering relativistic field theories in four dimensions. For now we will work with theories which are generally covariant and also possess $U(1)_V\times U(1)_A$ global symmetry up to anomalies. These symmetries imply the existence of Ward identities for the stress-energy tensor as well as the vector and axial currents. These Ward identities are independent of the state, and apply at finite temperature \cite{Itoyama:1982up}, and at finite chemical potential. To obtain them we first turn on background $U(1)_V\times U(1)_A$ gauge fields, which we label as $V_{\mu}$ and $A_{\mu}$ respectively, as well as a background metric $g_{\mu\nu}$. We then study the dependence of the generating functional $W=W[A,V,g]$ on those background fields. $W$ is related to the partition function as
\beq
W[A,V,g] = - i \ln Z[A,V,g].
\eeq
The symmetry currents and stress-energy tensor are defined through variations of $W$ with respect to the background fields. Denoting that variation as $\delta_s$ we have
\beq
\label{E:defCurrents}
\delta_s W = \int d^4x \sqrt{-g}\left[ \delta A_{\mu} J_A^{\mu}  + \delta V_{\mu} J_V^{\mu} + \frac{1}{2}\delta g_{\mu\nu} T^{\mu\nu}\right].
\eeq

Since the theory has $U(1)_V\times U(1)_A$ global symmetry as well as general covariance, it is invariant under independent $U(1)_V\times U(1)_A$ gauge transformations as well as diffeomorphisms. We collectively notate such a variation as $\delta_{\lambda}$, under which the sources transform as
\begin{align}
\begin{split}
\label{E:deltaLambda}
\delta_{\lambda} A_{\mu} &= \partial_{\mu} \Lambda_A + \Lie_{\xi} A_{\mu}, \\
\delta_{\lambda} V_{\mu} & = \partial_{\mu} \Lambda_V + \Lie_{\xi} V_{\mu}, \\
\delta_{\lambda} g_{\mu\nu} & = \Lie_{\xi} g_{\mu\nu},
\end{split}
\end{align}
where we have parametrized an infinitesimal diffeomorphism by $x^{\mu} \to x^{\mu} + \xi^{\mu}$ and $\Lie_{\xi}$ denotes the Lie derivative along the vector field $\xi^{\mu}$. Substituting the gauge and coordinate variation~\eqref{E:deltaLambda} into~\eqref{E:defCurrents} we find the variation of $W$,
\begin{align}
\begin{split}
\label{E:deltaW1}
\delta_{\lambda}W = -\int d^4x\sqrt{-g} &\left[ \Lambda_A\nabla_{\mu} J_A^{\mu} + \Lambda_V\nabla_{\mu} J_V^{\mu} \right.
\\
&\left.+ \xi_{\nu}\left(\nabla_{\mu} T^{\mu\nu} - F_V^{\nu\rho}J_{V,\rho}+V^{\nu}\nabla_{\rho}J_V^{\rho} - F_A^{\nu\rho}J_{A,\rho}+A^{\nu}\nabla_{\rho}J_A^{\rho}\right)\right]\,,
\end{split}
\end{align}
where we have defined $\nabla_{\mu}$ to be the covariant derivative with respect to the Levi-Civita connection
$\Gamma^{\mu}_{\nu\rho} = \frac{1}{2}g^{\mu\sigma}(\partial_{\nu}g_{\rho\sigma} + \partial_{\rho}g_{\nu\sigma} - \partial_{\sigma} g_{\nu\rho})$,
used the definition of the Lie derivative, and integrated by parts. We have also defined the curvatures of the axial and vector gauge fields as $F_{A,\mu\nu} = \partial_{\mu}A_{\nu}-\partial_{\nu}A_{\mu}$ and similarly for $F_V$. In the absence of anomalies, we demand that $\delta_{\lambda}W$~\eqref{E:deltaW1} vanishes for arbitrary $\Lambda_A,\Lambda_V,\xi_{\nu}$. This gives the usual Ward identities
\begin{align}
\begin{split}
\nabla_{\mu} J_A^{\mu} &= \nabla_{\mu}J_V^{\mu} = 0,
\\
\nabla_{\nu}T^{\mu\nu} & = F^{\mu\nu}_VJ_{V,\nu} + F^{\mu\nu}_A J_{A,\nu}.
\end{split}
\end{align}

Now suppose that our theory has anomalies, so that the generating functional is no longer invariant under gauge and coordinate transformations; see e.g.~\cite{BBook,Harvey:2005it,Bilal:2008qx} for reviews. In order for 
$W$ to obey the Wess-Zumino consistency condition~\cite{Wess:1971yu} , it turns out that the only possible anomalies are pure `flavor' anomalies involving three $U(1)$ currents, and mixed flavor-gravitational anomalies involving a single $U(1)$ current and two stress-energy tensors. In order to make contact with Standard Model physics, where we view $U(1)_V$ as (non-dynamical, for now) electromagnetism and $U(1)_A$ as a global (non-singlet) axial current, we consider $AVV$, $AAA$, and $ATT$ anomalies. The anomalous variation of $W$ may then be obtained from a differential six-form known as the anomaly polynomial $\mathcal{P}$. To write it down we parametrize the Levi-Civita connection as a matrix-valued one-form and the Riemann curvature as a matrix-valued two-form,
\begin{align}
\begin{split}
\Gamma\ab{\mu}{\nu} &= \Gamma^{\mu}_{\nu\rho}dx^{\rho}, \\
R\ab{\mu}{\nu} & = \frac{1}{2}R\ab{\mu}{\nu\rho\sigma}dx^{\rho}\wedge dx^{\sigma}.
\end{split}
\end{align}
In terms of forms, the Riemann curvature $R\ab{\mu}{\nu}$ is just the non-abelian curvature of $\Gamma\ab{\mu}{\nu}$,
\beq
R\ab{\mu}{\nu} = d\Gamma\ab{\mu}{\nu} + \Gamma\ab{\mu}{\rho}\wedge\Gamma\ab{\rho}{\nu}\,.
\eeq
We then parametrize $\mathcal{P}$ as
\beq
\mathcal{P} = c_g F_A\wedge F_V\wedge F_V  + \frac{\bar{c}_g}{3}F_A\wedge F_A\wedge F_A+ c_m F_A \wedge R\ab{\mu}{\nu}\wedge R\ab{\nu}{\mu}.
\eeq
The coefficients $c_g$ and $c_m$ quantify the strength of the flavor and mixed anomalies respectively. For a theory with chiral fermions $f$ and charges $q_{A,f}, q_{V,f}$ under the axial and vector gauge transformations, they are given by 
\begin{equation}
c_g = -\frac{1}{2!(2\pi)^2} \sum_{f} \chi_iq_{A,f} q_{V,f}^2\,,\qquad \bar{c}_g= -\frac{1}{2!(2\pi)^2} \sum_{f} \chi_fq_{A,f}^3\,,\qquad c_m = -\frac{1}{4!(8\pi^2)}\sum_{f}\chi_f q_{A,f}\,,
\end{equation}
where the sum is performed over fermion species and $\chi_f=\pm1$ indicates the chirality of the fermion (we assign right-handed fermions positive chirality).

The anomaly polynomial is closed, $d\mathcal{P} = 0$, and so may be written locally as the derivative of a five-form $I_{CS}$ which is defined up to a total derivative,
\beq
I_{CS} = c_g A\wedge F_V\wedge F_V +\frac{\bar{c}_g}{3}A\wedge F_A\wedge F_A+ c_m A \wedge R\ab{\mu}{\nu}\wedge R\ab{\nu}{\mu} + (\text{total derivative}).
\eeq
The gauge variation of the Chern-Simons form is exact,
\beq
\delta_{\lambda} I_{CS} = d G_{\lambda}.
\eeq
Taking the total derivative terms to vanish, $G_{\lambda}$ is simply given by
\beq
\label{E:G}
G_{\lambda} = \Lambda_A \left( c_g F_V\wedge F_V + \frac{\bar{c}_g}{3}F_A\wedge F_A+c_m R\ab{\mu}{\nu}\wedge R\ab{\nu}{\mu}\right),
\eeq
which is notably both $U(1)_V$ and coordinate invariant.

The anomalous variation of $W$ is related to the Chern-Simons form by
\beq
\delta_{\lambda} W = - \int G_{\lambda}.
\eeq
Using~\eqref{E:deltaW1} and~\eqref{E:G} we then obtain the anomalous Ward identities
\begin{align}
\label{E:ConWard1}
\nonumber
\nabla_{\mu} J_A^{\mu} & = \frac{1}{4}\epsilon^{\mu\nu\rho\sigma}\left[ c_g F_{V,\mu\nu}F_{V,\rho\sigma}+ \frac{\bar{c}_g}{3}F_{A,\mu\nu}F_{A,\rho\sigma}+ c_m R\ab{\alpha}{\beta\mu\nu}R\ab{\beta}{\alpha\rho\sigma}\right]\,,
\\
\nabla_{\mu} J_V^{\mu} & = 0\,,
\\
\nonumber
\nabla_{\nu}T^{\mu\nu} & = F^{\mu\nu}_V J_{V,\nu} + F^{\mu\nu}_A J_{A,\nu} - \frac{1}{4}A^{\mu}\epsilon^{\nu\rho\sigma\tau}\left[ c_g F_{V,\nu\rho}F_{V,\sigma\tau} +\frac{\bar{c}_g}{3}F_{A,\nu\rho}F_{A,\sigma\tau}+ c_m R\ab{\alpha}{\beta\nu\rho}R\ab{\beta}{\alpha\sigma\tau}\right]\,,
\end{align}
where the four-dimensional Levi-Civita tensor satisfies $\ep^{0123} = 1/\sqrt{-g}$. However, had we added a total derivative to $I_{CS}$,
$G_{\lambda}$ would be modified, and we would have obtained a different set of anomalous Ward identities. For instance, suppose we redefined the Chern-Simons 
term by
\beq
I'_{CS} = I_{CS} + d(c_s A\wedge V \wedge F_V),
\eeq
which is neither $U(1)_A$ nor $U(1)_V$ invariant. This redefinition leads to a modified $G_{\lambda}$,
\beq
G'_{\lambda} = G_{\lambda} + c_s \left[ -\Lambda_A F_V\wedge F_V + \Lambda_V F_A \wedge F_V\right],
\eeq
which in turn yields modified Ward identities for the currents,
\begin{align}
\begin{split}
\label{E:ConWard2}
\nabla_{\mu} J_A^{\mu}&= \frac{1}{4}\epsilon^{\mu\nu\rho\sigma} \left[ (c_g-c_s) F_{V,\mu\nu}F_{V,\rho\sigma} +\frac{\bar{c}_g}{3}F_{A,\mu\nu}F_{A,\rho\sigma}+ c_m R\ab{\alpha}{\beta\mu\nu}R\ab{\beta}{\alpha\rho\sigma}\right],
 \\
\nabla_{\mu} J_V^{\mu} & = \frac{c_s}{4}\epsilon^{\mu\nu\rho\sigma}F_{A,\mu\nu}F_{V,\rho\sigma}.
\end{split}
\end{align}
Note that depending on the choice of $c_s$, we can shift the $AVV$ anomaly from the axial current to the vector sector. There is a similar term which may be used to shift the mixed anomaly from the non-divergence of the $U(1)_A$ current to the non-divergence of the stress-energy tensor. However, the $AAA$ anomaly is not mixed and so there is no analogue of $c_s$ that may be used to eliminate the consistent anomaly for $\bar{c}_g$.

The parameter $c_s$ simply corresponds to the choice of a local contact term in the theory. Its effect on the Ward identities may be accounted for by redefining the generating functional by an additive term,
\beq
W' = W - c_s \int A \wedge V \wedge F_V.
\eeq
For theories with a functional integral description, this shift simply corresponds to redefining the action by the same additive term, which factors out of the functional integral since it is built out of background fields alone. As a result the parameter $c_s$ just corresponds to a contact term which is part of the definition of the theory. This is an example of a so-called Bardeen counterterm~\cite{Bardeen:1984pm}.

Thus far we have been discussing the dynamics of the \emph{consistent} currents which follow from varying $W$. They are named consistent because they obey the Wess-Zumino consistency condition~\cite{Wess:1971yu}. However, they are both anomalous and non-covariant under gauge and coordinate transformations. To see this, consider taking two successive variations of $W$, where the first variation is a gauge and coordinate transformation, and the second is a general variation of the background fields. That is we consider $\delta_s \delta_{\lambda} W = - \delta_s \int G_{\lambda}$. Since $G_{\lambda}$ generically depends on $F_A, F_V$, and $R\ab{\mu}{\nu}$ it follows that this double variation is nonzero. However, the variations commute by the Wess-Zumino consistency condition, giving
\beq
-\delta_s \int G_{\lambda} = \delta_{\lambda} \delta_s W =  \int d^4x \sqrt{-g} \left[ \delta A_{\mu} \delta_{\lambda}J_A^{\mu} + \delta V_{\mu} \delta_{\lambda}J_V^{\mu} + \frac{1}{2}\delta g_{\mu\nu} \delta_{\lambda}T^{\mu\nu}\right].
\eeq
Indeed, by integrating the gauge and coordinate variations of the currents and stress-energy tensor, one finds that they are covariant up to additive terms built out of the background fields. These terms are known as Bardeen-Zumino (BZ) polynomials~\cite{Bardeen:1984pm} and do not follow from the variation of any local four-dimensional functional.

We then have the freedom to redefine our symmetry currents and stress tensor in such a way as to subtract off these non-covariant parts. These new objects are the so-called \emph{covariant} currents and stress tensor, which unlike the \emph{consistent} currents are neither consistent nor follow from the variation of a generating functional. They are given by
\begin{align}
\begin{split}
\label{E:CovJ}
J_{A,cov}^{\mu}  & = \frac{1}{\sqrt{-g}}\frac{\delta W}{\delta A_{\mu}} + P_{BZ,A}^{\mu}\,,
\\
J_{V,cov}^{\mu} & = \frac{1}{\sqrt{-g}}\frac{\delta W}{\delta V_{\mu}} + P_{BZ,V}^{\mu}\,,
\\
T_{cov}^{\mu\nu} & = \frac{2}{\sqrt{-g}}\frac{\delta W}{\delta g_{\mu\nu}} + P_{BZ}^{\mu\nu}\,.
\end{split}
\end{align}
Ignoring the mixed anomaly for now, for the choice of $W$ we used in writing down the anomalous Ward identities~\eqref{E:ConWard2} the BZ polynomials for the currents become
\begin{align}
\begin{split}
\label{E:BZ}
P_{BZ,A}^{\mu} & = \epsilon^{\mu\nu\rho\sigma} \left[c_sV_{\nu}\partial_{\rho}V_{\sigma} + \frac{2\bar{c}_g}{3}A_{\nu}\partial_{\rho}A_{\sigma}\right]\,,
\\
P_{BZ,V}^{\mu} & =\epsilon^{\mu\nu\rho\sigma} \left[ 2(c_g - c_s) A_{\nu}\partial_{\rho}V_{\sigma} + c_s V_{\nu}\partial_{\rho}A_{\sigma}\right]\,.
\end{split}
\end{align}
By adding these polynomials to the consistent currents~\eqref{E:CovJ} and using the anomalous Ward identities~\eqref{E:ConWard2} for the consistent currents, we thereby find the anomalous Ward identities for the covariant currents. These are given by (see e.g.~\cite{Jensen:2012kj} for a derivation of the Ward identity for the stress tensor)
\begin{align}
\begin{split}
\label{E:covWard}
\nabla_{\mu} J_{A,cov}^{\mu} & = \frac{1}{4}\epsilon^{\mu\nu\rho\sigma}\left[ c_g F_{V,\mu\nu}F_{V,\rho\sigma} +\bar{c}_g F_{A,\mu\nu}F_{A,\rho\sigma}+ c_m R\ab{\alpha}{\beta\mu\nu}R\ab{\beta}{\alpha\rho\sigma}\right]\,,
\\
\nabla_{\mu} J_{V,cov}^{\mu} & = \frac{c_g}{2}\epsilon^{\mu\nu\rho\sigma} F_{A,\mu\nu}F_{V,\rho\sigma}\,,
\\
\nabla_{\nu}T_{cov}^{\mu\nu} & = F^{\mu}_{V,\nu}J_{V,cov}^{\nu} + F^{\mu}_{A,\nu} J_{A,cov}^{\nu} + \frac{c_m}{2}\nabla_{\nu}\left[ \epsilon^{\alpha\beta\gamma\delta} F_{A,\alpha\beta} R\ab{\mu\nu}{\gamma\delta}\right]\,.
\end{split}
\end{align}
The covariant Ward identities depend only on the parameters $c_g$, $\bar{c}_g$, and $c_m$ of the anomaly polynomial, and not on $c_s$ or any other local counterterm.

\subsection{Anomalous hydrodynamics}
\label{S:review2}
Now consider heating up the theories of the previous section to a temperature $T$ and possibly turning on chemical potentials $\mu_V$ and $\mu_A$ for the vector and axial currents respectively. We assume that the resulting thermal state is translationally and rotationally invariant. The long-wavelength dynamics of such a theory are often well-described by relativistic hydrodynamics~\cite{LL6}, which one may think of as the effective theory for the gapless collective modes describing the relaxation of conserved quantities.

In order to formulate hydrodynamics one begins with the parameters that label the equilibrium state -- the temperature $T$, the chemical potentials $\mu_V$ and $\mu_A$, and a local timelike velocity $u^{\mu}$ (normalized to $u^2 = -1$) -- and promotes them to become classical space-time fields. We remind the reader that the chemical potential $\mu_A$ is conjugate to the consistent (non-conserved) axial current. These are termed the hydrodynamic variables. We then subject the theory to $O(1)$ background gauge fields and an $O(1)$ metric, which possess gradients much longer than the inverse temperature. In the gradient expansion~\cite{Bhattacharyya:2008jc,Baier:2007ix}, a field strength is then $O(\partial)$ and the Riemann curvatures are $O(\partial^2)$. This is the correct scaling required to study the response of the fluid in the source-free equilibrium state. The next step is to express the one-point functions of the currents and stress tensor in a gradient expansion of the hydrodynamic variables and background fields. These are the constitutive relations of hydrodynamics. Third, one enforces the Ward identities as equations of motion, which uniquely determine the hydrodynamic variables up to boundary conditions and initial data.  Finally, one demands a local version of the second law of thermodynamics, namely the existence of an entropy current with positive divergence.

Coupling the theory to background fields is not necessary to study hydrodynamics, but it is eminently useful for two reasons. First, demanding consistency in the presence of background fields provides additional constraints on the source-free hydrodynamics. For instance, we will later see that the chiral vortical conductivity is constrained in just this way. Second, by turning on sources we may compute correlation functions in the hydrodynamic limit and so match hydrodynamics to field theory.

In real-time finite temperature field theory, there are different types of correlation functions with various time orderings. These may be described with the closed-time-path (CTP) formalism from which one defines the CTP generating functional $W_{CTP}$ (see~\cite{Wang:1998wg} for a review). In the CTP formalism, one extends the time contour by first going from $t_1\in (-\infty,+\infty)$ and then doubling back as $t_2\in(+\infty,-\infty)$. One then introduces sources on both infinite segments of the time contour $J_1$ and $J_2$, from which one defines the linear combinations $J_r = (J_1+J_2)/2$ and $J_a = J_1-J_2$. The $r$-type sources couple to $a$-type operators, while $a$-sources couple to $r$-operators. The fully retarded functions are the $ra..a$ functions, which are $n$-point functions with a single $r$ operator and the rest of $a$-type. These are the correlation functions that are directly accessible in hydrodynamics, see e.g.~\cite{Moore:2010bu}. We regard the one-point functions in the constitutive relations as the one-point functions of the $r$ currents and stress tensor, expressed in terms of the hydrodynamic variables (which we may interpret as auxiliary fields whose purpose is to give a local representation of the constitutive relations) and the background fields. We take the latter to be $r$-type sources.

In order to compute these correlation functions, one solves the hydrodynamic equations of motion in the presence of background fields. The solution gives the hydrodynamic variables as functionals of the sources, which may then be plugged back into the constitutive relations. We thus find the one-point functions of the $r$ currents and stress tensor in the presence of background fields. The $ra..a$ functions are then defined by variation.

When we have anomalies, we must specify which currents and therefore which Ward identities we study in hydrodynamics. In the literature, it has been standard practice to study the covariant currents. These obey the covariant Ward identities and, since they are covariant, may be expressed in terms of covariant constitutive relations. However since the covariant and consistent currents are simply related by the BZ polynomials Eq.~\eqref{E:BZ}, the consistent constitutive relations (which obey the consistent Ward identities) are simply the covariant ones minus the BZ polynomials.

For our theories, the most general constitutive relations for the covariant currents and stress tensor are
\begin{align}
\begin{split}
\label{E:onePointFns}
J_{A,cov}^{\mu} & = \mathcal{N}_A u^{\mu} + \nu_A^{\mu},
\\
J_{V,cov}^{\mu} & = \mathcal{N}_V u^{\mu} + \nu_V^{\mu},
\\
T^{\mu\nu}_{cov} & = \mathcal{E} u^{\mu} u^{\nu} + \mathcal{P} \Delta^{\mu\nu} + u^{\mu} q^{\nu}+u^{\nu} q^{\mu} + \tau^{\mu\nu},
\end{split}
\end{align}
where
\beq
u^{\mu}q_{\mu} = u^{\mu}\tau_{\mu\nu}=g^{\mu\nu}\tau_{\mu\nu}=0\,,
\eeq 
and we have defined the transverse projector $\Delta^{\mu\nu} = g^{\mu\nu}+u^{\mu}u^{\nu}$. We have also decomposed the currents and stress tensor into irreducible representations of the residual rotational invariance which fixes $u^{\mu}$. To first order in the gradient expansion, the scalars in~\eqref{E:onePointFns} are
\begin{subequations}
\label{E:constRel1}
\begin{align}
\mathcal{N}_A & = \rho_A, & \mathcal{N}_V & = \rho_V,& \mathcal{P} & = p - \zeta \nabla_{\mu}u^{\mu}, & \mathcal{E} & = -p +T s+ \mu_A \rho_A + \mu_V \rho_V,
\end{align}
where $p$ is the thermodynamic pressure, $s$ is the entropy density $(\partial p/\partial T)|_{\mu_A,\mu_V}$, and the charge densities are  $(\rho_A=\partial p/\partial \mu_A)|_{T,\mu_V}$ and $(\rho_V = \partial p/\partial \mu_V)|_{T,\mu_A}$. Collectively denoting the axial and vector currents with an index $a,b=A,V$, the vectors are\footnote{As discussed below, in writing~\eqref{E:constRel1} we are choosing a particular hydrodynamic frame which is the natural one that follows from the generating functional of zero-frequency correlation functions~\cite{Jensen:2012jh}.}
\begin{align}
\begin{split}
\label{E:constV}
\nu_a^{\mu} &= \sigma_{ab} \left( E_b^{\mu} - T \Delta^{\mu\nu}\partial_{\nu}\left(\frac{\mu_b}{T}\right)\right) +\chi_{E,ab}E_b^{\mu} + \chi_{T,a}\Delta^{\mu\nu}\partial_{\nu}T+ \xi_{ab} B_b^{\mu} + \xi_a w^{\mu},
\\
q^{\mu} & = \xi_{a} B_a^{\mu} + \xi w^{\mu}.
\end{split}
\end{align}
Finally the only tensor is
\beq
\tau^{\mu\nu} = - 2\eta \,\sigma^{\mu\nu}.
\eeq
\end{subequations}
In the expressions above we have implicitly defined
\begin{subequations}
\label{E:hydroDefs}
\begin{align}
\sigma^{\mu\nu} & = \frac{\Delta^{\mu\rho}\Delta^{\nu\sigma}}{2}\left( \nabla_{\rho}u_{\sigma}+\nabla_{\sigma}u_{\rho} - \frac{2}{3}g_{\rho\sigma}\nabla_{\alpha}u^{\alpha}\right),
& w^{\mu} = \epsilon^{\mu\nu\rho\sigma}u_{\nu}\nabla_{\rho}u_{\sigma},
\\
E_a^{\mu} & = F_a^{\mu\nu}u_{\nu}, & B_a^{\mu} = \frac{1}{2}\epsilon^{\mu\nu\rho\sigma}u_{\nu}F_{a,\rho\sigma},
\end{align}
\end{subequations}
where $w^{\mu}$ is the local vorticity of the plasma and $E_a^{\mu}$ and $B_a^{\mu}$ are the electric and magnetic fields in the local rest frame. 

Demanding the existence of an entropy current with positive divergence further restricts the coefficients in the constitutive relations. It gives the equality-type relations~\cite{Son:2009tf,Neiman:2010zi,Banerjee:2012iz,Jensen:2012jy} (note that our anomaly coefficient $c_{abc}$ is related to that in Son \& Surowka \cite{Son:2009tf} by $c_{abc} = -  C^{abc}_{S/S}/2$)
\begin{align}
\begin{split}
\label{E:equalityCon}
\chi_{T,a} &=0\,, \qquad \chi_{E,ab}  =0\,, \\
\xi_{ab} &= - 2 c_{abc}\mu_c + \tilde{c}_{ab} T\,, \\
\xi_a & = - c_{abc} \mu_b \mu_c +\tilde{c}_{ab}\mu_a T +  \tilde{c}_a T^2\,, \\
\xi & = -\frac{2}{3} c_{abc}\mu_a\mu_b\mu_c +\tilde{c}_{ab}\mu_a\mu_b T+ 2\tilde{c}_a \mu_a T^2+\tilde{c}T^3\,,
\end{split}
\end{align}
where $c_{abc}$ is the totally symmetric anomaly coefficient built out of $c_g$ and $\bar{c}_g$. It has nonzero components $c_{AVV}=c_{VAV}=c_{VVA} = c_g$ and $c_{AAA} = \bar{c}_g$. The $\tilde{c}_{ab}, \tilde{c}_a$, and $\tilde{c}$ are (at this stage) undetermined constants.
The first equality in the first line is well known~\cite{LL6}, while the second equality was only recently established in~\cite{Son:2009tf}. The terms proportional to $c_{abc}$ in $\xi_a$ and $\xi$ were first discovered in hydrodynamics via AdS/CFT~\cite{Erdmenger:2008rm,Banerjee:2008th}; these (along with the terms $\propto c_{abc}$ in $\xi_{ab}$) were later understood more generally from an entropy analysis in hydrodynamics~\cite{Son:2009tf}. The existence of the constants $\tilde{c}_a$ and $\tilde{c}$ was noted in~\cite{Neiman:2010zi}, and the constant $\tilde{c}_{ab}$ was identified in~\cite{Banerjee:2012iz,Jensen:2012jy}.

Most of the $\tilde{c}$'s violate CPT. To see this we consider the transformation properties of the various fields under C, P, and T in Table~\ref{T:CPT}. It follows that $\tilde{c}_{ab}$ and $\tilde{c}$ are CPT-violating, while the $\tilde{c}_a$ are CPT-preserving. We will set the CPT-violating constants $\tilde{c}_{ab}=\tilde{c}=0$ for the rest of this section.

\begin{table}
\begin{center}
\begin{tabular}{|c|c|c|c|}
\hline
field & C & P & T \\
\hline
$T,g_{00}$ & + & + & + \\
\hline
$\mu_V,V_0$ & - & + & + \\
\hline
$V_i$ & - & - & - \\
\hline
$\mu_A,A_0$ & + & - & + \\
\hline
$A_i$ & + & + & - \\
\hline
$u^i,g_{0i}$ & + & - & - \\
\hline
$g_{ij}$ & + & + & + \\
\hline
\end{tabular}
\end{center}
\caption{\label{T:CPT} Transformation properties of the hydrodynamic variables and sources under C, P, and T.}
\end{table}

Due to the presence of the anomaly coefficient $c_{abc}$ in $\xi_{ab},\xi_a$, and $\xi$, the latter are sometimes referred to as describing anomaly-induced transport. Note that at this stage the anomaly-induced transport is described by six \emph{a priori} independent coefficients $\{\xi_{AA},\xi_{AV}=\xi_{VA},\xi_{VV},\xi_A,\xi_V,\xi\}$ which are in fact determined by four numbers, the AVV and AAA anomaly coefficients $c_g$ and $\bar{c}_g$, the CPT-preserving constants $\tilde{c}_a$, and thermodynamic quantities.

After imposing~\eqref{E:equalityCon}, the divergence of the entropy current $s^{\mu}$ is~\cite{Son:2009tf}
\beq
\label{E:divS}
\nabla_{\mu} s^{\mu} = \frac{\zeta}{T}\theta^2 + \frac{\sigma_{ab}}{T}V^{\mu}_a V_{b,\mu} + \frac{\eta}{T}\sigma^{\mu\nu}\sigma_{\mu\nu},
\eeq
where we have defined
\beq
\label{E:defsThetaV}
\theta = \nabla_{\mu}u^{\mu}, \qquad V^{\mu}_a = E^{\mu}_a - T \Delta^{\mu\nu}\partial_{\nu}\left(\frac{\mu_a}{T}\right).
\eeq
Since both $V^{\mu}_a$ and $\sigma^{\mu\nu}$ are spacelike tensors, their squares are positive definite. In order for the right-hand-side of~\eqref{E:divS} to be positive, which we interpret as the positivity of entropy production, we must enforce the standard inequality-type constraints on the remaining transport coefficients
\beq
\zeta \geq 0, \qquad ||\sigma_{ab}|| \geq 0, \qquad \eta\geq 0,
\eeq
where by the second entry we mean that $\sigma_{ab}$ must be a positive-definite matrix. It is then clear that the quantities $\xi_{ab}, \xi_a,$ and $\xi$ are dissipationless parameters, while $\zeta, \sigma_{ab}$, and $\eta$ are dissipative transport coefficients. More precisely, the symmetric part of $\sigma_{ab}$ is dissipative, while the antisymmetric part is dissipationless.\footnote{We pause to note something which we have not seen previously discussed in the literature. Namely, the antisymmetric part of $\sigma_{ab}$ is an interesting object: it is a dissipationless quantity which moreover characterizes real-time, out-of-equilibrium transport. In these ways it is somewhat akin to the Hall viscosity or the anomalous Hall conductivity, which also characterize dissipationless out-of-equilibrium transport in $2+1$-dimensions~\cite{Jensen:2011xb}. However, by generalizing Onsager's relations we find that the antisymmetric part of $\sigma_{ab}$ is somewhat more complicated than say the Hall viscosity. It violates T but preserves C and P, and so violates CPT. This is similar to but distinct from a chemical potential, which violates C but preserves T and P. In contrast the Hall viscosity preserves C, but violates P and T, and so it preserves CPT and thus may be nonzero in a source-free parity-violating phase.} The quantities $\zeta$ and $\eta$ are the usual bulk and shear viscosities, while $\sigma_{ab}$ is the matrix of conductivities. 

\subsection{Anomalous hydrostatics}
\label{S:review3}
Recently the equality-type constraints~\eqref{E:equalityCon} that follow from demanding an entropy current were obtained without the use of an entropy current or even of hydrodynamics~\cite{Jensen:2012jh,Jensen:2012jy,Banerjee:2012iz}. The major step in that work is the study of zero-frequency, low-momentum correlation functions. That is, it is useful to study theories in the \emph{hydrostatic} limit. Normally, nonzero temperature leads to a finite static correlation length and thus screening.\footnote{The notable exceptions are theories in a superfluid phase, which have a propagating Goldstone mode, theories with dynamical $U(1)$ gauge fields like those we study later in this work, and theories tuned to a critical point.} That is, static correlation functions of all operators fall off exponentially at long distance, which in momentum space corresponds to the statement that zero-frequency functions are analytic at zero momentum. It then follows that the generating functional of zero-frequency correlation functions $\Wstatic$ may be written as a local functional in a derivative expansion,
\begin{equation*}
\Wstatic = \sum_n W_n\ + W_{anom}\,.
\end{equation*}
We collectively notate the contributions to this functional with $n$ derivatives as $W_n$, where the $W_n$'s are invariant under all symmetries and $W_{anom}$ reproduces the anomalous variation. This expansion will of course only have at best a finite radius of convergence, up to momenta corresponding to the inverse screening length. The resulting object is proportional to the Euclidean generating functional evaluated for stationary background fields.

Theories which are gauge and diffeomorphism invariant will have a generating functional involving local gauge and diffeomorphism-invariant scalars built out of the background fields. In addition to the background fields themselves, those scalars may depend on quantities that involve a timelike vector field $K^{\mu}$ which covariantly defines what we mean by time. More precisely, we consider backgrounds where the Lie derivative of $K$, $\Lie_K$, annihilates the background gauge fields and metric. As a result $K$ generates a timelike isometry. We define time through the integral curves of $K$. Since the background is time-independent, we may define a thermal partition function in the usual way after Euclideanizing time and compactifying the resulting time circle with coordinate periodicity $\beta$.

There are then additional gauge-invariant tensors involving $K$. We define the suggestively named
\beq
T^{-1} = \int_0^{\beta} d\tau \sqrt{-K^2(\tau)}, \qquad u^{\mu} = \frac{K^{\mu}}{\sqrt{-K^2}}, \qquad \mu_a = T \int A_a,
\eeq
where the holonomy in the last expression is around the time circle and $\tau$ is an affine parameter along the time circle. These quantities are independent but their derivatives are not. These satisfy some differential interrelations which follow from the fact that $K$ generates a symmetry of the background,
\beq
\nabla_{\mu}u_{\nu} = - u_{\mu}a_{\nu} + \omega_{\mu\nu}, \qquad (\nabla_{\mu}+a_{\mu})T=0, \qquad (\nabla_{\mu}+a_{\mu})\mu_a = E_{a,\mu},
\eeq
where we have defined local acceleration and local vorticity
\beq
a_{\mu} = u^{\nu}\nabla_{\nu}u_{\mu}, \qquad \omega_{\mu\nu} = \frac{\Delta_{\mu\rho}\Delta_{\nu\sigma}}{2}(\nabla^{\rho}u^{\sigma} - \nabla^{\sigma}u^{\rho}).
\eeq
Note that the tensor structures which correspond to dissipation -- the shear tensor $\sigma^{\mu\nu}$,  expansion $\nabla_{\mu}u^{\mu}$, and the vector combinations $(\nabla_{\mu}+a_{\mu})T$ and $E^{\mu}_a - T\Delta^{\mu\nu}\partial_{\nu}\left( \frac{\mu_a}{T}\right)$ -- all vanish, corresponding to the fact that we are indeed studying the theory in a stationary equilibrium. The most general gauge-invariant tensor is built out of these quantities, the curvatures $F_{A,\mu\nu}$ and $F_{V,\mu\nu}$, the Riemann tensor $R\ab{\mu}{\nu\rho\sigma}$, and covariant derivatives thereof.

By varying $W_n$ with respect to sources we obtain the one-point functions of operators as a functional of background fields with terms that include $n$ derivatives. However, since we are studying real-time finite temperature field theory we should be careful to specify the correlation functions that are computed from $W_n$, whether $ra..a$ functions or otherwise. Fortunately, it turns out that this caution is unnecessary for the following reason. The correlation functions computed from $W_n$ lead to zero-frequency functions upon Fourier transform. Any such zero-frequency function, whether the fully retarded $ra..a$ functions or the fully symmetrized $r..r$ functions, is proportional to the corresponding Euclidean zero-frequency function~\cite{Evans:1991ky}.

As a result we may regard the variations of $W_n$ as giving $ra..a$ functions at zero frequency with $n$ factors of momentum. These same correlation functions are computed in hydrodynamics and so we may match the two, thereby relating parameters of $W_n$ to $n^{th}$ order hydrodynamic coefficients. From the perspective of hydrodynamics, the resulting one-point functions express the constitutive relations in a specific hydrodynamic {\it frame} (a definition of hydrodynamic variables such as $T$, $u^\mu$, see e.g.~\cite{Bhattacharya:2011tra}). This frame has been termed the {\it thermodynamic frame}~\cite{Jensen:2012jh}, and exhibits several important properties.  One is the  fact that coefficients that appear in $W_n$ encode thermodynamic (or hydrostatic) response coefficients in the constitutive relations with $n$ (and only $n$) derivatives~\cite{Jensen:2012jh}. Since we study the response of the source-free thermal state to long-wavelength background sources, this implies a direct matching between the derivative expansion of the generating functional to the derivative expansion of hydrodynamics. Furthermore, the anomaly-induced response described by $\Wstatic$ (see $W_{anom}$ below in (\ref{E:Wanom})), which includes terms from the abelian anomaly with one derivative and the mixed anomaly with three derivatives, matches terms in the constitutive relations with exactly one~\cite{Banerjee:2012iz,Loganayagam:2011mu} and three derivatives~\cite{Jensen:2012kj}. In other hydrodynamic frames, e.g. the Landau frame, the parameters appearing in $W_n$ will appear at $n^{th}$ and generally {\it all} higher orders in the gradient expansion~\cite{Kharzeev:2011ds,Bhattacharyya:2013ida}. We implicitly work in the thermodynamic frame for the rest of this subsection.

At zeroth order in derivatives, the only gauge-invariant scalars are the local temperature $T$ and chemical potentials $\mu_a$, and so the most general gauge-invariant scalar is an arbitrary function of these which we call $p(T,\mu_a)$. At first order in derivatives, it turns out that the the only gauge-invariant scalars are terms which are analogous to Chern-Simons terms in that their gauge and coordinate variation is a total derivative. Only keeping the CPT-preserving one-derivative terms, we have
\beq
\label{E:W0W1}
W_0 = \int d^4x\sqrt{-g}\, p(T,\mu_a), \qquad W_1 = \int d^4x \sqrt{-g}\left( \tilde{c}_A A_{\mu} + \tilde{c}_V V_{\mu}\right) T^2 w^{\mu},
\eeq
where $w^{\mu}$ is constructed from derivatives of $u_{\mu}$ in the same way as in~\eqref{E:hydroDefs} and the $\tilde{c}_a$ are suggestively named constants. Remarkably, if we pick a coordinate and gauge choice in which $K = \partial_t$ and the background fields are explicitly time-independent,\footnote{Note that we study theories subjected to real background fields in Lorentzian signature. The corresponding Euclideanized background fields are necessarily complex.}
\begin{align}
\begin{split}
\label{E:fields}
A & = A_0(x)(dt+\mathfrak{a}_i(x)dx^i) + \hat{A}_i(x)dx^i,
\\
V & = V_0(x)(dt+\mathfrak{a}_idx^i) + \hat{V}_i(x)dx^i,
\\
g & = - e^{2\mathfrak{s}(x)}(dt+\mathfrak{a}_idx^i)^2 + \hat{g}_{ij}(x)dx^idx^j,
\end{split}
\end{align}
then we can write down a \emph{local} functional whose gauge and coordinate variation gives the correct anomalous variation of $\Wstatic$. For the definition of consistent currents in~\eqref{E:ConWard2}, that functional is~\cite{Banerjee:2012iz}
\beq
\label{E:Wanom}
W_{anom} = -2 \int (dt) \wedge \hat{A} \wedge \left[ c_g V_0\left( d\hat{V} + \frac{V_0}{2}d\mathfrak{a}\right) +\frac{\bar{c}_g}{3}A_0\left(d\hat{A}+\frac{A_0}{2}d\mathfrak{a}\right)\right]-c_s\int A\wedge V \wedge F_V + W_{grav},
\eeq
where $(dt)=-i\beta$ as we mentioned in the Introduction, and $W_{grav}$ is a complicated functional with three derivatives. Its precise expression is given in~\cite{Jensen:2012kj} and is (thankfully) unimportant for this work. Before going on, we note that the functional $W_{anom}$ reproduces the correct anomalous variation \emph{independently} of the gradient expansion and so goes beyond the hydrostatic limit. It is an exact part of the zero-frequency generating functional. 

This gauge and coordinate choice also manifests that the terms in $W_1$~\eqref{E:W0W1} which involve the $\tilde{c}_a$ are rather special. They may be written as Chern-Simons forms on the spatial slice~\cite{Banerjee:2012iz},
\beq
\label{E:W1asCS}
W_1 =- \frac{1}{\beta^2}\int (dt) \wedge (\tilde{c}_V \hat{V}+\tilde{c}_A \hat{A})\wedge d\mathfrak{a}\,,
\eeq
where the factors of $\beta$ are required by dimensional analysis. The reader may note that there are other Chern-Simons terms we may have added to $W_1$, proportional to $\hat{A}\wedge d\hat{A}, \hat{V}\wedge d\hat{V}, \hat{A} \wedge d\hat{V}$, and $\mathfrak{a}\wedge d\mathfrak{a}$. However all of these terms violate \emph{four-dimensional} CPT, and we will set them to zero in this Section. Despite the fact that these terms violate CPT, we note that they are still related to the entropy current analysis. Indeed a short calculation shows that these Chern-Simons terms correspond precisely to the CPT-violating coefficients $\tilde{c}_{ab}$ and $\tilde{c}$ in~\eqref{E:equalityCon}. 

The covariant currents which follow from the variation of $W_0+W_1+W_{anom}$ by~\eqref{E:CovJ} are precisely those~\eqref{E:onePointFns} we discussed earlier in hydrodynamics. They have expressions of the form~\eqref{E:constRel1} after setting the dissipative tensor structures to vanish, i.e. taking $\{\theta,V^{\mu}_a,(\nabla_{\mu}+a_{\mu})T,\sigma^{\mu\nu}\}\to 0$ (see~\eqref{E:defsThetaV} for the definitions of $\theta$ and $V^{\mu}_a$). Most importantly, the remaining coefficients in~\eqref{E:constRel1} are related to the parameters in $W_0+W_1+W_{anom}$ by the \emph{same} equality-type relations~\eqref{E:equalityCon} that originally came from demanding the existence of an entropy current. More simply, the hydrostatic generating functional independently derives the equality-type relations, including those involving the anomalies, without reference to hydrodynamics.

We can characterize the anomaly-induced response coefficients $\xi_{ab}, \xi_a$, and $\xi$ via Kubo formulae. That is, by computing the appropriate correlation functions in hydrodynamics or by varying the generating functional, we may evaluate $\{\xi_{ab}, \xi_a,\xi\}$ in a given theory. One useful set of Kubo formulae for these coefficients is given by simply varying the generating functional twice to obtain zero-frequency two-point functions, that is by studying the two-point functions of the consistent currents. 

However, the usual two-point functions in the literature involve the variation of a \emph{covariant} current with respect to background fields, which gives the mixed two-point function of a covariant current with a consistent one. For instance we have
\beq
\langle J_{a,cov}^i(k) J_b^j(-k)\rangle =\frac{1}{\sqrt{-g}}\frac{\delta \langle J_{a,cov}^i(k)\rangle}{\delta A_{b,j}(k)}\,,
\eeq
Computing two-point functions of this type in hydrostatics leads to Kubo formulae for the chiral conductivities (see also~\cite{Amado:2011zx})
\begin{align}
\begin{split}
\lim_{k\to 0} \frac{i \epsilon^{ijk}k_k}{2k^2}\langle J_{a,cov}^i(k)J_{b}^j(-k)\rangle &= \xi_{ab} ,
\\
\lim_{k\to 0} \frac{i \epsilon^{ijk}k_k}{2k^2}\langle J_{a,cov}^i(k) T^{0j}(-k)\rangle&= \xi_a,
\\
\lim_{k\to 0} \frac{i \epsilon^{ijk}k_k}{2k^2}\langle T_{cov}^{0i}(k)J^j(-k) \rangle&= \xi_a,
\\
\lim_{k\to 0}\frac{i\epsilon^{ijk}k_k}{2k^2}\langle T_{cov}^{0i}(k)T^{0j}(-k)\rangle &= \xi.
\end{split}
\end{align}
The two-point functions of consistent currents are slightly different, owing to the variation of the BZ polynomials. We have instead
\begin{align}
\begin{split}
\label{E:KuboCon}
\lim_{k\to 0} \frac{i \epsilon^{ijk}k_k}{2k^2}\langle J_{a}^i(k)J_{b}^j(-k)\rangle &= \xi_{ab} +\delta \xi_{ab},
\\
\lim_{k\to 0} \frac{i \epsilon^{ijk}k_k}{2k^2}\langle J_{a}^i(k) T^{0j}(-k)\rangle &= \xi_a,
\\
\lim_{k\to 0}\frac{i\epsilon^{ijk}k_k}{2k^2}\langle T^{0i}(k)T^{0j}(-k)\rangle &= \xi,
\end{split}
\end{align}
where $\delta \xi_{ab}$ is the symmetric matrix
\beq
\delta \xi_{ab} = \left(\begin{array}{cc}   \frac{2\bar{c}_g}{3}A_0& c_s V_0 \\ c_s V_0 & 2(c_g-c_s)A_0  \end{array}\right),
\eeq
and we take the ordering to be $a,b=A,V$. Note that the two-point function $\langle T^{0i}(k)J_a^j(-k)\rangle$ is related to $\langle J_a^i(k)T^{0j}(-k)\rangle$ by $i\leftrightarrow j$, $k\to -k$. The quantities $V_0$ and $A_0$ are the background values of the time-components of $V_{\mu}$ and $A_{\mu}$, which in a flat-space equilibrium are related to $\mu_V$ and $\mu_A$ by $V_0=\mu_V$ and $A_0=\mu_A$.

We can summarize the anomaly-induced response in terms of a $3\times 3$ matrix of zero-frequency `conductivities', characterized by the correlators
\beq
\left(\begin{array}{ccc}
 \langle J_V^i(k)J_V^j(-k)\rangle & \langle J_V^i(k) J_A^j(-k)\rangle & \langle J_V^i(k) T^{0j}(-k)\rangle
 \\
 \langle J_A^i(k) J_V^j(-k)\rangle & \langle J_A^i(k)J_A^j(-k)\rangle & \langle J_A^i(k) T^{0j}(-k)\rangle
 \\
 \langle T^{0i}(k) J_V^j(-k)\rangle & \langle T^{0i}(k)J_A^j(-k)\rangle & \langle T^{0i}(k) T^{0j}(-k)\rangle
 \end{array}\right).
\eeq
The ${\cal O}(k)$ terms form a symmetric matrix with six coefficients, which determine the six response parameters $\{\xi_{AA}, \xi_{AV}=\xi_{VA}, \xi_{VV}, \xi_A, \xi_V, \xi\}$. 

As discussed earlier, the relations \eqref{E:equalityCon} link the response parameters to anomaly coefficients. However, there are also two CPT-preserving coefficients, $\tilde{c}_V$ and $\tilde{c}_A$ which appear unconstrained. Recently, calculations at weak~\cite{Landsteiner:2011cp} and strong~\cite{Landsteiner:2011iq} coupling indicated
that the parameters $\tilde{c}_a$ were in fact proportional to the mixed flavor-gravitational anomaly coefficients. In the present instance, the relation is
\beq
\label{E:4dMatch}
\tilde{c}_V = 0, \qquad \tilde{c}_A = - 8 \pi^2 c_m\,.
\eeq

It has proven surprisingly difficult to understand the origin of these relations and the circumstances under which they hold. Both $\tilde{c}_A$ and $c_m$ appear in the zero-frequency two-point function of the axial current with the stress tensor, and substituting~\eqref{E:4dMatch} we have 
\beq
\label{E:JTconjecture}
\langle J_A^i(k) T^{0j}(-k)\rangle = \cdots +8 \pi^2 c_mT^2 O(k) + c_m O(k^3).
\eeq
With the identification~\eqref{E:4dMatch}, $c_m$ apparently contributes to both the $O(k)$ and $O(k^3)$ parts of the zero-frequency functions respectively, and there is a relative transcendental factor of $\sim\pi^2$ between them. The methods discussed thus far -- demanding the existence of an entropy current or studying the hydrostatic generating functional -- treat each order in momenta independently and furthermore lead to algebraic, rather than \emph{transcendental}, relations between response coefficients. Indeed, the two terms involving $c_m$ in~\eqref{E:JTconjecture} are comparable at a momentum scale $k\sim 2\pi T$, which is outside of the hydrostatic regime. This suggests that a proof of~\eqref{E:4dMatch} must go beyond the hydrodynamic limit. 

There are currently two independent proofs of~\eqref{E:4dMatch}. One~\cite{Jensen:2012kj} involves studying the Euclidean theory on a conical geometry which interpolates between the thermal cylinder and the vacuum. This generalizes the Cardy formula~\cite{Bloete:1986qm,Affleck:1986bv} in two-dimensional conformal field theory (CFT), which relates the pressure of a 2d CFT to its central charge. The second~\cite{Golkar:2012kb} involves a direct integration of a Weyl fermion in the Matsubara formalism. The resulting tower of Dirac Matsubara modes in the dimensionally reduced spatial theory provides a one-loop shift of the Chern-Simons terms, and again leads to~\eqref{E:4dMatch}. We refer the reader to these references for further details of the caveats and assumptions relevant for each derivation.\footnote{Curiously,~\eqref{E:4dMatch} seems to hold in theories which do not contain fields of spin greater than $1$, which are sensitive to topology. For instance, the partition function of a theory of free gravitinos on $\mathbb{R}^{2,*}$ does not agree with the partition function of the same theory on $\mathbb{R}^2$ due to the Killing spinors broken by deleting the origin~\cite{Solodukhin:2011gn}. Correspondingly, the relation~\eqref{E:4dMatch} does not hold for chiral gravitinos~\cite{Loganayagam:2012zg}.}

From this discussion it is clear that when applied to Standard Model physics, the relation~\eqref{E:4dMatch} as well as all of the anomaly-induced response in~\eqref{E:equalityCon} may be modified. The reason is that all of the results above were obtained assuming that the anomalies are shared between {\it global} symmetries. As a result, when taking one of the $U(1)$ symmetries to be weakly gauged all of the anomaly-induced response may in principle be subject to radiative corrections. We undertake a systematic study of these corrections in the rest of this work. 

\section{Weak gauging and dynamical photons}
\label{S:effectiveS}

The discussion in Section~\ref{S:review} focused on theories with global $U(1)_V$ and $U(1)_A$ symmetries, which are broken by $AVV$, $AAA$, and $ATT$ anomalies in the presence of non-dynamical background fields $V_\mu$ and $A_\mu$, and $g_{\mu\nu}$. When $V_\mu$ is dynamical, the generating functional and the hydrodynamic description need to be modified. In this and the following Section we discuss the effects of the dynamical photon field $V_\mu$, under the assumption of weak gauging, i.e. a perturbatively small gauge coupling.

\subsection{Anomalous conservation laws}
\label{S:dyn1}
When $V_\mu$ is the dynamical field, the full generating functional $W[V,A,g]$ of Section~\ref{S:review1} becomes the effective action for the photon, so that we define
\beq
\label{eq:Z31}
\mathcal{Z}[J_{\rm ext}, A, g] = \int [dV_{\mu}] \exp\left[ i W[V, A, g] + i \int\!\! d^4x\, \sqrt{-g}\, V_{\mu} J_{\rm ext}^{\mu} \right]\,.
\eeq
There is now a functional integral over the photon field which we couple to an external conserved current, $\nabla_\mu J_{\rm ext}^\mu=0$. We will use $J_{\rm ext}^{\mu}$ to compute correlation functions of $V_{\mu}$ as well as to ensure that the equilibrium at nonzero $\mu_V$ is stable~\cite{kapusta81}. When $U(1)_V$ is dynamical, the generating functional $W[V,A,g]$ must ensure that the corresponding consistent current $J^\mu_V$ of Section~\ref{S:review1} is conserved, $\nabla_\mu J^\mu_V = 0$. This amounts to choosing $c_s=0$ in Eq.~(\ref{E:ConWard2}), so that $W[V,A,g]$ is gauge-invariant under $U(1)_V$.
The generating functional of the theory with a dynamical $V_{\mu}$ is given in the usual way by
\beq
\label{E:dynW}
\mathscr{W}[J_{\rm ext}, A, g] = - i \ln \mathcal{Z}[J_{\rm ext}, A, g]\,.
\eeq
The one-point functions of the energy-momentum tensor, the axial current, and the photon field in the presence of external sources may be defined by the usual variational procedure,
\beq
\label{eq:TJV31}
\mathcal{T}^{\mu\nu}\equiv \frac{2}{\sqrt{-g}}\frac{\delta \mathscr{W}}{\delta g_{\mu\nu}}\,, \qquad 
\mathcal{J}_A^{\mu} \equiv \frac{1}{\sqrt{-g}}\frac{\delta \mathscr{W}}{\delta A_{\mu}}\,, \qquad 
\mathcal{V}_{\mu} \equiv \frac{1}{\sqrt{-g}} \frac{\delta \mathscr{W}}{\delta J_{\rm ext}^{\mu}}\,.
\eeq
As a result of conservation of $J^\mu_{\rm ext}$, physical quantities are invariant under $\mathcal{V}_\mu \to \mathcal{V}_\mu + \partial_\mu \Lambda(x)$, for an arbitrary $\Lambda(x)$. We can also write
\beq
  \mathcal{T}^{\mu\nu} = \langle T^{\mu\nu}\rangle
   + g^{\mu\nu} \langle V_\lambda J^\lambda_{\rm ext}\rangle\,, \qquad
  \mathcal{J}_A^\mu = \langle J^{\mu}_A\rangle\,, \qquad
  \mathcal{V}_{\mu} = \langle V_\mu \rangle\,,
\eeq
where $T^{\mu\nu}$ and $J_{A}^{\mu}$ are the stress tensor and axial current that follow from variation of $W$ as in~(\ref{E:defCurrents}), and the brackets denote averaging over the photon field configurations. The anomalous conservation laws for the consistent stress tensor and current $\mathcal{T}^{\mu\nu}$ and $\mathcal{J}_A^{\mu}$ are similar to (\ref{E:ConWard1}),
\begin{align*}
 \nabla_{\!\mu}\mathcal{J}_A^{\mu} &= \frac{1}{4}\epsilon^{\mu\nu\rho\sigma}\left[ c_g \langle F_{V,\mu\nu} F_{V,\rho\sigma} \rangle + \frac{\bar{c}_g}{3}F_{A,\mu\nu}F_{A,\rho\sigma} + c_m R\ab{\alpha}{\beta\mu\nu}R\ab{\beta}{\alpha\rho\sigma}\right]\,,
\\
\nabla_{\!\nu} \mathcal{T}^{\mu\nu} &= F_A^{\mu\nu}\mathcal{J}_{A,\nu} -\frac{1}{4}A^{\mu}\epsilon^{\nu\rho\sigma\tau}\left[ c_g \langle F_{V,\nu\rho} F_{V,\sigma\tau} \rangle +\frac{\bar{c}_g}{3}F_{A,\nu\rho}F_{A,\sigma\tau} + c_m R\ab{\alpha}{\beta\nu\rho}R\ab{\beta}{\alpha\sigma\tau}\right] .
\end{align*}
Compared to (\ref{E:ConWard1}), there is no $F_V^{\mu\nu} {J}_{V,\nu}$ term in the right-hand side, as it is already contained in the divergence of the energy-momentum tensor $\mathcal{T}^{\mu\nu}$.

\subsection{Anomalous (magneto-)hydrodynamics}
\label{S:dyn2}
The main modification to hydrodynamics is that $V_\mu$ now needs to be included as one of the hydrodynamic variables. This is based on the familiar statement that static $U(1)$ magnetic fields are not screened, hence for excitations with sufficiently low frequency, one adds the magnetic field to the set of hydrodynamic variables. This leads to magneto-hydrodynamics, or MHD, a description where magnetic fields are treated classically, on par with other hydrodynamic variables such as $T$ and $u^\mu$, see e.g.~\cite{DavidsonMHD} (anomalies in MHD have also been discussed in~\cite{Neiman:2010zi}). Taking quantum fluctuations of $V_\mu$ into account for low-frequency collective excitations requires a treatment that goes beyond classical hydrodynamics. 

If the gauge coupling of $U(1)_V$ is sufficiently small, we may consider classical configurations of the photon field, which we call $v_\mu$, which solve the classical equations of motion, extremizing the exponent in (\ref{eq:Z31}), 
\beq
\label{E:Veom0}
J_v^\mu + J_{\rm ext}^{\mu} = 0\,.
\eeq
Here $J_v^\mu = \frac{1}{\sqrt{-g}} \frac{\delta W}{\delta V_\mu}\big|_{V=v}$ is the conserved current obtained by the variation of $W[V,A,g]$. 
The hydrodynamic variables are thus $T$, $\mu_V$, $\mu_A$, $u^\mu$, and $v_\mu$, where $\mu_{V}$ and $\mu_A$ are defined so as to match the Euclidean temporal holonomies of 
$V_\mu$ and $A_\mu$ in equilibrium. As noted earlier, $\mu_A$ as defined is not conjugate to a conserved charge when $V_{\mu}$ is dynamical. However, the non-conservation of $J_A^{\mu}$ is gradient-suppressed, and we will continue to include $\mu_A$ as a hydrodynamic variable.

Eq.~(\ref{E:Veom0}) should be viewed as an analogue of Maxwell's equations, specifying the dynamics of $v_\mu$. 
Note that upon solving~\eqref{E:Veom0}, the current $J_v^{\mu}$ is conserved, so that $\nabla_{\mu}J_v^{\mu}=0$ is not a new hydrodynamic equation. Such classical treatment neglects photon loops, and in a slight abuse of terminology we will call this effective description magneto-hydrodynamics, or MHD.

The other hydrodynamic equations are the anomalous conservation laws of the axial current and the energy-momentum tensor,
\begin{align}
\begin{split}
 \nabla_{\!\mu} {J}_A^{\mu} &= \frac{1}{4}\epsilon^{\mu\nu\rho\sigma}\left[ c_g F_{v,\mu\nu} F_{v,\rho\sigma} + \frac{\bar{c}_g}{3}F_{A,\mu\nu}F_{A,\rho\sigma} + c_m R\ab{\alpha}{\beta\mu\nu}R\ab{\beta}{\alpha\rho\sigma}\right]\,,
\\
\nabla_{\!\nu} {T}^{\mu\nu} &= F_A^{\mu\nu}J_{A,\nu} -\frac{1}{4}A^{\mu}\epsilon^{\nu\rho\sigma\tau}\left[ c_g F_{v,\nu\rho} F_{v,\sigma\tau}  +\frac{\bar{c}_g}{3}F_{A,\nu\rho}F_{A,\sigma\tau} + c_m R\ab{\alpha}{\beta\nu\rho}R\ab{\beta}{\alpha\sigma\tau}\right] .
\end{split}
\label{eq:JT32}
\end{align}
where $J_A^\mu$ and $T^{\mu\nu}$ are $\mathcal{J}_A^{\mu}$ and $\mathcal{T}^{\mu\nu}$, evaluated when $V_\mu$ is treated as a classical field
solving (\ref{E:Veom0}).  
In four spacetime dimensions, there are 9 hydrodynamic equations (\ref{E:Veom0}), (\ref{eq:JT32}), and 10 hydrodynamic variables $T$, $\mu_V$, $\mu_A$, $u^\mu$, $v_\mu$. The $U(1)_V$ gauge freedom can be used to eliminate the extra degree of freedom in $v_\mu$.

In the thermodynamic frame, the constitutive relations for $J_A^\mu$ and $T^{\mu\nu}$ in MHD are exactly the same as in Section~\ref{S:review2}, expressed in terms of $T$, $\mu_V$, $\mu_A$, $u^\mu$, $v_\mu$, $A_\mu$ and $g_{\mu\nu}$. As in Section~\ref{S:review2}, we adopt the scaling such that the gauge fields and the metric are ${\cal O}(1)$ in the derivative expansion, hence $F_{v,\mu\nu}$ and
$F_{A,\mu\nu}$ are ${\cal O}(\partial)$, and $R^\mu_{\ \nu\alpha\beta}$ is ${\cal O}(\partial^2)$.  This scaling does not allow us to consider a background  magnetic field at zeroth order in the derivative expansion.

In order to find $J_v^\mu[v,A,g]$, one in principle needs to know $W[V,A,g]$, which is a complicated non-local functional. However, $W[V,A,g]$ is the generating functional in the theory with the global $U(1)_V$; hence the relation
$$
  \frac{1}{\sqrt{-g}} \frac{\delta W}{\delta V_\mu} = J_V^\mu,
$$
can be viewed as providing an expression for $\frac{1}{\sqrt{-g}} \frac{\delta W}{\delta V_\mu}$ in the hydrodynamic limit, when the right-hand side, expressed as $J_V^\mu[V,A,g]$, is determined by the hydrodynamic equations in the theory with global $U(1)_V$. Thus one has to solve the hydrodynamic equations in the theory with global $U(1)_V$, express the current $J_V^\mu$ in terms of $V_\mu$, $A_\mu$, and $g_{\mu\nu}$ (which all act as sources when $U(1)_V$ is global), and use the resulting $J_v^\mu = J_V^\mu[V,A,g]_{V=v}$ in~\eqref{E:Veom0} in order to find $v_\mu[A,g,J_{\rm ext}]$. Once the dynamics of $v_\mu$ is determined, the remaining equations~\eqref{eq:JT32} can be used to express the hydrodynamic variables $T$, $\mu_V$, $\mu_A$, $u^\mu$ in terms of the sources, and eventually find $J_A^\mu[A,g, J_{\rm ext}]$ and $T^{\mu\nu}[A,g, J_{\rm ext}]$.

Any solution to the MHD equations (\ref{E:Veom0}), (\ref{eq:JT32}) is also a solution to the hydrodynamic equations (\ref{E:ConWard1}) in the theory with global $U(1)_V$ (obviously, $J_v^\mu + J_{\rm ext}^{\mu} = 0$ as fixed by the dynamics of the photon field implies $\nabla_{\!\mu}(J_v^\mu + J_{\rm ext}^{\mu}) = 0$). Therefore the entropy current with non-negative divergence in the hydrodynamic theory with global $U(1)_V$ will have a non-negative divergence when evaluated on the solutions to MHD. This shows that an entropy current in MHD can be taken to be exactly the same as the entropy current in the theory with global $U(1)_V$. It is not clear from this argument, however, whether this provides the unique MHD entropy current. We hope to return to this question in the future. 

We now turn to the question of MHD correlation functions. The prescription is almost identical to the variational method in hydrodynamics outlined in Section~\ref{S:review2}. Namely, the MHD equations (\ref{E:Veom0}), (\ref{eq:JT32}) need to be solved in order to find the hydrodynamic variables in terms of the sources $A_\mu$, $g_{\mu\nu}$, and $J_{\rm ext}^\mu$, which upon using the constitutive relations give $v_\mu[A,g,J_{\rm ext}]$, $J_A^\mu[A,g, J_{\rm ext}]$ and $T^{\mu\nu}[A,g, J_{\rm ext}]$. Varying with respect to the sources then allows one to compute (retarded) correlation functions of $J_A^{\mu}, T^{\mu\nu}$, and $V_{\mu}$.

The correlation functions of the vector current are more subtle, as the on-shell value of $J_v^\mu$ is just $-J_{\rm ext}^\mu$, which does not depend on the $A$ and $g$ sources. This implies in particular that the correlation functions defined by varying $J_v^\mu$ with respect to $A_\nu$ and $g_{\rho\sigma}$ vanish identically in MHD. However, one should keep in mind that the current $J_V^\mu$  does not in general coincide with the expectation value of the conserved current operator whose charge generates the $U(1)_V$ symmetry. We call the latter current operator $\JV^\mu$. For example, in flat-space quantum electrodynamics $\langle \JV^\mu \rangle$ and $J_V^\mu$ differ by a term proportional to $\partial_{\nu}F_V^{\mu\nu}$ which is identically conserved~\cite{Collins:2005nj}. Within the hydrodynamic description with classical $v_\mu$, we will write
\beq  
  J_v^{\mu} = \langle \JV^{\mu} \rangle - \C^{\mu}\,, 
\eeq
where $\C^{\mu}$ is an identically conserved vector built out of hydrodynamic variables and background sources. Setting $J_{\rm ext}^\mu = 0$ for simplicity, so that $\langle \JV^\mu \rangle = \C^\mu$ according to \eqref{E:Veom0}, we parametrize $\C^{\mu}$ in the following way, to second order in derivatives,
\begin{align}
\label{E:JVmatch}
\C^{\mu} = \C_{(1)}^{\mu} + \C_{(2)}^{\mu} + \cdots
\end{align}
where
\begin{align}
\C_{(1)}^{\mu} = &  
   \nabla_{\!\nu}\!\left[ \j_0  u^{[\mu}A^{\nu]}\right] + 
   \epsilon^{\mu\nu\rho\sigma}\nabla_{\!\nu}\!\left[ \tj_0 u_{\rho}A_{\sigma}\right] 
   \\[5pt]
\begin{split}
\C_{(2)}^{\mu} =  & \nabla_{\!\nu}\!\left[  \left( \j_1 F_v^{\mu\nu} + \j_2 F_A^{\mu\nu} + \j_3 \omega^{\mu\nu} +\j_4u^{[\mu} E_v^{\nu ]} + \j_5u^{[\mu} E_A^{\nu]}  +\j_6 u^{[\mu}a^{\nu]}\right.\right.
\\[5pt]
& \hspace{1in} \left. \left. + \j_7 u^{[\mu}\partial^{\nu]}\mu_V + \j_8 u^{[\mu}\partial^{\nu]}\mu_A + \j_9 u^{[\mu}\partial^{\nu]}T)\right) \right] 
\\[5pt]
 + & \epsilon^{\mu\nu\rho\sigma}\nabla_{\!\nu}\! \left[ \tj_1 F_{v,\rho\sigma} + \tj_2 F_{A,\rho\sigma} + \tj_3 \omega_{\mu\nu} + \tj_4 u_{\rho}E_{v,\sigma}+ \tj_5 u_{\rho}E_{A,\sigma} + \tj_6 u_{\mu}a_{\nu} \right. 
\\[5pt]
& \hspace{1in} \left. + \tj_7 u_{\rho} \partial_{\sigma}\mu_V + \tj_8 u_{\rho}\partial_{\sigma}\mu_A + \tj_9 u_{\rho} \partial_{\sigma} T\right] \\[5pt]
+ & (U(1)_A \textrm{-violating}).
\end{split}
\end{align}
The brackets indicate antisymmetrization, $T^{[\mu\nu]} = \frac12(T^{\mu\nu}-T^{\nu\mu})$. The vector $\C^{\mu}$ is invariant under $U(1)_V$, but not under $U(1)_A$, as the latter is explicitly broken by the anomaly.  We have neglected $U(1)_A$-violating terms at ${\cal O}(\partial^2)$ as one can show that they are unimportant for our analysis later in this article.  The parameters $\j_i$ and $\tj_i$ are functions of $T,\mu_V,\mu_A, u^{\mu}A_{\mu}$, and $A_\mu A^\mu$. Expressing $\C^{\mu}$ in terms of the sources $A_\mu$ and $g_{\mu\nu}$ upon solving the hydrodynamic equations allows one to compute correlation functions of one $\JV^\mu$ with multiple $J_A^\mu$ and $T^{\mu\nu}$. These correlation functions will be given in terms of the parameters $\j_i$ and $\tj_i$ which need to be determined by matching to the microscopic theory.

To compute correlation functions involving more than one vector current, it may be helpful to reformulate the classical MHD equations as tree-level perturbation theory. For example, in order to evaluate the two-point function of $\JV^\mu$, one needs all tree-level diagrams that connect two factors of $\C^{\mu}$, expressed in terms of the tree-level propagators of the hydrodynamic variables. The MHD described above requires a small gauge coupling for $U(1)_V$, and as a classical theory it neglects loop of both photons and other collective excitations. We leave the detailed study of MHD correlation functions for future work.

\subsection{Anomalous (magneto-)hydrostatics}
\label{S:AMHs}

\subsubsection{The (magneto-)hydrostatic effective theory}
\label{S:331}
In Section~\ref{S:review3} we reviewed how the hydrostatic response of a theory where the photon is non-dynamical may be calculated directly from the generating functional $\Wstatic$ defined on the spatial slice. In a gauge and coordinate choice where the background fields are explicitly time-independent, $\Wstatic[V,A,g]$ is a local functional (unlike the full $W[V,A,g]$), thanks to the finite static correlation length. When the photon is dynamical, the unscreened static magnetic field will make the corresponding generating functional $\mathscr{W}[J_{\rm ext}, A, g]$ non-local even in the static limit, making a derivative expansion of $\mathscr{W}$ impractical. Instead, a convenient static low-momentum effective description can given by a dimensionally reduced Euclidean field theory, as in~\cite{Braaten:1995cm, Braaten:1995jr, Andersen:1995ej}. The effective theory describes the Matsubara zero mode of the four-dimensional photon, below the momentum cutoff scale $\Lambda\lesssim T$ whose exact value is determined by the masses of the other fields in the microscopic theory. The cutoff is such that the zero mode of the photon is the lightest degree of freedom. Let us now write down the action of this three-dimensional effective theory, taking into account $AAA$, $AVV$, and $ATT$ anomalies.

We turn on the external sources $A$, $g$, and $J_{\rm ext}$ which are time-independent, with momenta below the cutoff. We will write the action $\Seff[V;A,g]$ in the derivative expansion,
\beq
\label{E:Seffn}
\Seff = \Seff^{(0)} + \Seff^{(1)} + \Seff^{(2)}+ {\cal O}(\partial^3)\,,
\eeq
where $\Seff^{(n)}$ denotes the terms in $\Seff$ with $n$ derivatives. 
In principle, $\Seff$ may be obtained by starting with the partition function~\eqref{eq:Z31}, continuing to Euclidean (compact) time, integrating out all of the nonzero Matsubara modes of $V_{\mu}$, integrating out all spatial momenta above the scale $\Lambda$, and then continuing the photon field back to real time.
At weak $U(1)_V$ gauge coupling, the resulting $\Seff$ will be equal to the hydrostatic generating functional $\Wstatic$ of Section~\ref{S:review3}, plus perturbative corrections that come from loops with momenta above the cutoff. If we neglect photon loops, as in MHD, then $\Seff$ is precisely $\Wstatic$.

The action $\Seff$ must be both $U(1)_V$ and diffeomorphism-invariant, but it need not be $U(1)_A$-invariant. The $U(1)_A$-violating terms have two distinct origins: (i.) there are the anomalous $U(1)_A$-violating terms in $\Wstatic$ given by Eq.~\eqref{E:Wanom}, and (ii.) there are terms which come from integrating out higher Matsubara modes and momenta above the cutoff. The latter are perturbatively suppressed under our weak gauging assumption, and we will explicitly account for this in parametrizing the operator coefficients in $\Seff$.

In terms of the time-independent fields~\eqref{E:fields}, the zero-derivative piece is
\beq
\label{E:Seff0}
\Seff^{(0)} = \int \!(dt)\, d^3x\, e^{\MFs}\sqrt{\hat{g}}\,\left( \p(T,\mu_V,\mu_A)+\delta \p +V_{\mu}J_{\rm ext}^{\mu}\right)
\eeq
where $p$ is the pressure of the theory with $V_\mu$ non-dynamical, and the second term $\delta \p=\delta\p(T,\mu_V,\mu_A,\hat{A}^2)$ arises from integrating out the photon field with momenta above the cutoff. It depends on
\beq
T = e^{-\mathfrak{s}}\beta^{-1}\,, \qquad \mu_V = e^{-\mathfrak{s}}V_0\,, \qquad \mu_A = e^{-\mathfrak{s}}A_0\,, \qquad \hat{A}^2 = \hat{g}^{ij}\hat{A}_i\hat{A}_j\,.
\eeq
After matching the anomaly-induced terms, which we can separate out according to our weak-gauging assumption, 
the functional form of the one-derivative effective action is
$$
  \Seff^{(1)} = W_{anom} + W_1 + \dots\,,
$$
where $W_{anom}$ and $W_1$ are given by~\eqref{E:Wanom} and~\eqref{E:W1asCS}, respectively. Specifically, 
\begin{align}
\label{E:Seff1}
\nonumber
\Seff^{(1)} =  & -\int \!(dt)\, d^3x\, \tilde{\epsilon}^{\,ijk} \left[ \left(2c_g V_0+\frac{\tilde{c}_{AV}}{\beta}-\delta \tf_{AV}\right) \hat{A}_i \partial_j \hat{V}_k + \left(\frac{2\bar{c}_g}{3}A_0+\frac{\tilde{c}_{AA}}{2\beta}-\delta\tf_{AA}\right) \hat{A}_i \partial_j \hat{A}_k \right. \\
& \qquad\qquad\qquad  + \left. \left(c_g V_0^2+\frac{\bar{c}_g}{3}A_0^2+\frac{\tilde{c}_A}{\beta^2}-\delta \tf_{Aa}\right) \hat{A}_i \partial_j \MFa_k\right.
\\
\nonumber
& \qquad \qquad \qquad \qquad \left. - \frac{\tilde{c}_{VV}}{2\beta} \hat{V}_i \partial_j \hat{V}_k + \frac{\tilde{c}_V}{\beta} \hat{V}_i \partial_j \MFa_k - \frac{\tilde{c}}{2\beta^3} \MFa_i\partial_j \MFa_k \right] + \delta \Seff^{(1)},
\end{align}
where $\tilde{c}_{AA}$, $\tilde{c}_{AV}$, and $\tilde{c}_{A}$ are constants which respect $U(1)_A$. The coefficients $\tilde{c}_{VV}$, $\tilde{c}_V$, and $\tilde{c}$ are constants, due to $U(1)_V$ and $U(1)_{KK}$ gauge invariance. 
Since they are Chern-Simons couplings in the three-dimensional theory, they only receive corrections at one-loop order from charged fields~\cite{Coleman:1985zi}. Indeed, the diagrammatic analysis of Coleman and Hill \cite{Coleman:1985zi} indicates, under fairly general assumptions, that radiative corrections to abelian Chern-Simons coefficients arise only from fermions at one-loop order. There are no fermions in the effective theory, and thus the constant coefficients should be equal to their values before gauging $U(1)_V$. In particular, $\tilde{c}_V=0$ and, due to the $ATT$ anomaly, $\tilde{c}_A=-8 \pi^2 c_m$~\cite{Jensen:2012kj,Golkar:2012kb}. Nonperturbative arguments against higher loop corrections to the $\tilde{c}$'s can also be made by exploiting analyticity of the Wilsonian effective action \cite{Closset:2012vp}. As mentioned in Section~\ref{S:introduction}, we retain the CPT-violating constants to facilitate contact with physics at nonzero $\mu_5$.

The functions $\delta \tf$ depend on $T, \mu_V, \mu_A,$ and $\hat{A}^2$ 
and come from integrating out non-zero Matsubara modes of all the fields in the theory, plus the high momentum component of the photon zero modes. The remaining $U(1)_A$-violating corrections $\delta \Seff^{(1)}$ are also induced perturbatively, 
\begin{align}
\label{S1axial}
\begin{split}
\delta \Seff^{(1)}
& = \int\! (dt)\, d^3x\, e^{\MFs}\sqrt{\hat{g}} \left[ ( \delta\! \f_{dA}\,\hat{g}^{ij}+\delta\! \f_{AAdA}\,\hat{A}^i\hat{A}^j)\hat{\nabla}_i\hat{A}_j \right.\\
& \qquad\qquad + \left. \hat{g}^{ij}\left( \delta\! \f_{AV}\,\hat{A}_i\partial_j V_0 + \delta\! \f_{AA}\, \hat{A}_i \partial_j A_0 + \delta\! \f_{As}\,\hat{A}_i\partial_j\MFs\right)\right]\,.
\end{split}
\end{align}
The full expression for the two-derivative part $\Seff^{(2)}$ is rather long. For the purpose of computing the leading perturbative corrections to two-point functions of spatial currents and the momentum density, the only relevant terms in $\Seff^{(2)}$ are $U(1)_A$-invariant terms which do not involve gradients of $\mathfrak{s}, A_0,$ and $\hat{g}_{ij}$. The $U(1)_A$-violating terms may be shown to contribute to the two-point functions at higher order in the $U(1)_V$ gauge coupling than we consider (in QED they contribute at order $e^4$ and higher). We summarize the scalars and pseudoscalars which appear in $\Seff^{(2)}$ and are relevant for us in Table~\ref{T:2ndOrderScalars}. For completeness, the second-order $U(1)_A$-invariant terms which are irrelevant for us are
\begin{align}
\nonumber
\text{scalars}:\,\, & \hat{R}\,,\, (\partial\MFs)^2\,,\, f_{ij}f^{ij}\,,\, (\partial A_0)^2\,,\, f_{ij}\hat{F}_A^{ij}\,,\, \hat{F}_{A,ij}\hat{F}^{ij}_A\,,\, \partial_i \MFs \partial^i A_0\,,\, \partial_i \MFs \partial^i V_0\,,\, \partial_i A_0 \partial^i V_0\,,\, \\
\nonumber
\text{pseudoscalars}: \,\, & \epsilon^{ijk}\partial_i\MFs\partial_j\MFa_k, \epsilon^{ijk}\partial_i \MFs\partial_j \hat{A}_k, \epsilon^{ijk}\partial_i \MFs\partial_j \hat{V}_k, \epsilon^{ijk} \partial_i A_0 \partial_j \MFa_k, \epsilon^{ijk} \partial_i A_0 \partial_j \hat{A}_k, \epsilon^{ijk}\partial_i A_0 \partial_j \hat{V}_k\,,\,
\end{align}
where $\hat{R}$ is the Ricci curvature scalar constructed from the spatial metric $\hat g$, $f_{ij}=\partial_i\MFa_j-\partial_j\MFa_i$, and $\hat{F}_V$ and $\hat{F}_A$ are the field strengths constructed from $\hat{V}_i$ and $\hat{A}_i$. The epsilon tensor on the spatial slice is $\epsilon^{ijk} = \tilde{\epsilon}^{\,ijk}/\sqrt{\hat g}$, with $\tilde{\epsilon}^{\,123}=1$. We then have
\beq
\label{E:Seff2}
\Seff^{(2)} = \int\! (dt)\,d^3x\, e^{\MFs}\sqrt{\hat{g}} \left[ \sum_{i=1}^4 \a_i s_i + \sum_{i=1}^3 \ta_i \tilde{s}_i \right] + (\text{irrelevant terms})\,,
\eeq
where the $\a_i$'s and $\ta_i$'s are functions of $T, \mu_V,$ and $\mu_A$, and the scalars $s_i$ and pseudoscalars $\tilde{s}_i$ are defined in Table~\ref{T:2ndOrderScalars}. As we will see, to accurately compute the chiral conductivities to low order in perturbation theory, we also require a single parity-violating term with three derivatives. This term is singled out relative to the other two- and three-derivative terms in that it contributes to the spatial photon propagator at low order in momentum. It is
\beq
\Seff^{(3)} =\int \!(dt)\, d^3x\, e^{\MFs}\sqrt{\hat{g}}\, \ta_4\,\epsilon^{ijk} (\hat{F}_V)_{ij}\hat{\nabla}_l(\hat F_V)_k^{\phantom{k}l} + (\text{irrelevant terms})\,.
\eeq
\begin{table}[tc]
\begin{center}
\renewcommand{\arraystretch}{1.2}
\begin{tabular}{|c|c|c|c|c|}
\hline
$i$ & 1 & 2 & 3 & 4 \\
\hline
scalars ($s_i$) & $(\partial V_0)^2$ & $\hat{F}_{V,ij}\hat{F}_V^{ij}$ & $f_{ij}\hat{F}_V^{ij}$ & $\hat{F}_{A,ij}\hat{F}_V^{ij}$ \\
\hline
pseudoscalars ($\tilde{s}_i$) & $\epsilon^{ijk}\partial_i V_0 \partial_{j}\hat{V}_k$ & $\epsilon^{ijk}\partial_i V_0 \partial_j \hat{A}_k$ & $\epsilon^{ijk}\partial_i V_0 \partial_j \MFa_k$ &   \\
\hline
\end{tabular}
\caption{\label{T:2ndOrderScalars}The scalars and pseudoscalars with two derivatives which may appear in $\Seff^{(2)}$ and are relevant for the calculation of the chiral conductivities in Section~\protect\ref{S:pert}.}
\end{center}
\end{table}%
We may use the effective theory (\ref{E:Seffn}) to calculate in the hydrostatic limit the correlation functions of the operators
\beq
\label{E:TJAstat}
T^{\mu\nu} = \frac{2}{\sqrt{-g}} \frac{\delta \Seff}{\delta g_{\mu\nu}}\,, \qquad J_A^{\mu} = \frac{1}{\sqrt{-g}} \frac{\delta \Seff}{\delta A_{\mu}} \,.
\eeq
The correlation functions of the vector current are more subtle: ${\delta \Seff}/{\delta V_{\mu}}$ is the ``equation of motion'' operator with constrained correlation functions, rather than the vector current whose charge generates $U(1)_V$. Both the subtlety and its resolution are virtually identical to the situation in MHD, Section~\ref{S:dyn2}. The vector current operator $\JV^{\mu}$ of the microscopic theory must be matched to a conserved current operator in the effective theory. Since there are no fundamental fields charged under $U(1)_V$ in $\Seff$, the most general conserved vector current in the effective theory is conserved identically. We parameterize the vector current in the effective theory as in \eqref{E:JVmatch},
\beq
\label{E:JVstat}
\JV^{\mu} = \sum_n (\JV)^{\mu}_{(n)}\,,
\eeq 
where the subscript refers to the order in the derivative expansion of the current.
As the fields in the effective theory are time-independent, the current is conserved provided $\partial_{i} (\sqrt{-g}\, \JV^i) {=}0$. We are studying states that have no fixed background magnetic field, and the vector current has to be expressed in terms of the sources and the dynamical photon field.

The most general one-derivative part of $\JV^i$ is
\beq
\label{E:JVstat1}
(\JV)^i_{(1)} =  e^{-\MFs} \,{\epsilon}^{ijk}\partial_j \!\left[ \delta\tj_0 \hat{A}_k + \frac{\delta \tilde{C}_V}{\beta} \hat{V}_k + \frac{\delta \tilde{C} }{\beta^2}\MFa_k \right]\,.
\eeq
The parameter $\delta\tj_0$ is a function of $T,\mu_V,\mu_A$, and $\hat{A}^2$. The term involving $\delta\tj_0$ is $U(1)_A$-violating. It arises due to the integration of momenta down to the cutoff scale and so is perturbatively small.
The factors of $\beta$ are inserted so that $\delta\tilde{C}_V$ and $\delta\tilde{C}$ are dimensionless, and gauge invariance under $U(1)_V$ and $U(1)_{KK}$ further implies that they are constant. Under four-dimensional CPT, $\delta\tilde{C}_V$ is CPT-violating while $\delta\tilde{C}$ is CPT-preserving. There is no identically conserved and $U(1)_V$-invariant covariant four-current which realizes either $\epsilon^{ijk}\partial_j\hat{V}_k$ or $\epsilon^{ijk}\partial_j\MFa_k$ in the hydrostatic limit. It follows that in the absence of Lorentz and CPT violating sources, the $\delta \tilde{C}$'s vanish in the classical limit, when our hydrostatic theory is just the stationary limit of MHD. Since they are necessarily constant, it seems unlikely that they will be generated in the full theory. However, we retain them for generality, keeping in mind the possibility of Lorentz and CPT violating sources such as $\mu_5$.

The most general two-derivative piece of $\JV^i$ is
\begin{align}
\label{E:JVstat2}
\nonumber
(\JV)^i_{(2)} = &\frac{1}{\sqrt{-g}}\partial_j\!\left[  \sqrt{-g}\left( \j_0 \hat{F}_V^{ij} + \j_1 \hat{F}_A^{ij} +\j_2 f^{ij}\right)\right] + e^{-\MFs}\,\epsilon^{ijk}\partial_j\left[ \tj_1\partial_k V_0 + \tj_2 \partial_k A_0 + \tj_3 \partial_k \MFs\right] 
\\ &+ (U(1)_A\text{-violating})\,,
\end{align}
where $\j_i$ and $\tj_i$ are functions of $T,\mu_V,$ and $\mu_A$. The $U(1)_A$-violating terms will contribute to two-point functions of spatial currents beyond the leading order in perturbation theory.  
Using the expressions (\ref{E:TJAstat}) and (\ref{E:JVstat}) for the currents along with the photon propagators, hydrostatic correlation functions may be computed by the usual method of diagrammatic perturbation theory.

The effective action simplifies if we work in flat space at constant $T$ and $\mu_V$, and the only external sources are the constant $\mu_A=A_0$ and $J_{\rm ext}^0$.
To third order in derivatives,
\begin{align}
\nonumber
\Seff = \int \!(dt)\, d^3x &\left[ \bar{\p}(V_0) +\frac{\tilde{c}_{VV}}{2\beta^2}\epsilon^{ijk}\hat{V}_i\partial_j\hat{V}_k + \a_1(V_0) (\partial V_0)^2 + \a_2(V_0)\hat{F}_V^2 + \ta_1(V_0) \epsilon^{ijk}\partial_i V_0\partial_j \hat{V}_k\right.
\\
\label{E:SeffTrivial}
&\hspace{.1in}\left.+\ta_4(V_0)\epsilon^{ijk} \hat F_{ij}\, \partial_l (\hat{F}_{V})_k^{\phantom{k}l}
\right] + {\cal O}(\partial^4)\,.
\end{align}
The effective pressure $\bar{\p}\equiv p+\delta p+V_0 J_{\rm ext}^0$ has a local minimum at $V_0=\bar V_0$ which must be identified as $\bar V_0=\mu_V$. 
Around the minimum,
\beq
\bar{\p}(V_0)=\bar{\p}+\coeff12\bar{\p}^{\,\prime\prime} \delta V_0^2 + ..\,, \qquad \a_i(V_0) = \a_i + \a_i'\, \delta V_0 + ..\,, \qquad \ta_i(V_0) = \ta_i + \ta_i'\, \delta V_0 + ..\,,
\eeq
where $\delta V_0 = V_0 {-} \bar V_0$, and the prime indicates a derivative evaluated at $\bar V_0$. 
From~\eqref{E:SeffTrivial} we see that, ignoring the higher-derivative terms, the tree-level photon propagator is 
\begin{align}
\begin{split}
\label{E:propagator}
\langle \delta V_0(k)\, \delta V_0(-k)\rangle& = -\frac{1+O(k^2)}{2\a_1k^2 + \bar{\p}^{\,\prime\prime}}\,,
\\
\langle \hat{V}_i(k)\hat{V}_j(-k)\rangle &= \frac{i(\tilde{c}_{VV}T + 2\ta_4 k^2)\epsilon^{ijk}k_k + 4\a_2\delta_{ij}k^2+O(k^4)}{k^2(\tilde{c}_{VV}^2T^2-4(4\a_2^2-\ta_4 \tilde{c}_{VV}T)k^2)}\,,
\end{split}
\end{align}
where we have chosen Feynman gauge, and $\langle \delta V_0(k)\, \hat V_i(-k)\rangle=0$.
The temporal photon has a mass $m^2 = \bar{\p}^{\,\prime\prime}/2\a_1$, interpreted as electric screening. Without the CPT-violating constant $\tilde{c}_{VV}$ the spatial photon has no mass, which is interpreted as the absence of magnetic screening. The Chern-Simons term in the effective action generates magnetic screening when $\tilde{c}_{VV}T > 4\a_2^2/\ta_4$, and magnetic anti-screening when $\tilde{c}_{VV}T < 4\a_2^2/\ta_4$. In the perturbative regime, we expect that the contribution $\ta_4$ is subleading compared to $\a_2^2$, leading to anti-screening \cite{Redlich:1984md,Niemi:1985ir}. Recall that turning on the constant axial chemical potential $\mu_5$ (as distinct from $\mu_A$) will generate a contribution to $\tilde{c}_{VV}$. In the full theory the anti-screening visible in the spatial effective theory will likely lead to an instability (see e.g. the recent discussion in \cite{Khaidukov:2013sja}). We comment further on magnetic anti-screening in Section~\ref{S:discuss}.

\subsubsection{The thermal effective action and chiral conductivities}
\label{S:Gamma}

In order to directly encode the retarded correlators, and thus the Euclidean Kubo formulae, it is useful to consider the formal result of integrating over the remaining Matsubara zero-modes of the photon. The effective theory $\Seff$ then leads to the generating functional $\mathscr{W}[J_{\rm ext},A,g]$ in Eq.~\eqref{E:dynW} where the sources are time-independent and slowly varying. The 1PI effective action $\Gamma[\mathcal{V},A,g]$ (that we will refer to here as the ``thermal" effective action)
is  related to the generating functional by a functional Legendre transform,
\beq
\label{E:GammaW}
\Gamma[\mathcal{V},A,g] = \mathscr{W}[J_{\rm ext}[\mathcal{V},A,g],A,g] - \int d^4x \sqrt{-g}\,\mathcal{V}_{\mu} J_{\rm ext}^{\mu}[\mathcal{V},A,g]\,,
\eeq 
where $\mathcal{V}_{\mu}$ is now the expectation value and we solve
\beq
\label{E:VJext}
  \frac{1}{\sqrt{-g}}\frac{\delta\mathscr{W}[J_{\rm ext},A,g]}{\delta J_{\rm ext}^\mu} 
  = \mathcal{V}_{\mu} \,, \ \ \textrm{or}\ \ \ 
  \frac{1}{\sqrt{-g}}\frac{\delta \Gamma[\mathcal{V},A,g]}{\delta{\cal V}_\mu} = - J_{\rm ext}^\mu
\eeq
for ${\cal V}[J_{\rm ext},A,g]$, or equivalently for $J_{\rm ext}[{\cal V}, A, g]$. Since we are only Legendre transforming in the $J_{\rm ext},\mathcal{V}$ variables and not in the other sources, $\Gamma$ is the 1PI effective action in the presence of fixed background $A_\mu$ and $g_{\mu\nu}$.

In a zero-temperature theory with a mass gap $m_{\rm gap}$ and slowly varying sources (with gradients $\lambda$ longer than the inverse gap, $\lambda m_{\rm gap}\gg 1$), both the generating functional and effective action may be expressed locally in a derivative expansion. The derivative expansion effectively accounts for the response of the vacuum to background fields, neglecting non-localities over the scale of the inverse gap, so that the small expansion parameter is $\sim 1/\lambda m_{\rm gap}$. In a theory with massless fields, both $\mathscr{W}$ and $\Gamma$ are generically nonlocal, reflecting the infinite correlation length.

Similar statements hold for thermal field theories subjected to time-independent background fields. At momenta and Euclidean energies well below the temperature, the theory dynamically dimensionally reduces on the thermal circle to give a three-dimensional effective theory in which the temperature dependence of the full theory is fully encoded in the mass and coupling parameters. The inverse static screening length plays the role of a mass gap. So a thermal field theory with finite static screening length will possess a hydrostatic generating functional which may be expressed locally in a derivative expansion as argued in~\cite{Jensen:2012jh,Banerjee:2012iz}. The hydrostatic $W$ effectively describes the response of the flat space \emph{thermal} state to background fields. 

When the static screening length is infinite, both $\mathscr{W}$ and $\Gamma$ will generically be nonlocal. When the magnetic photon is massless, the derivative couplings enforced by gauge invariance soften the infrared behaviour. 
The three-dimensional gauge field $\hat{V}_i$ may be dualized to a compact massless scalar, which crucially as a consequence of gauge invariance is derivatively coupled to itself and to the massive $V_0$.\footnote{The same words apply to a superfluid phase, as studied in~\cite{Bhattacharyya:2012xi}. As a result our arguments should also apply to superfluids.} 
Such a theory should not possess infrared divergences, and so we expect that $\Gamma$ (though not $\mathscr{W}$) may in this case be written locally in a derivative expansion. This expansion can be at best asymptotic as it was for the hydrostatic generating functional.

Since the thermal effective action is local, we can parameterize it to low order in derivatives in terms of the photon expectation value $\mathcal{V}_{\mu}$ and the sources $A_{\mu}$ and $g_{\mu\nu}$. By the symmetries of the problem, we must impose $U(1)_V$ gauge invariance as well as diffeomorphism invariance. As a result $\Gamma$ has the same structure as the effective field theory action $\Seff$, however the various coefficients that appear in it will differ from those of the Wilsonian effective action. To first order in derivatives it is
\begin{align}
\Gamma^{(0)} =& \int\! (dt)\, d^3x\, e^{\MFs}\sqrt{\hat{g}}\, \P\,,
\\
\nonumber
\Gamma^{(1)} = & \int \!(dt)\,d^3x\, \tilde{\epsilon}^{\,ijk} 
 \Big[ \tF_{AA}\hat{A}_i \partial_j\hat{A}_k + \tF_{AV} \hat{A}_i \partial_j \hat{\mathcal{V}}_k + \tF_{Aa} \hat{A}_i \partial_j\MFa_k 
\\[5pt]
\nonumber
& \qquad\qquad\qquad\qquad   + \frac{\tilde{c}_{VV}}{2\beta}\hat{\mathcal{V}}_i \partial_j\hat{\mathcal{V}}_k - \frac{\tilde{c}_V}{\beta^2} \hat{\mathcal{V}}_i \partial_j\MFa_k + \frac{\tilde{c}}{2\beta^3}\MFa_i \partial_j\MFa_k \Big]
\\[5pt]
\nonumber
+& \int \!(dt)\, d^3x\, e^{\MFs}\sqrt{\hat{g}} \left[\left( \F_{dA} \hat{g}^{ij} + \F_{AAdA}\hat{A}^i\hat{A}^j\right) \hat{\nabla}_i\hat{A}_j + \F_{AV}\hat{A}^i\partial_i \mathcal{V}_0 + \F_{AA}\hat{A}^i\partial_i A_0 + \F_{As}\hat{A}^i\partial_i\MFs\right],
\end{align}
where the coefficients may be functions of $\mu_V=e^{-\MFs}\mathcal{V}_0,\mu_A,T$, and $\hat{A}^2$. Again, the Chern-Simons constants $\tilde{c}_{VV}$, $\tilde{c}_{V}$, and $\tilde{c}$ can only receive corrections at one-loop order from charged matter, and therefore should stay the same as in~\eqref{E:Seff1}. There are also some two and three derivative terms which will be relevant when calculating the chiral conductivities. We have
\begin{align}
\begin{split}
\Gamma^{(2)} &= \int \!(dt)\, d^3x\, e^{\MFs}\sqrt{\hat{g}} \left( \sum_{i=1}^4 \A_i s_i +\sum_{i=1}^3 \tilde{\A}_i \tilde{s}_i\right) + (\text{irrelevant terms})\,,
\\
\Gamma^{(3)} & = \int \!(dt)\, d^3x\, e^{\MFs}\sqrt{\hat{g}}\,\tA_4\, \epsilon^{ijk} (\hat{F}_{\mathcal{V}})_{ij}\hat{\nabla}_l(\hat{F}_{\mathcal{V}})_k^{\phantom{k}l} + (\text{irrelevant terms})\,.
\end{split}
\end{align}
The one-point functions of the axial current and the energy-momentum tensor are given by the variation of $\mathscr{W}[J_{\rm ext}, A,g]$ in~\eqref{eq:TJV31}. Using~\eqref{E:VJext}, they may be equivalently written in terms of the variation of $\Gamma[\mathcal{V},A,g]$,
\beq
\label{E:TJAstat2}
     \langle T^{\mu\nu} \rangle
  = \frac{2}{\sqrt{-g}} \left( \frac{\delta \Gamma}{\delta g_{\mu\nu}}\right)_{{\cal V},A}\,, 
    \qquad 
     \langle J_A^{\mu} \rangle
  = \frac{1}{\sqrt{-g}} \left( \frac{\delta \Gamma}{\delta A_{\mu}} \right)_{{\cal V},g} \,.
\eeq
The vector current, on the other hand, can not be derived in a similar manner because of the constraint~\eqref{E:VJext}. Instead, the vector current $\JV^{\mu}$ should be parameterized in a derivative expansion in terms of the classical photon $\mathcal{V}_{\mu}$ and the sources, as in Eq.~\eqref{E:JVstat}. The expression for $\JV^{\mu}$ must be conserved, $U(1)_V$ gauge-invariant, and transform as a vector under coordinate re-parametrization. As a result it will have the same structure as the hydrostatic expressions for $\JV^i$ that we wrote down in Eqs.~\eqref{E:JVstat1}, \eqref{E:JVstat2}. To second order in derivatives it is 
\begin{align}
\label{E:JV33}
\begin{split}
\langle \JV^{i}\rangle = 
&    e^{-\MFs}\epsilon^{ijk}\partial_j \!\left(
    \tJ_0 \hat{A}_k 
   + \frac{\tilde{C}_V}{\beta} \hat{\mathcal{V}}_k + \frac{\tilde{C}}{\beta^2}\MFa_k\right) 
   + \frac{1}{\sqrt{-g}}\partial_j\!\left[ \sqrt{-g}\left( \J_0\hat{F}_{\mathcal{V}}^{ij}
   +\J_1\hat{F}_A^{ij}+\J_2 f^{ij}\right) \right] 
\\
&   + e^{-\MFs} \epsilon^{ijk} \partial_j \!\left[ 
     \tJ_1\partial_k \mathcal{V}_0
    +\tJ_2\partial_k A_0 + \tJ_3\partial_k \MFs \right] 
    + (O(\partial^2),  U(1)_A \text{-violating terms})\,,
\end{split}
\end{align}
where the $\J$'s and $\tJ$'s are functions of $\mu_V = e^{-\MFs}\mathcal{V}_0, \mu_A,T$, and $\hat{A}^2$, and the $\tilde{C}$'s are constants.

The two-point functions of the currents may be evaluated diagrammatically from the effective action $\Seff$, and expressed in terms of the parameters of $\Gamma$. The two-point functions of the axial current and the energy-momentum tensor come from the second variation of $\mathscr{W}[J_{\rm ext}, A, g]$, or equivalently from the first variation of $\langle J_A^\mu\rangle [J_{\rm ext}, A, g]$ and $\langle T^{\mu\nu}\rangle [J_{\rm ext}, A, g]$,
\begin{subequations}
\begin{align}
\label{E:GIJAT}
	G_{AA}^{\mu,\nu}(x)  & =
	\frac{1}{\sqrt{-g}} 
	\left(\frac{\delta \langle J_A^{\mu}(x)\rangle }{\delta A_{\nu}(0)}\right)_{\!J_{\rm ext},\,g}\,,
	&
	G_{TA}^{\mu\nu,\sigma}(x) & =
	\frac{1}{\sqrt{-g}} 
	\left(\frac{\delta \langle T^{\mu\nu}(x)\rangle }{\delta A_{\sigma}(0)}\right)_{\!J_{\rm ext},\,g}\,, \\[5pt]
	\nonumber
	G_{AT}^{\sigma,\mu\nu}(x)  & =
	\frac{2}{\sqrt{-g}}
	\left(\frac{\delta \langle J_A^{\sigma}(x)\rangle }{\delta h_{\mu\nu}(0)}\right)_{\!J_{\rm ext},\,A}\,, 
	&
	G_{TT}^{\sigma\tau,\mu\nu}(x)  &= 
	\frac{2}{\sqrt{-g}}
	\left(\frac{\delta \langle T^{\sigma\tau}(x) \rangle }{\delta h_{\mu\nu}(0)}\right)_{\!J_{\rm ext},\,A}\,.
\end{align}
After the variation, we set $g_{\mu\nu}=\eta_{\mu\nu}$, $A_i=0$, $A_0=\mu_A$, $J_{\rm ext}^i =0$, while $J_{\rm ext}^0$ is determined by the neutrality constraint $\partial P/\partial\mu_V = 0$. If $\langle J_A^\mu \rangle$ and $\langle T^{\mu\nu}\rangle$ are derived from the variation of $\Gamma[{\cal V},A,g]$ as in (\ref{E:TJAstat2}), the correlation functions can be written in the same way (\ref{E:GIJAT}), where now $\langle J_A^\mu \rangle = \langle J_A^\mu \rangle [{\cal V}[J_{\rm ext}, A, g], A,g]$ and $\langle T^{\mu\nu}\rangle = \langle T^{\mu\nu}\rangle [{\cal V}[J_{\rm ext}, A, g], A,g]$.
The mixed correlation functions of the vector current may also be defined by the variational procedure, given $\langle \JV^{\mu}\rangle [{\cal V}[J_{\rm ext}, A, g], A,g]$, such as Eq.~(\ref{E:JV33}),
\begin{equation}
\label{E:GIJVA}
  G_{VA}^{\mu,\nu}(x) = 
  \frac{1}{\sqrt{-g}} 
  \left( \frac{\delta \langle \JV^{\mu}(x)\rangle }{\delta A_{\nu}(0)} \right)_{\!J_{\rm ext},\,g}\,, \qquad 
  G_{VT}^{\sigma,\mu\nu}(x) = 
  \frac{2}{\sqrt{-g}} 
  \left( \frac{\delta \langle \JV^{\sigma}(x)\rangle }{\delta h_{\mu\nu}(0)} \right)_{\!J_{\rm ext},\,A}\,. 
\end{equation}
\end{subequations}
In order to have a variational prescription for $G_{AV}^{\mu,\nu}$, $G_{TV}^{\mu\nu,\sigma}$ and $G_{VV}^{\mu\nu}$, we need to introduce a source for the vector current operator (\ref{E:JVstat}) in the effective theory, $\Seff \to \Seff + \int\! \sqrt{-g}\, X_\mu \JV^\mu$. Integrating over the photon field will give rise to $\mathscr{W}[J_{\rm ext}, A, g, X]$, and the corresponding $\Gamma[{\cal V}, A, g, X]$, where ${\cal V}_\mu [J_{\rm ext},A,g,X]$ is determined by (\ref{E:VJext}). Similar to (\ref{E:TJAstat2}) we have
$$
    \langle \JV^{\mu} \rangle
  = \frac{1}{\sqrt{-g}} \frac{\delta \Gamma}{\delta X_{\mu}} \,,
$$
which is solved by $\Gamma = \Gamma[X{=}0] + \int\!\sqrt{-g}\,X_\mu \langle \JV^\mu \rangle$, as $\langle \JV^\mu\rangle$ does not depend on $X$. If the vector current is linear in $\delta V_0$ and $\hat V_i$, the effect of $X$ is to shift $J_{\rm ext}$ by an amount proportional to $X$. In this case the correlation functions of the vector current obtained through the variation with respect to $X$ can be built in terms of the photon propagators obtained through the variation with respect to $J_{\rm ext}$.

To compute correlation functions in this equivalent tree-level description, we expand the currents $\langle \JV^i\rangle$, $\langle J_A^i\rangle $, and $\langle T^{0i}\rangle $ to linear order in the $\mathcal{V}_{\mu}$ and the background fields $A_i$ and $g_{0i}$ (while keeping $A_0=\mu_A, g_{00}=-1$, and $g_{ij}=\delta_{ij}$ fixed as before), and connect the ${\cal V}$'s by exact photon propagators. Two-point functions receive contributions from two types of diagrams: (i.) those with a single photon running between two operator insertions, and (ii.) contact diagrams with no intermediate photons. In terms of the three-dimensional fields, we have
\beq
A_0 = \mu_A\,, \qquad \hat{A}_i = A_i + \frac{\mu_A}{2}g_{0i}\,, \qquad \MFs=0\,, \qquad \MFa_i = -\frac{1}{2}g_{0i}\,, \qquad \hat{g}_{ij} = \delta_{ij}+\frac{1}{4}g_{0i}g_{0j}\,.
\eeq
To second order in derivatives, the axial and momentum currents follow from variation of $\Gamma$ and to linear order in the photon, $A_i$, and $g_{0i}$ are given by
\begin{subequations}
\label{E:GammaCurrents}
\begin{align}
\begin{split}
\langle J_A^i\rangle  =&  2\frac{\partial \P}{\partial\hat{A}^2}\left(A_i+\frac{\mu_A}{2} g_{0i}\right) + \epsilon^{ijk}\left[2\tF_{AA}\partial_j\left( A_k+\frac{\mu_A}{2}g_{0k}\right) + \tF_{AV}\partial_j \hat{\mathcal{V}}_k - \frac{\tF_{Aa}}{2}\partial_j g_{0k}\right] 
\\
& + \F_{AV}\partial_i \delta\mathcal{V}_0+ 2\A_4 \partial_j \hat{F}_{\mathcal{V}}^{ij} + (\text{irrelevant two-derivative terms})\,,
\\
\langle T^{0i}\rangle  = & \frac{1}{2}\P g_{0i} + \mu_A \langle J_A^i\rangle  +\epsilon^{ijk}\left[ -\tF_{Aa}\partial_j\left( A_k+\frac{\mu_A}{2}g_{0k}\right) + \frac{\tilde{c}_V}{\beta^2}\partial_j\hat{\mathcal{V}}_k +\frac{\tilde{c}}{2\beta^3} \partial_j g_{0k} \right] 
\\
& -2 \A_3 \partial_j \hat{F}_{\mathcal{V}}^{ij}+ (\text{irrelevant two-derivative terms})\,,
\end{split}
\end{align}
where the omitted terms will not contribute to the two-point functions of spatial currents to $O(k)$. 
Meanwhile the vector current is
\begin{align}
\begin{split}
\langle \JV^i\rangle  =& \epsilon^{ijk}\partial_j\left[\tJ_0 \left(A_k+\frac{\mu_A}{2}g_{0k}\right)+\frac{\tilde{C}_V}{\beta} \hat{\mathcal{V}}_k - \frac{\tilde{C}}{2\beta^2} g_{0k}\right] + \partial_j (\J_0 \hat{F}_V^{ij}) 
\\
&+ (\text{irrelevant two-derivative terms})\,.
\end{split}
\end{align}
\end{subequations}
The photon propagator is 
\begin{align}
\begin{split}
\langle \delta V_0(k)\, \delta V_0(-k)\rangle &= -\frac{1+O(k^2)}{2\A_1k^2+\frac{\partial^2 \P}{\partial\mu_V^2} }\,, 
\\
 \langle \hat{V}_i(k)\hat{V}_j(-k)\rangle &= \frac{i (\tilde{c}_{VV}T +2\tA_4 k^2)\epsilon^{ijk}k_k + 4\A_2\delta_{ij}k^2+O(k^4)}{k^2(\tilde{c}_{VV}^2T^2-4(4\A_2^2-\tA_4\tilde{c}_{VV}T) k^2) }\,.
\end{split}
\end{align}
The zero-frequency two-point functions in momentum space are
$$
  G_{IJ}(k) = -i\beta\! \int\! d^3x\, e^{-ikx} G_{IJ}(x)\,,
$$
where $G_{IJ}(x)$ are variational correlation functions, such as (\ref{E:GIJAT}), (\ref{E:GIJVA}).
The low-momentum behavior of the two-point functions varies depending on whether the CPT-violating constant $\tilde{c}_{VV}$ vanishes. If it vanishes, then the spatial photon is massless, the photon propagator is $\sim \delta_{ij}/k^2$, and diagrams with a single photon contribute to the two-point functions at low order in momentum. For example, diagrams of this type contribute to the $O(k)$ part of two-point functions when the spatial photon runs from an $O(\partial^2)$ vertex to a $O(\partial)$ vertex. If $\tilde{c}_{VV}$ does not vanish, then the spatial photon receives a topological mass so that the low-momentum propagator is $\sim \epsilon_{ijk}k^k /k^2$, in which case the single-photon diagrams contribute at higher order in momentum. We therefore treat these cases separately.

For $\tilde{c}_{VV}=0$, we find the low-momentum two-point functions to be
\begin{subequations}
\label{E:Kubo1}
\begin{flalign}
%\langle \JV^i(k)\JV^j(-k)\rangle 
&\hspace{1cm} G_{VV}^{ij}(k) 
=  - \frac{\tilde{C}_V^2T^2}{4\A_2}\left[ \delta^{ij}-\frac{k^ik^j}{k^2}\right] + i \left[\frac{\tilde{C}_VT (4\J_0\A_2-\tilde{C}_V T\tA_4 )}{8\A_2^2} \right]\epsilon^{ijk}k_k + O(k^2)\,, &
\end{flalign}

\begin{flalign}
%\langle \JV^i(k) J_A^i(-k)\rangle  
&\hspace{1cm} G_{VA}^{ij}(k)
=  -\frac{\tilde{C}_VT\tF_{AV} }{4\A_2}\left[ \delta^{ij}-\frac{k^ik^j}{k^2}\right] & \nonumber\\
&\hspace{2.5cm} -i \left[\tJ_0-\frac{ (2\J_0 \A_2- \tilde{C}_VT\tA_4)\tF_{AV} + 4\tilde{C}_VT \A_2\A_4}{8\A_2^2}\right] \epsilon^{ijk}k_k +O(k^2)\,,&
\end{flalign}

\begin{flalign}
%\langle \JV^i(k)T^{0j}(-k)\rangle 
&\hspace{1cm} G_{VT}^{i,0j}(k)
 = \mu_A %\langle \JV^i(k)J_A^j(-k)\rangle 
   G_{VA}^{ij}(k)
   -\frac{\tilde{C}_V\tilde{c}_VT^3}{4\alpha_2}\left[ \delta^{ij}-\frac{k^ik^j}{k^2}\right] &\nonumber\\
&\hspace{2.5cm}  +i\left[ \tilde{C}T^2+\frac{(2\J_0\A_2- \tilde{C}_VT\tA_4)\tilde{c}_VT^2-4\tilde{C}_VT \A_2\A_3}{8\A_2^2}\right] \epsilon^{ijk}k_k + O(k^2)\,,&
\end{flalign}

\begin{flalign}
%\langle J_A^i(k)J_A^j(-k)\rangle  
&\hspace{1cm} G_{AA}^{ij}(k)
=  2 \frac{\partial \P}{\partial\hat{A}^2}\delta^{ij} -\frac{\tF_{AV}^2}{4\A_2}\left[\delta^{ij}-\frac{k^ik^j}{k^2}\right] &\nonumber\\
&\hspace{2.5cm}  - i \left[ 2\tF_{AA} - \frac{\tF_{AV}(8\A_4\A_2-\tF_{AV}\tA_4)}{8\A_2^2} \right] \epsilon^{ijk}k_k + O(k^2)\,,&
\end{flalign}

\begin{flalign}
%\langle J_A^i(k)T^{0j}(-k)\rangle  
&\hspace{1cm} G_{AT}^{i,0j}(k) 
=   \mu_A %\langle J_A^i(k)J_A^j(-k)\rangle 
   G_{AA}^{ij}(k)
   - \frac{\tF_{AV}\tilde{c}_VT^2}{4\A^2}\left( \delta^{ij}-\frac{k^ik^j}{k^2}\right)&
\nonumber\\
&\hspace{2.5cm}  + i \left[ \tF_{Aa} + \frac{\tilde{c}_V^2(4\A_2\A_4-\tF_{AV}\tA_4) - 4\tF_{AV}\A_2\A_3}{8\A_2^2}\right]\epsilon^{ijk}k_k+O(k^2)\,,&
\end{flalign}

\begin{flalign}
%\langle T^{0i}(k)T^{0j}(-k)\rangle  
&\hspace{1cm} G_{TT}^{0i,0j}(k)
=  2\mu_A %\langle J_A^i(k)T^{0j}(-k)\rangle 
  G_{AT}^{i,0j}(k) 
  - \mu_A^2 %\langle J_A^i(k)J_A^j(-k)\rangle 
  G_{AA}^{ij}(k)
  + P\delta^{ij} - \frac{\tilde{c}_V^2T^4}{4\A_2}\left[ \delta^{ij}-\frac{k^ik^j}{k^2}\right] &
\nonumber\\
&\hspace{2.5cm} - i \left[ \tilde{c}T^3 + \frac{\tilde{c}_VT^2(8\A_2\A_3+\tA_4\tilde{c}_VT^2)}{8\A_2^2}\right]\epsilon^{ijk}k_k+O(k^2)\,.&
\end{flalign}
\end{subequations}
The terms with an inverse factor of $\A_2$ all come from one-photon diagrams. Note that $\tA_4$, which parameterizes a three-derivative term in $\Gamma$, contributes via the one-photon diagrams to the conductivities. 

Note the appearance of infrared finite contact terms, proportional to $(\de^{ij} - k^ik^j/k^2)$, in several correlators. These arise from one-photon diagrams with both vertices built from the Chern-Simons current (magnetic field), $\tilde\ep^{ijk}\ptl_j {\cal V}_k = \langle B^i\rangle$. Since the only long-range interaction in the effective theory with $\tilde{c}_{VV}=0$ is due to the static magnetic field, it is not surprising that these infrared-finite terms correspond to the magnetic dipole-dipole two point function, $\langle B^i(r) B^j(0) \rangle \propto 1/|r|^3 (\de^{ij} - 3r^ir^j/r^2)$, up to a contact term. The anomalies and other CPT-violating coefficients then determine the leading appearance of this long-range interaction in the various current and energy-momentum correlators.

Before considering the case with nonzero $\tilde{c}_{VV}$, we see that even in classical MHD there will be new contributions to the two-point functions relative to the result~\eqref{E:KuboCon} in ordinary hydrodynamics~\eqref{E:KuboCon}. Completely neglecting photon loops, the parameters in $\Gamma$ are just the corresponding parameters of $\Wstatic$, giving 
\beq
\tF_{AA} = -\frac{2\bar{c}_g}{3}A_0\,,\hspace{.2in} \tF_{AV} = -2c_g V_0\,,\hspace{.2in} \tF_{Aa} =- c_g V_0^2 -\frac{\bar{c}_g}{3} A_0^2 - \frac{\tilde{c}_A}{\beta^2}\,,
\eeq
along with $\tilde{c}_{AA}=\tilde{c}_{AV}=\tilde{c}_V=\tilde{c}=\tilde{C}_V=\tilde{C}=0, \tJ_0=0$. The remaining response parameters $\A_2, \tA_4$ are unconstrained functions of state. Plugging these values into~\eqref{E:GammaCurrents} and~\eqref{E:Kubo1} gives equilibrium one- and two-point functions in classical MHD.

The two-point functions are somewhat simpler when $\tilde{c}_{VV}\neq 0$, 
\begin{subequations}
\label{E:Kubo2}
\begin{flalign}
%\langle \JV^i(k)\JV^j(-k)\rangle  
&\hspace{1cm}G_{VV}^{ij}(k)
= i \left[ \frac{\tilde{C}_V^2T}{\tilde{c}_{VV}}\right]\epsilon^{ijk}k_k + O(k^2)\,,&
\end{flalign}

\begin{flalign}
%\langle \JV^i(k)J_A^j(-k)\rangle  
&\hspace{1cm}G_{VA}^{ij}(k)
= - i \left[ \tJ_0 -\frac{\tilde{C}_V \tF_{AV} }{\tilde{c}_{VV}}\right]\epsilon^{ijk}k_k + O(k^2)\,,&
\end{flalign}

\begin{flalign}
%\langle \JV^i(k)T^{0j}(-k)\rangle  
&\hspace{1cm}G_{VT}^{i,0j}(k)
= \mu_A %\langle \JV^j(k)J_A^j(-k)\rangle 
  G_{VA}^{ij}(k)
  +i\left[ \tilde{C}T^2+\frac{\tilde{c}_V\tilde{C}_VT^2}{\tilde{c}_{VV}}\right]\epsilon^{ijk}k_k+O(k^2)\,,&
\end{flalign}

\begin{flalign}
%\langle J_A^i(k)J_A^j(-k)\rangle  
&\hspace{1cm}G_{AA}^{ij}(k)
=2\frac{\partial \P}{\partial\hat{A}^2}\delta^{ij}
- i\left[2 \tF_{AA}-\frac{\tF_{AV}^2}{\tilde{c}_{VV}T}\right]\epsilon^{ijk}k_k + O(k^2)\,,&
\end{flalign}

\begin{flalign}
%\langle J_A^i(k)T^{0j}(-k)\rangle  
&\hspace{1cm}G_{AT}^{i,0j}(k)
= \mu_A %\langle J_A^i(k)J_A^j(-k)\rangle 
  G_{AA}^{ij}(k)
  +i\left[ \tF_{Aa}+\frac{\tF_{AV}\tilde{c}_VT}{\tilde{c}_{VV}}\right]\epsilon^{ijk}k_k+O(k^2)\,,&
\end{flalign}

\begin{flalign}
%\langle T^{0i}(k)T^{0j}(-k)\rangle  
&\hspace{1cm}G_{TT}^{0i,0j}(k)
=  2\mu_A %\langle J_A^i(k)T^{0j}(-k)\rangle
  G_{AT}^{i,0j}(k)
  - \mu_A^2 %\langle J_A^i(k)J_A^j(-k)\rangle
  G_{AA}^{ij}(k)
%\nonumber\\
 +P\delta^{ij} - i\left[ \tilde{c}T^3-\frac{\tilde{c}_VT^3}{\tilde{c}_{VV}}\right]\epsilon^{ijk}k_k+O(k^2)\,.&
\end{flalign}
\end{subequations}
These relations are akin to Kubo formulae. By computing these correlation functions in the effective hydrostatic theory and matching to~\eqref{E:Kubo1} or~\eqref{E:Kubo2}, we obtain the parameters  in $\Gamma$ and $\langle \JV\rangle $. We will do so in the next Section to one-loop order for the case when all of the $\tilde{c}$'s and $\tilde{C}$'s vanish. 

\section{Perturbative corrections to chiral correlators}
\label{S:pert}

In Section~\ref{S:AMHs} we constructed the hydrostatic effective theory on the spatial slice relevant for field theories at $T>0$ with a dynamical photon. In this Section we will use that effective description to compute the chiral conductivities to low order in perturbation theory. We will consider theories where the CPT-violating constants $\tilde{c}_{VV}$, $\tilde{c}_{AA}$, $\tilde{c}_{AV}$ and $\tilde{c}$ vanish, and the Chern-Simons couplings $\tilde{c}_V$ and $\tilde{c}_A$ assume the values obtained when $U(1)_V$ is a global symmetry, namely $\tilde{c}_V=0,\tilde{c}_A = - 8 \pi^2 c_m$. However, we must discuss some generalities before proceeding to calculate correlation functions.

Previously, we ordered terms in the hydrostatic generating functional $\Wstatic$ by the number of derivatives. That sort of power counting is appropriate in hydrodynamics where the natural dimensionless small parameter is $k\lambda_{\rm mfp}$ for $k$ an inverse gradient and $\lambda_{\rm mfp}$ the mean free path. But in an effective theory of the photon the more natural power counting is the usual one where we order terms in $\Seff$ by their operator dimension. Since the hydrostatic effective theory is three-dimensional, the free-field power counting is slightly different than for a four-dimensional effective theory. To deterimine it we turn off background fields, in which case the effective action simplifies to~\eqref{E:SeffTrivial}. The kinetic terms for $V_0$ and $\hat{V}_i$ are canonically normalized if we rescale these fields by factors of $\sqrt{T}$ so that we assign them to have the three-dimensional free-field dimension $1/2$. Similarly, we assign $A_{i}$ to have dimension $1/2$ and the $\MFa_i$ to have dimension $1/2$.

On reinserting the background fields, the irrelevant operators will contribute to loop corrections to the relevant and marginal interactions, in particular renormalizing the Chern-Simons-like couplings $\tf$'s which contribute to the chiral conductivities. Irrelevant interactions give UV-sensitive loop corrections, and we will observe this feature below. The UV cutoff for the spatial effective theory is $\Lambda \lesssim T$, corresponding to the first Matsubara mode, and thus these UV sensitive terms must be fixed by matching to a given UV theory at this scale. However, the effective theory does capture the physics of the Matusbara zero-modes, and will be able to compute finite corrections to chiral conductivities associated with these IR effects.

In addition to counting operator dimensions properly, we must employ some sort of perturbative expansion in order to obtain consistent results for correlation functions to a given order in perturbation theory. While we have QED-like theories in mind in what follows, we do not specialize to a particular microscopic theory. As a result we must be a little schematic in our assumptions about the relative sizes of the various coefficients. We detail our assumptions in the next Subsection.

\subsection{Feynman rules}
We proceed by first evaluating the relevant propagators and interaction vertices in the effective theory. Since we are interested in computing correlation functions of the $J_A^i, \JV^i$, and $T^{0i}$ we want to compute the response of the effective action to perturbations of $A_i$ and $g_{0i}$ in a background where $A_0=\mu_A, g_{00}=-1$, and $g_{ij}=\delta_{ij}$. We also employ an external current $J_{\rm ext}^0$ to enforce charge neutrality at nonzero $\mu_V$. The fluctuation $V_0$ denotes a fluctuation around this state. The three-dimensional external sources are then
\label{E:flucs}
\beq
A_0 = \mu_A\,, \qquad \hat{A}_i = A_i + \frac{\mu_A}{2}g_{0i}\,, \qquad \MFs=0\,, \qquad \MFa_i = -\frac{1}{2}g_{0i}\,, \qquad \hat{g}_{ij} = \delta_{ij}+\frac{1}{4}g_{0i}g_{0j}\,.
\eeq
To quadratic order in the background fields and including low-dimension operators relevant for the chiral conductivities, the effective action is
\begin{equation}
 \label{E:Seff41}
 \Seff = \int \!(dt) d^3x \,\mathcal{L}_{\rm eff}\,, \qquad \mathcal{L}_{\rm eff} = \mathcal{L}_0 + \mathcal{L}_{1} +\hdots\,,
 \end{equation}
 where the subscript indicates the number of derivatives in the term. The zeroth and one-derivative pieces are
\begin{subequations}
 \label{E:SeffQuad}
 \begin{align}
 \mathcal{L}_0 &= \left(1-\frac{{g}_{0i}{g}_{0i}}{8}\right)\sum_{n=2}^6 \frac{1}{n!}\frac{\partial^n \bar\p}{\partial V_0^n}V_0^n\,,
\\
\nonumber
 \mathcal{L}_{1} &= -\left( 2c_g (\mu_V+V_0)-\epsilon^2\delta \tf_{AV}\right)\epsilon^{ijk}\left(A_i+\frac{\mu_A}{2}g_{0i}\right)\partial_j \hat{V}_k 
\\
\label{E:L1}
&\hspace{.3in}- \left( \frac{2\bar{c}_g}{3}\mu_A-\epsilon^2\delta \tf_{AA}\right) \epsilon^{ijk}\left( A_i+\frac{\mu_A}{2}g_{0i}\right)\partial_j \left( A_k+\frac{\mu_A}{2}g_{0k}\right)
\\
\nonumber
&\hspace{.3in} + \left( c_g(\mu_V+V_0)^2 + \frac{\bar{c}_g}{3}\mu_A^2 - \frac{8\pi^2c_m}{\beta^2}-\epsilon^2\delta\tf_{Aa}\right) \epsilon^{ijk}\left( A_i + \frac{\mu_A}{2}g_{0i}\right)\partial_j \frac{g_{0k}}{2}\,,
\end{align}
where in writing $\mathcal{L}_1$ we have omitted the parity-preserving terms in the first line of~\eqref{S1axial} on the grounds that those terms are irrelevant for the calculation to follow. 
We have introduced a perturbation theory counting parameter $\epsilon$ which parametrizes the strength of the photon interactions, and in the case of QED can be taken as the electromagnetic coupling $e$. The perturbative counting is implemented by the appropriate rescaling the parameters of the effective theory in Section~\ref{S:331}, such as $\delta\tf_{AV}\to \epsilon^2 \delta\tf_{AV}$ and $\delta\tj_0 \to \epsilon^2 \delta\tj_0$.
Working with QED-like theories, we assume that the terms in $\mathcal{L}_0$ as well as the anomaly coefficients are $O(\epsilon^0)$ in the perturbative expansion. The $\delta\tf$'s are generated by high-momentum loops, and so we assume that they are accompanied by factors of at least $O(\epsilon^2)$. The term with two derivatives is
\begin{align}
\label{E:L2}
\begin{split}
 \mathcal{L}_2 & = \left(\frac{\a_1}{\epsilon^2}+\a_1'V_0\right) \left(\delta_{ij}+\frac{2{g}_{0i}{g}_{0j}-\delta_{ij}{g}_{0k}{g}_{0k}}{8}\right)\partial^i V_0\partial^j V_0
\\
& \;\;\;\;\;\;+ \left(\frac{\a_2}{\epsilon^2}+\a_2'V_0\right)\left( \delta_{ij} + \frac{4{g}_{0i}{g}_{0j}-\delta_{ij}{g}_{0k}g_{0k}}{8} \right) \hat{F}_V^{ik} \hat{F}_V{\ab{j}{k}}
\\
& \;\;\;\;\;\;+ \left[\frac{\a_3+\a_3'V_0}{\epsilon^2}f_{ij} +(\a_4+\a_4'V_0) \hat{F}_{A,ij}\right]\hat{F}_V^{ij} 
\\
& \;\;\;\;\;\;+ \epsilon^{ijk}\partial_iV_0\left[-\frac{1}{2}(\ta_1+\ta_1'V_0)\partial_j g_{0k} + (\ta_2+\ta_2' V_0)\partial_j A_k + (\ta_3+\ta_3'V_0)\partial_j \hat{V}_k \right] + \cdots\,.
\end{split}
 \end{align}
The indices are raised and lowered with the flat Minkowski metric, $a_i' = \ptl a_i/\ptl  V_0$ etc,  and the dots refer to the two-derivative operators that do not contribute to the correlation functions of interest (for example, the $f_{ij}f^{ij}$ term necessarily gives two factors of the external momenta and so cannot contribute to the $O(\partial)$ part of $n$-point functions) as well as all higher-derivative operators. In writing $\mathcal{L}_2$ we have also introduced a ``natural'' scaling with $\epsilon$. In order to determine the approximate perturbative scalings in QED-like theories, consider the Maxwell term for the four-dimensional photon. In a general hydrostatic background, it decomposes as
\beq
\label{E:L3}
-\frac{\sqrt{-g}}{4e^2}F_{V,\mu\nu}F_V^{\mu\nu} = \frac{e^{\MFs}\sqrt{\hat{g}}}{4e^2}\left( 2e^{-2\MFs}(\partial V_0)^2 - \hat{F}_{V,ij}\hat{F}_V^{ij} -2 V_0\hat{F}_{V,ij}f^{ij} -2V_0^2f_{ij}f^{ij}\right).
\eeq
We see that the couplings $\a_1,\a_2,\a_3$, and $\a_3'$ are of order $O(e^{-2})$ in the hydrostatic effective action for QED. With this motivation, we assume that $\a_1,\a_2,\a_3$, and $\a_3'$ are $O(\epsilon^{-2})$ in our perturbative expansion, while taking the other couplings in $\mathcal{L}_2$ to be of $O(\epsilon^0)$. 

Finally, we will need the unique three-derivative term which contributes to the spatial photon propagator at low order in momentum. It is
\beq
\mathcal{L}_3 = \ta_4 \epsilon^{ijk}\partial_i \hat{V}_k\partial_l(\hat{F}_V)_k^{\phantom{k}l}+\cdots\,,
\eeq
\end{subequations}
where the dots indicate terms which are unimportant for us.
The coefficient $\ta_4$ is taken as $O(\epsilon^0)$.

Our perturbative power counting in $\epsilon$ is slightly different from the usual one in which perturbative couplings are introduced into interaction vertices and the kinetic terms are canonically normalized. Instead we have kept the perturbative parameter $\epsilon$ in the kinetic terms. In order to translate from our conventions to the ones with perturbative vertices, one simply recales $V_0\to \epsilon V_0, \hat{V}_i\to\epsilon\hat{V}_i,\mu_V\to \epsilon \mu_V$.

Setting the background fields to vanish, the relevant part of the effective action is
\begin{align}
\begin{split}
\Seff = \int\! (dt) d^3x &\left[ \mathcal{L}_0 + \left(\epsilon^{-2}\a_1+\a_1'V_0\right)(\partial V_0)^2 + \left(\epsilon^{-2}\a_2+\a_2'V_0\right)\hat{F}_V^2 \right.
\\
&\qquad \left.+ \epsilon^{ijk}\partial_i V_0 \left( (\ta_3+\ta_3'V_0)\partial_j\hat{V}_k\right) + \ta_4 \epsilon^{ijk}\partial_i \hat{V}_k \partial_l (F_V)_k^{\phantom{k}l}\right] + \cdots\,.
\end{split}
\end{align}
The couplings $\ta_3$ and $\ta_3'$ multiply total derivative, and so do not contribute perturbatively to correlation functions. The terms in $\mathcal{L}_0$ give a mass term and scalar potential for $V_0$, and there are addition cubic irrelevant interactions parameterized by $\a_1'$ and $\a_2'$. Neglecting higher-dimension operators, we see that parity is violated only through the coupling $\ta_4$. Ignoring cubic and higher interactions, we find the photon propagator to be
\beq
  \langle V_0(k)V_0(-k)\rangle 
  = -\frac{\epsilon^2}{\epsilon^2 \bar{p}'' + 2\a_1 k^2}\,, \qquad 
  \langle \hat{V}_i(k)\hat{V}_j(-k)\rangle 
  = - \epsilon^2 \frac{2\a_2\delta^{ij} + i \epsilon^2\ta_4 \epsilon^{ijk}k_k}{8\a_2^2 k^2}\,,
\eeq
where we have chosen Feynman gauge for the spatial photon. Note that in our perturbative expansion, the scalar mass $m^2\equiv \frac{\epsilon^2}{2 a_1}\bar{p}''$ is $O(\epsilon^2)$ and thus perturbatively small.%
\footnote{
	This is precisely the case in QED, where the mass here is of the order 
	of the Debye mass $m_D\propto |e|T$.
} 
Varying the effective action \eqref{E:Seff41} with respect to the ${A}_i$, and ${g}_{0i}$, we obtain the axial and momentum currents,
\begin{subequations}
\label{E:currents}
\begin{align}
\label{E:JATpert}
J_A^i  = &  -(2c_g(\mu_V+V_0)-\epsilon^2\delta\tf_{AV})\epsilon^{ijk}\partial_j\hat{V}_k - 2\left( \frac{2\bar{c}_g}{3}\mu_A-\epsilon^2\delta\tf_{AA}\right)\epsilon^{ijk}\partial_j A_k 
\\
\nonumber
& + \left( c_g(\mu_V+V_0)^2 - \bar{c}_g\mu_A^2 - \frac{8\pi^2c_m}{\beta^2} - \epsilon^2\delta\tf_{Aa} + 2\epsilon^2\mu_A \delta \tf_{AA}\right)\epsilon^{ijk}\partial_j\frac{g_{0k}}{2}
\\
\nonumber
& + 2\partial_j\left[(\a_4+\a_4'V_0)\hat{F}_{{V}}^{ij}\right]\,,
\\
\nonumber
T^{0i} = &\frac{1}{2}\bar\p\, \delta^{ij}+ \mu_A J_A^i + \epsilon^{ijk}\partial_j\left[\left( c_g(\mu_V+V_0)^2  + \frac{\bar{c}_g}{3}\mu_A^2 -\frac{8\pi^2 c_m}{\beta^2}-\epsilon^2\delta\tf_{Aa}\right)\left(A_k + \frac{\mu_Ag_{0k}}{2}\right)  \right]
\\
\nonumber
& -2\epsilon^{-2} \partial_j \left[ (\a_3 +\a_3'V_0) \hat{F}_V^{ij}\right] \,.
\end{align}

As we discussed in Section~\ref{S:AMHs}, correlation functions of the vector current $\JV^i$ may be computed by expressing $\JV^i$ in terms of the background fields and the photon in a derivative expansion as in~\eqref{E:JVstat}-\eqref{E:JVstat2}. Expanding $\JV^i$ to linear order in the background fields $A_i$ and $g_{0i}$ as well as keeping terms of low operator dimension, we have
\beq
\label{E:JVpert}
\JV^i = \epsilon^2\epsilon^{ijk}\partial_j\left( \delta \tj_0 A_k + \frac{\delta \tilde{C}_V}{\beta} \hat{V}_k  + \left(\delta\tj_0\mu_A - \frac{\delta \tilde{C}}{\beta^2}\right)\frac{g_{0k}}{2}\right) + \partial_j \left[ (\j_0+\j_0' V_0)\hat{F}_V^{ij}\right]+\hdots\,,
\eeq
\end{subequations}
where the dots indicate terms which will be unimportant for us, and we remind the reader that by the discussion near~\eqref{E:JVstat1} the $\delta\tilde{C}$'s are constant, must be at least perturbatively small in the gauge coupling, and most likely vanish in the absence of four-dimensional Lorentz and CPT-violating background sources. 
We retain them here for completeness, but enforce the perturbative scaling that $\delta \tj_0,\delta\tilde{C}_V,\delta\tilde{C}$ are accompnaied by factors of at least $\epsilon^2$, while $\j_0,\j_0'$ may be $O(\epsilon^0)$.

\subsection{Perturbative computations}
\label{S:calc}

Rather than compute the chiral conductivities directly, we will compute the parameters appearing in the 1PI effective action $\Gamma$ we discussed in Section~\ref{S:Gamma}. By inserting the relevant expressions for the parameters into the one and two-point functions,~\eqref{E:GammaCurrents} and~\eqref{E:Kubo1} respectively, we obtain the quantum-corrected response. We will content ourselves to compute the leading quantum corrections to the parameters $\tJ_0,\tF_{AA}, \tF_{AV},$ and $\tF_{Aa}$. We stop there because that is sufficient to demonstrate that all anomaly-induced response receives quantum corrections when $U(1)_V$ is gauged. As a byproduct we obtain the leading corrections to the chiral vortical conductivity; the leading order ${\cal O}(e^2)$ term has recently been computed in QED in~\cite{Golkar:2012kb,Hou:2012xg}. We discuss the matching between the two in Section~\ref{4D}.

We work to low order in perturbation theory in the former expansion parameter $\epsilon$. By the power counting we imposed in the previous Subsection, we see that intermediate spatial photons introduce a factor of $\epsilon^2$, while intermediate $V_0$'s introduce a factor of $\epsilon$ that depends on whether $V_0$ appears in a tree-level exchange or inside of a loop. 
When $V_0$ appears in a tree-level exchange, its propagator is $O(\epsilon^0)$ at low momentum. Inside a loop, the $V_0$ propagator introduces at least one additional factor of $\ep$.

Let us now determine the $\epsilon$-counting for diagrams which contribute to the two-point functions of the axial, momentum, and vector currents~\eqref{E:currents}. Since the interaction vertices $\a_1',\a_2',\ta_4$ are $O(\epsilon^0)$, a diagram with $n_T$ tree-level $V_0$'s, $n_0$ intermediate $V_0$'s inside a loop, $n_V$ spatial photons, and $O(\epsilon^0)$ operator insertions then scales at minimum with a factor of $\epsilon^{n_0+2n_V}$. Insertions with a $\delta\tf,\delta\tj_0$ or $\delta\tilde{C}$ introduce an extra factor of $O(\epsilon^2)$, while insertions with a factor of $\a_3$ or $\a_3'$ introduce a factor of $O(\epsilon^{-2})$.

\subsubsection{Corrections to $\tJ_0$}

The parameters $\delta\tj_0$ and $\tJ_0$ appearing in~\eqref{E:JVstat1}, \eqref{E:GammaCurrents} control the response of the vector current to the source for the axial current. As they explicitly violate $U(1)_A$ symmetry, they are both at least $O(\epsilon^2)$ in our formal power counting, which upon rescaling $\delta\tj_0$ becomes 
\beq
\tJ_0=\epsilon^2 \,\delta\tj_0 + O(\epsilon^4)\,.
\eeq 
The leading quantum corrections to $\tJ_0$ then come from two one-loop diagrams. Since we are computing $O(k)$ terms in the effective action, we only consider the $O(k)$ parts of the relevant diagrams. The first loop has a $V_0$ and $\hat{V}_i$ in the virtual intermediate state connecting the vertices
\beq
\JV^i = \ldots + \partial_j (\j_0'V_0\hat{F}_V^{ij})+ \dots\,, \qquad J_A^i = \ldots -2 c_g V_0 \epsilon^{ijk}\partial_j\hat{V}_k+ \dots\,.
\eeq
The second diagram at this order is a $V_0 V_0$ bubble arising from the vertex $\epsilon^2 \delta\tj_0''$ appearing in the vector current.

\begin{figure}[t]
\centerline{\includegraphics[trim={2cm 18.5cm 2cm 2cm},clip,width=0.5\textwidth]{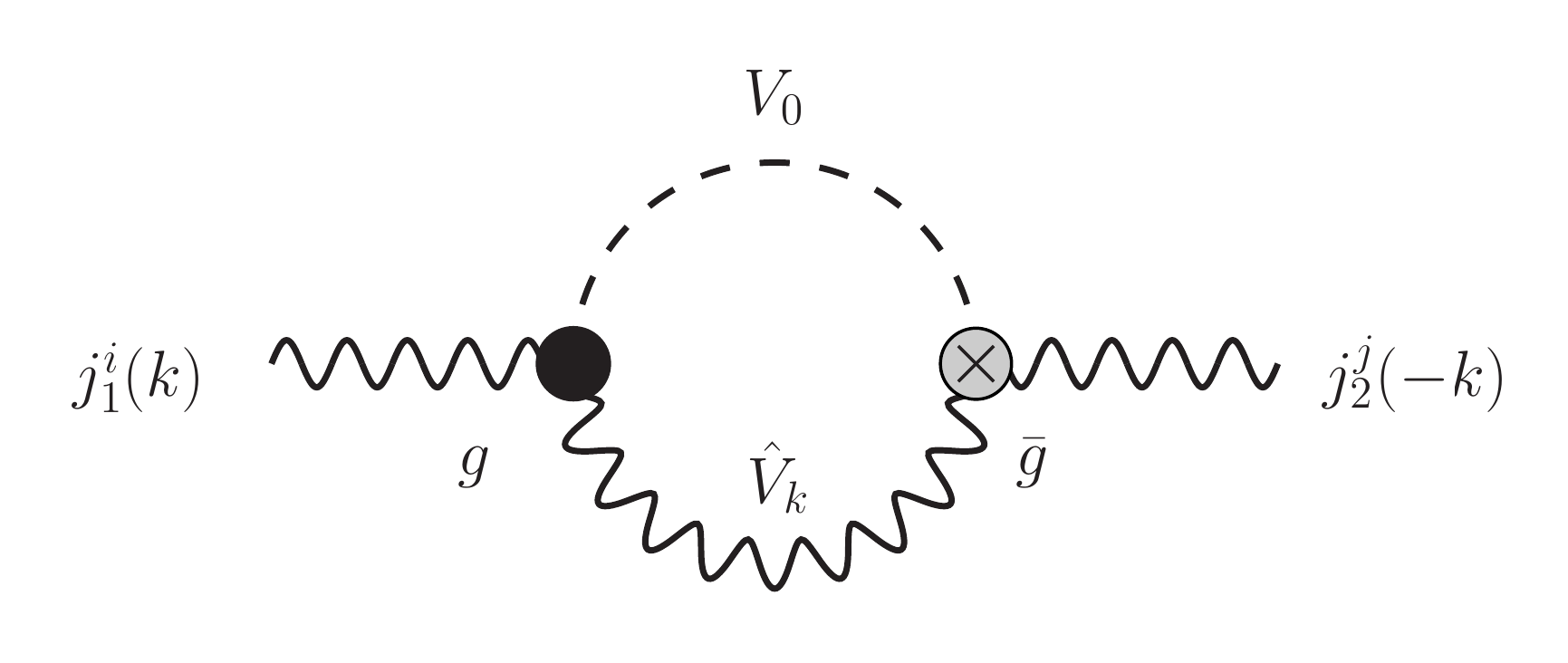}\includegraphics[trim={2cm 18.5cm 3cm 2cm},clip,width=0.5\textwidth]{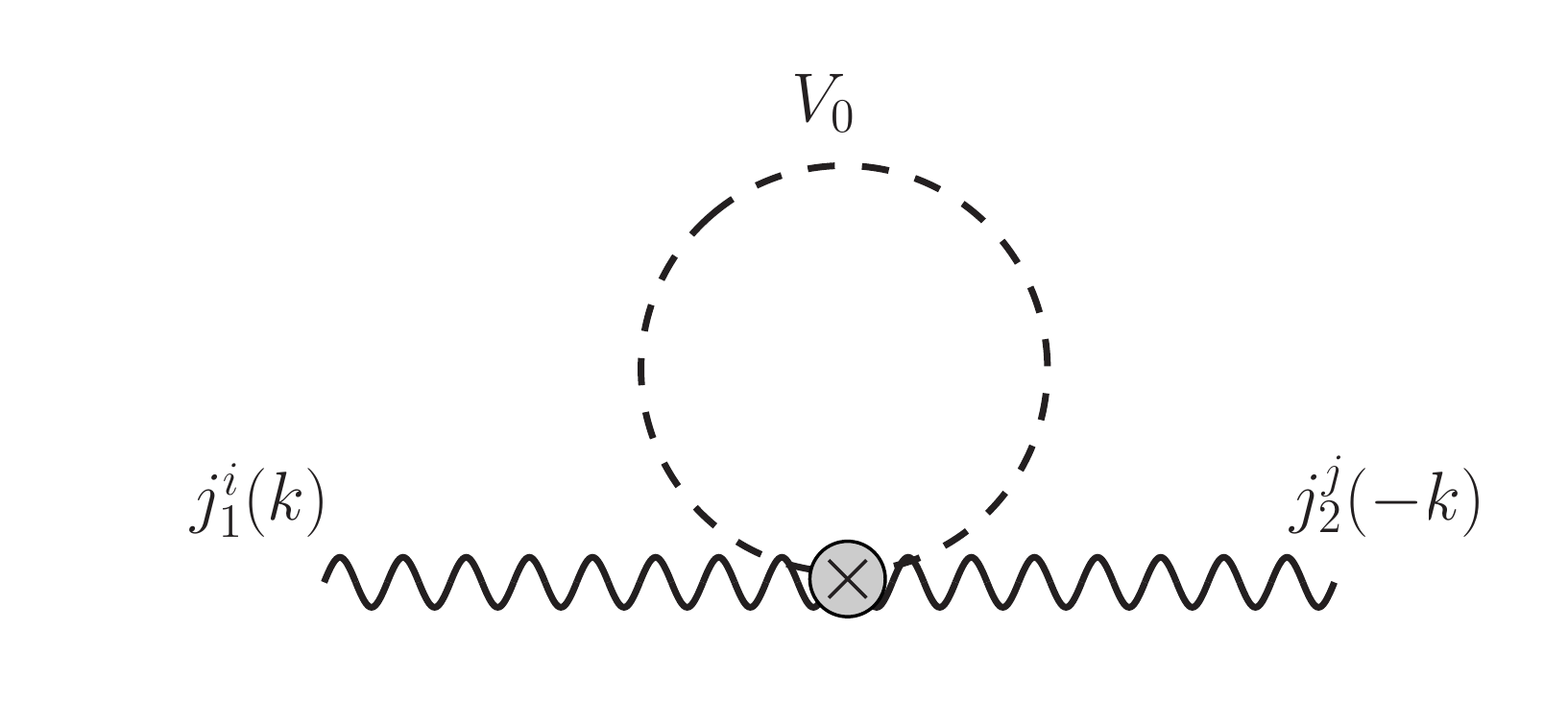}}
\caption{\footnotesize \label{F:VA} The basic one-loop diagrams correcting the various chiral conductivities. The cross denotes the P-violating vertex $\tilde{g}$ in~\protect\eqref{E:j1j2}, and the solid dot indicates the P-preserving vertex $g$ as discussed in the text.}
\end{figure}

These are the essential one-loop diagrams that appear when computing $\tJ_0$ as well as the $\tF$'s, albeit with different couplings in each case. Both diagrams are depicted in Figure~\ref{F:VA}. As a result we evaluate the loops with arbitrary couplings and then insert the relevant couplings when evaluating the parameters of $\Gamma$. Beginning with the $V_0 \hat{V}_i$ loop we consider two currents $j_1$ and $j_2$ with
\beq
\label{E:j1j2}
j_1^i = \partial_j( g V_0 \hat{F}_V^{ij})\,, \qquad j_2^i = \tilde{g}V_0\epsilon^{ijk}\partial_j\hat{V}_k\,.
\eeq
At low momentum we have
\begin{align}
\nonumber
\langle j_1^i(k)j_2^j(-k)\rangle_{V_0 \hat{V}} 
&= i\epsilon^4\frac{ g\tilde{g}}{8\a_1\a_2\beta}\int \frac{d^3p}{(2\pi)^3} \frac{1}{p^2}\frac{1}{p^2+m^2}\left( p^i \epsilon^{jkl}p_j - \epsilon^{ijk}p_k p^l\right)k_l + O(\epsilon^6,k^2)\,,
\\
& = -i \epsilon^4\frac{ g\tilde{g}}{12\a_1\a_2\beta}\epsilon^{ijk}k_k \left( \int \frac{d^3p}{(2\pi)^3}\frac{1}{p^2+m^2}\right) + O(\epsilon^6,k^2)\,,
\end{align}
where $m^2=\frac{\epsilon^2}{2\a_1} \bar p''$ is the scalar mass. The terms of $O(\epsilon^6)$ and higher come from perturbative corrections to the propagators involving $\ta_4$ and higher derivative operators. The integral in the second line is linearly divergent and so we regulate it with a hard UV cutoff $\Lambda$, giving%
\footnote{
	Although a fixed momentum cutoff is generally inconsistent with the $U(1)_V$ vector Ward identity, it is acceptable here as we will not need to shift the loop momentum in the integral. We retain the power-like divergences to highlight the appearance of terms which, after matching to the four-dimensional theory, are associated with the nonzero Matsubara modes and can arise at leading order in perturbation theory.
}
\beq
\label{E:j1j2Loop}
\langle j_1^i(k)j_2^j(-k)\rangle_{V_0 \hat{V}} = - i\epsilon^4\frac{g\tilde{g}T}{24\pi^2\a_1\a_2}\epsilon^{ijk}k_k\left( \Lambda - \frac{\pi|m|}{2} + O\left(\frac{m^2}{\Lambda}\right) \right)+ O(\epsilon^6,k^2)\,,
\eeq
Note that since $m\sim O(\epsilon)$ the subleading corrections in the $\Lambda\to\infty$ limit are $O(\epsilon^6)$, and so may be consistently neglected to this order. 

The $V_0 V_0$ bubble arises from a vertex
\beq
\label{E:j1i4}
j_1^i = \epsilon^{ijk} \partial_j \left( \frac{1}{2} \tilde{g}'' V_0^2 A_k\right)\,,
\eeq
where we treat $A_k$ as a source for $j_2^k$. The resulting bubble gives
\begin{align}
\begin{split}
\label{E:j1j2Bubble}
\langle j_1^i(k)j_2^j(-k)\rangle_{V_0 V_0} &= i\frac{\epsilon^2 \tilde{g}''}{4\a_1\beta}\epsilon^{ijk}k_k\int \frac{d^3p}{(2\pi)^3}\frac{1}{p^2+m^2} +O(k^2,\epsilon^4)\,,
\\
&= i \frac{\epsilon^2\tilde{g}''T}{8\pi^2 \a_1}\epsilon^{ijk}k_k \left( \Lambda - \frac{\pi |m|}{2}\right) + O(k^2,\epsilon^4)\,,
\end{split}
\end{align}
where we have again imposed a hard UV cutoff. For the case at hand we substitute $g=\j_0', \tilde{g}=-2c_g$, and $\tilde{g}'' = \epsilon^2 \delta \tj_0''$,  and upon noting how $\tJ_0$ appears in~\eqref{E:Kubo1}, we find
\beq
\tJ_0 = \epsilon^2\delta\tj_0 -\frac{ \epsilon^4T}{4\pi^2 \a_1 }\left[ \frac{\j_0'c_g}{3\a_2} + \frac{\delta \tj_0''}{2}\right]\left(\Lambda-\frac{\pi|m|}{2}\right) + O(\epsilon^6)\,.
\eeq
The two contributions in~\eqref{E:j1j2Loop} can be interpreted as follows. The power-like UV divergence is a consequence of the irrelevant interactions, and reflects the physics of the nonzero Matsubara modes that are integrated out in obtaining the spatial effective theory. This term needs to be matched to a specific UV completion. The second finite term, proportional to $|m|$ is non-analytic in the coupling and reflects the IR physics described by the effective theory. It is independent of the precise UV completion, other than the matching required to fix the infrared parameters.

\subsubsection{Corrections to $\tF_{AA}$}

The parameter $\tF_{AA}$, which controls the response of the axial current to the source for the axial current, is
\beq
\tF_{AA} = - \frac{2\bar{c}_g}{3}\mu_A + \epsilon^2\delta\tf_{AA}  +(\text{loops})\,.
\eeq
The leading loop corrections to it arise from a $V_0 \hat{V}_i$ loop of the form~\eqref{E:j1j2Loop} as well as a $V_0 V_0$ bubble. Actually, the $V_0 \hat{V}_i$ loop is twice~\eqref{E:j1j2Loop} (owing to the fact that that the $j_1=j_2=J_A$) with the identification $g=2\a_4'$ and $\tilde{g}=-2c_g$. There is also a $V_0 V_0$ bubble of the form~\eqref{E:j1j2Bubble} with $\tilde{g}'' = 2\epsilon^2 \delta\tf_{AA}''$. Combining the loops and noting how $\tF_{AA}$ appears in the axial-axial two-point function~\eqref{E:Kubo1} we find
\beq
\tF_{AA} =  - \frac{2\bar{c}_g}{3}\mu_A + \epsilon^2\delta\tf_{AA} -\frac{ \epsilon^4 T}{2\pi^2 \a_1}\left[ \frac{\a_4'c_g }{3\a_2}+\frac{\delta\tf_{AA}''}{4}\right]\left(\Lambda - \frac{\pi|m|}{2}\right) + O(\epsilon^6)\,.
\eeq

\subsubsection{Corrections to $\tF_{AV}$}

The response of the axial current to a magnetic field is controlled by the parameter $\tF_{AV}$, which is given by
\beq
\tF_{AV} = -2c_g \mu_V + \epsilon^2 \delta \tf_{AV} + (\text{loops})\,.
\eeq
Just as for $\tJ_0$ and $\tF_{AA}$, the leading corrections arise from a $V_0 \hat{V}_i$ loop of the form~\eqref{E:j1j2Loop} as well as the $V_0 V_0$ bubble. In terms of the loops~\eqref{E:j1j2Loop} and~\eqref{E:j1j2Bubble} we identify $g=4\a_2',\tilde{g}=-2c_g$ and $\tilde{g}'' = \epsilon^2 \delta\tf_{AV}$. Putting the loops together we find
\beq
\tF_{AV} = -2c_g \mu_V + \epsilon^2 \delta \tf_{AV} - \frac{\epsilon^4 T}{\pi^2 \a_1}\left[ \frac{\a_2'c_g }{3\a_2}+\frac{\delta\tf_{AV}''}{8}\right]\left(\Lambda - \frac{\pi|m|}{2}\right) + O(\epsilon^6)\,.
\eeq

\subsubsection{Corrections to $\tF_{Aa}$}

At zero axial chemical potential, the parameter $\tF_{Aa}$ governs the response of the axial current to vorticity. It is
\beq
\tF_{Aa} = - c_g \mu_V^2 - \frac{\bar{c}_g}{3}\mu_A^2-8\pi^2c_mT^2 +\epsilon^2\delta \tf_{Aa} +(\text{loops})\,.
\eeq
Like the leading corrections to $\tJ_0$ and the other $\tF$'s, the lowest order quantum corrections arise from a $V_0 \hat{V}_i$ loop of the form~\eqref{E:j1j2Loop} as well as a $V_0 V_0$ bubble. Unlike the corrections above which were at least $O(\epsilon^4)$, these are $O(\epsilon^2)$. There are two $V_0 \hat{V}_i$ loops: in the first we identify $j_1^i=J_A^i,j_2^j=T^{0j}$ along with $g=2\a_4'$ and $\tilde{g}=-2c_g\mu_A$, and in the second we identify $j_1^i=T^{0i},j_2^j=J_A^j$ along with $g=-2(\epsilon^{-2}\a_3'-\a_4'\mu_A)$ as well as $\tilde{g}=-2c_g$. The first loop is $O(\epsilon^4)$ while the second is $O(\epsilon^2)$. The $V_0 V_0$ loop is of the form~\eqref{E:j1j2Bubble} with an coupling $\tilde{g}''=2c_g$. The end result is
\beq
\tF_{Aa} = - c_g \mu_V^2 - \frac{\bar{c}_g}{3}\mu_A^2-8\pi^2c_mT^2 +\epsilon^2\left( \delta \tf_{Aa} - \frac{c_gT}{2\pi^2 \a_1}\left[ \frac{\a_3'}{3\a_2}-\frac{1}{2}\right]\left( \Lambda - \frac{\pi|m|}{2}\right)\right) + O(\epsilon^4)\,.
\eeq

\subsubsection{Summary of the one-loop corrections}

In Section~\ref{S:Gamma} we discussed the thermal effective action $\Gamma$ and 1PI vector current. We have calculated the leading quantum corrections to the one-derivative parts of $\Gamma$ and the vector current that involve the epsilon-tensor. From the previous Subsection we find
\begin{align}
\label{E:1LoopSummary}
\nonumber
\tJ_0 &= \epsilon^2\delta\tj_0 -\frac{ \epsilon^4T}{4\pi^2 \a_1 }\left[ \frac{\j_0'c_g}{3\a_2} + \frac{\delta \tj_0''}{2}\right]\left(\Lambda-\frac{\pi|m|}{2}\right) + O(\epsilon^6)\,,
\\
\nonumber
\tF_{AA} &=  - \frac{2\bar{c}_g}{3}\mu_A + \epsilon^2\delta\tf_{AA} -  \frac{\epsilon^4 T}{2\pi^2 \a_1}\left[ \frac{\a_4'c_g }{3\a_2}+\frac{\delta\tf_{AA}''}{4}\right]\left(\Lambda - \frac{\pi|m|}{2}\right) + O(\epsilon^6)\,,
\\
\tF_{AV} &= -2c_g \mu_V + \epsilon^2 \delta \tf_{AV} - \frac{ \epsilon^4 T}{\pi^2 \a_1}\left[ \frac{\a_2'c_g }{3\a_2}+\frac{\delta\tf_{AV}''}{8}\right]\left(\Lambda - \frac{\pi|m|}{2}\right) + O(\epsilon^6)\,,
\\
\nonumber
\tF_{Aa} &= - c_g \mu_V^2 - \frac{\bar{c}_g}{3}\mu_A^2-8\pi^2c_mT^2 +\epsilon^2\left( \delta \tf_{Aa} - \frac{c_g T}{2\pi^2 \a_1}\left[ \frac{\a_3'}{3\a_2}-\frac{1}{2}\right]\left( \Lambda - \frac{\pi|m|}{2}\right)\right) + O(\epsilon^4)\,.
\end{align}
Since the effective theory has no fermionic matter and the theory is IR finite, the Chern-Simons couplings are uncorrected to all orders in perturbation theory,
\beq
\tilde{c}_{AA}=\tilde{c}_{AV}=\tilde{c}_{VV}=\tilde{c}_V=\tilde{c} = \tilde{C}_V=\tilde{C}=0\,.
\eeq
We observe that the structure of~\eqref{E:1LoopSummary} indicates that the leading corrections to $\tJ_0, \tF_{AA},$ and $\tF_{AV}$ are in fact $O(\epsilon^4)$ rather than $O(\epsilon^2)$. Upon inserting the 1PI parameters~\eqref{E:1LoopSummary} back into~\eqref{E:Kubo1}, we obtain the leading corrections to the two-point functions  of the vector, axial, and momentum currents.

\subsection{Matching to the four-dimensional theory}
\label{4D}
The loop corrections to the chiral vortical conductivity contain power-like UV divergences at $O(\epsilon^2)$, which need to be matched at the cutoff scale ($\Lambda \lesssim T$) with the full four-dimensional theory. The spatial EFT can only be used to reliably compute the non-analytic subleading corrections of $O(\epsilon^2|\epsilon|)$. In this Subsection, we relate our perturbative calculations of the CVE in the EFT to four-dimensional calculations in QED with $N_f$ Dirac fermions. The leading quantum corrections to the CVE in QED to $O(e^2)$ at $\mu_V=\mu_A=0$ have been discussed in~\cite{Golkar:2012kb,Hou:2012xg}. In this case the anomaly coefficients are
\beq
c_g = -\frac{N_f}{4\pi^2}\,, \qquad \bar{c}_g = -\frac{N_f}{4\pi^2}\,, \qquad c_m = -\frac{N_f}{96\pi^2}\,.
\eeq

We will now compute the chiral vortical conductivity $\sigma_{CVE}$ in QED, defined through
\be
 \frac{i}{2} \lim_{k\rightarrow 0}\frac{\epsilon_{ijl} k^l}{k^2}
 \langle J_A^i(k) T^{0j}(-k) \rangle = \sigma_{CVE}. \label{siV}
\ee
For simplicity we will also work at zero chemical potentials, $\mu_V=\mu_A=0$. In~\eqref{E:Kubo1} we showed how $\sigma_{CVE}$ is related to the parameters appearing in the thermal effective action $\Gamma$ and the 1PI vector current. To proceed we require knowledge of the $\tF$'s, $\A_2,\A_3,\A_4$, and $\tA_4$. For QED we have
\beq
\A_2 = - \frac{1}{4e^2} + O(e^0)\,,
\eeq
and the combination of charge conjugation and parity symmetry sets
\beq
\A_3 =\A_4=\tA_4=0\,,
\eeq
which by~\eqref{E:Kubo1} gives
\beq
\sigma_{CVE} = - \tF_{Aa}\,.
\eeq
In~\eqref{E:1LoopSummary} we expressed $\tF_{Aa}$ in terms of the parameters of the Wilsonian effective action $\Seff$, which upon using
\beq
\a_1 = \frac{1}{2e^2}+O(e^0)\,, \qquad \a_2 = - \frac{1}{4e^2}+O(e^0)\,, \qquad \a_3' = -\frac{1}{2e^2}+O(e^0)\,,
\eeq
gives
\beq
\label{E:cve}
\sigma_{CVE}=-\tF_{Aa} = -\frac{ N_fT^2}{24} -e^2\delta \tf_{Aa} - \frac{e^2 N_f T}{24\pi^4}\left( \Lambda - \frac{\pi |m|}{2}\right) + O(e^4)\,.
\eeq
The $O(e^2)$ part of the CVE depends on the UV cutoff and the parameter $\delta\tf_{Aa}$, and so is determined by matching to the UV theory in question, in this case QED. The term proportional to $|m|$ is an IR-sensitive $O(e^2|e|)$ term which is independent of the details of the UV completion. 

It is interesting to explore the origin of this correction directly in four dimensions. Indeed, the corresponding ${\cal O}(e^2)$ contributions to $\delta \tf_{Aa}$ were recently studied in~\cite{Golkar:2012kb,Hou:2012xg}, and arise via two-loop contributions to $\langle J_A^i T^{0j}\rangle$. In the Matsubara formalism, to compare with the effective theory calculation we can isolate the zero-modes $n=0$, 
\be
 \de \si_{CVE} = \de \si_{CVE}^{n=0} + \de \si_{CVE}^{n\neq 0}. 
\ee
In the analysis of~\cite{Golkar:2012kb,Hou:2012xg}, the relevant diagram factorizes into a triangle anomaly ($AVV$) sub-diagram, and the use of an effective anomaly-induced axial vertex reduces the computation to two one-loop photon bubble diagrams. However, only one of these has a nonzero contribution from the zero modes, involving a $V_0  V_j$ loop. The effective axial current vertex for the zero modes used in~\cite{Golkar:2012kb,Hou:2012xg} takes the form $\Ga_{0ji} = - \Ga_{j0i} = c_g \ep_{ijk}p^k$, where $\vec{p}$ is the spatial loop momentum. At leading order the zero mode contribution is $T$-independent, and the ${\cal O}(e^2)$ contribution to $\delta \tf_{Aa}$ arises as expected only from the nonzero modes $\de \si_{CVE}^{n\neq0}$.  However, the IR contribution of the zero mode becomes nontrivial on resumming the $\langle V_0V_0\rangle$ propagator to incorporate the effect of Debye screening. In particular, if we resum the photon self-energy, 
the propagator for the zero mode of $V_0$ is modified as $1/p^2 \rightarrow 1/(p^2+m^2)$,
and incorporates the Debye mass $m^2$. 
Since $m^2 = {\cal O}(e^2T^2) \ll T^2$ it is consistent to resum the zero mode propagator, and neglect the 
corresponding corrections to the nonzero modes. With this resummation, we find
\begin{align}
 \de \si_{CVE}^{n=0} &= -\frac{e^2N_fT}{12\pi^2} \int \frac{d^3\vec{p}}{(2\pi)^3} \frac{1}{\vec{p}^2+m^2}
     = -\frac{e^2N_fT}{24\pi^4} \left[  \La - m \arctan(\La/m)\right]_{\La\rightarrow\infty}   \nonumber\\
      &=   \mathcal{O}(\Lambda) + \frac{e^2N_fT}{48\pi^3} |m|. \label{res0}
\end{align}
The cutoff-dependent piece, together with $\delta\sigma_{CVE}^{n\neq 0}$ gives (upon renormalization) a finite $O(e^2T^2)$ correction to $\sigma_{CVE}$~\cite{Golkar:2012kb,Hou:2012xg}. The $O(e^2|e|)$ term precisely matches the result~\eqref{E:cve} from the effective theory. 

\begin{figure}[t]
\centerline{\includegraphics[trim={2cm 17.5cm 0cm 2.5cm},clip,width=0.5\textwidth]{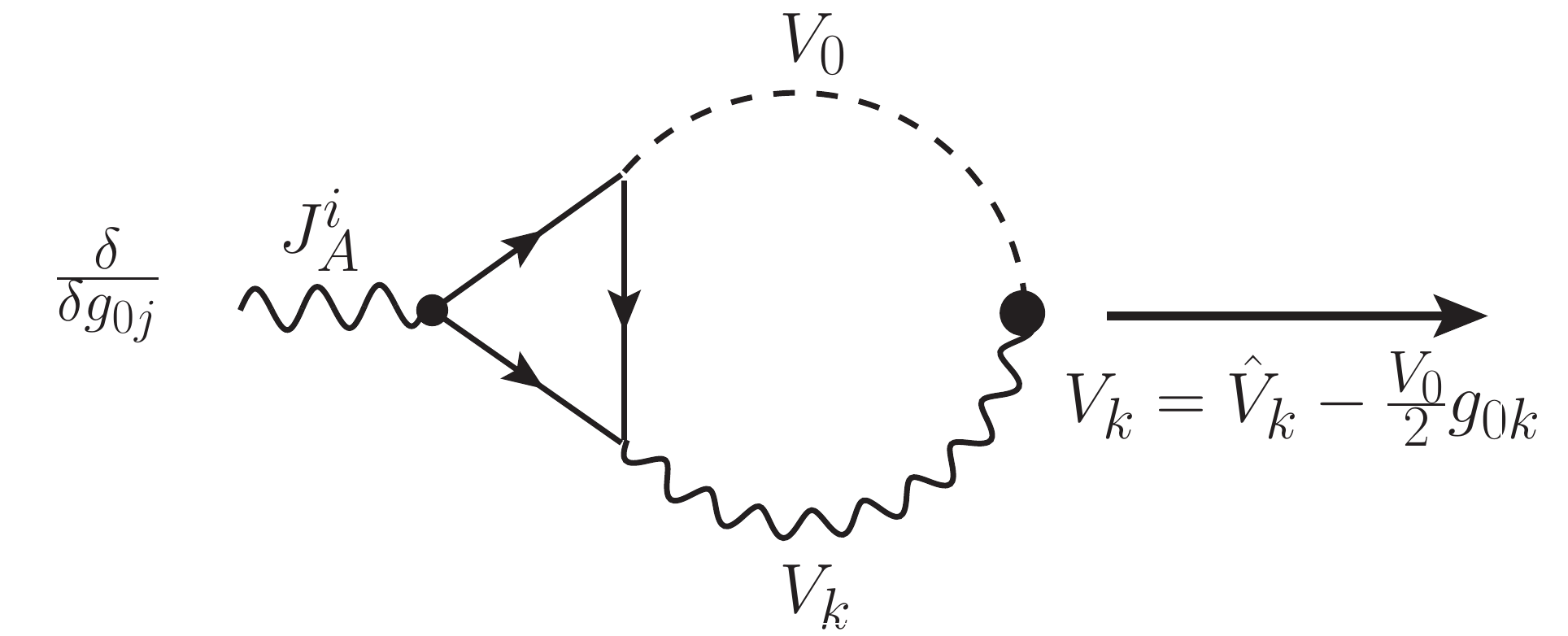}\includegraphics[trim={2.3cm 18.3cm 0cm 2cm},clip,width=0.5\textwidth]{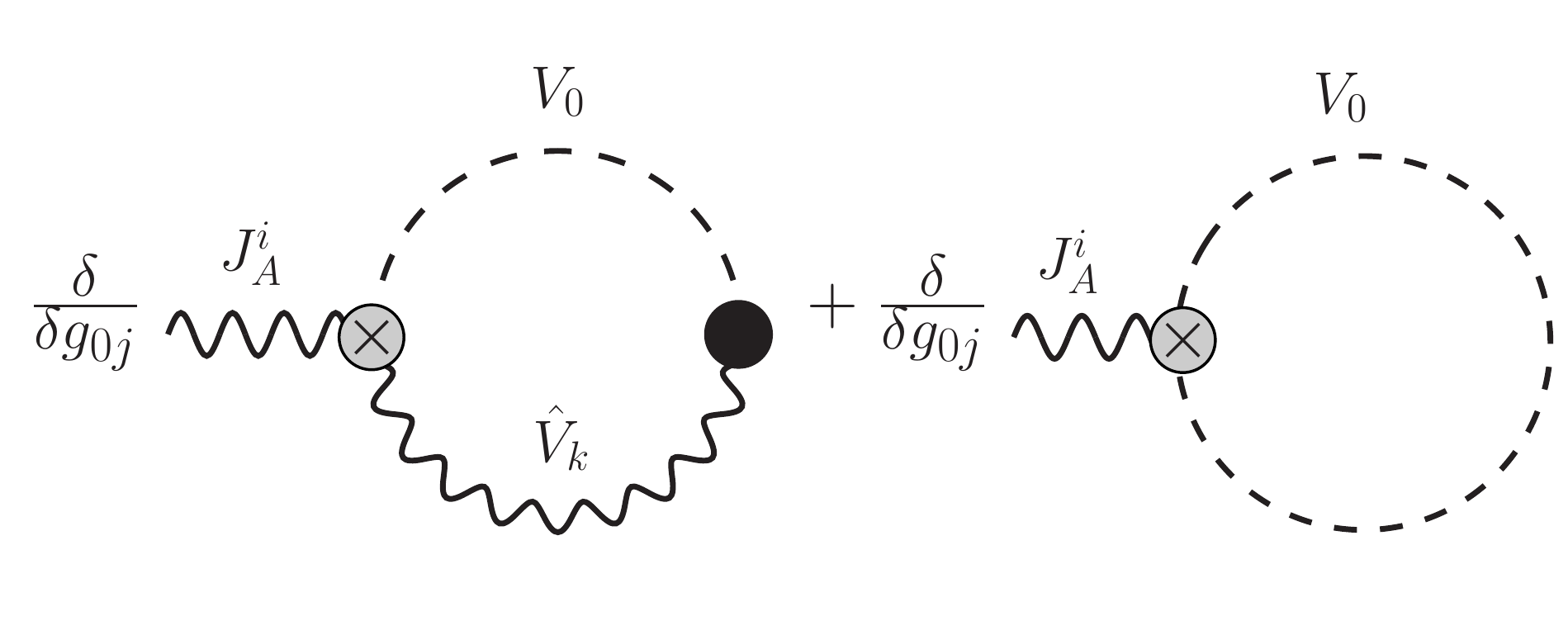}}
\caption{\footnotesize \label{F:2loop} We can associate the zero-mode contribution to the two-loop thermal diagram in four dimensions on the left with the two one-loop diagrams in the spatial effective theory on the right, where the crossed vertex represents the fermion loop. The $V_k  V_0$ mixing, denoted by a solid vertex on the left, is induced by a single $g_{0k}$ metric perturbation. The three-dimensional decomposition of the spatial vector $V_k = \hat{V}_k - \frac{V_0}{2}g_{0k}$ then leads, at linear order in $g_{0k}$, to the two diagrams on the right. The metric variations compute the two-point function via an insertion of $T^{0j}$, and the diagrams on the right are then equivalent to Fig.~\protect\ref{F:VA}).}
\end{figure}

The matching to the 3D effective theory is facilitated by studying the one-point function $\langle J_A^i\rangle$ to first order in the metric perturbation $g_{0k}$, with the two-point function following by variation. The relevant contribution is shown on the left of Fig.~\ref{F:2loop}, where the photon propagator in the bubble has been expanded to first order in $g_{0k}$, which induces the mixing between $V_0$ and $V_k$. Within the spatial effective theory, it is more natural to work with fields defined on the spatial slice, so we decompose $V_k = \hat{V}_k - \frac{V_0}{2}g_{0k}$, leading to the two diagrams on the right of Fig.~\ref{F:2loop}. These two diagrams contribute with relative factors of $1/3$ and $-1$, precisely as in the earlier discussion. 

We note that the calculations in \cite{Hou:2012xg,Golkar:2012kb} make use of a local piece of the anomalous AVV vertex which saturates the axial Ward identity, and it argued that the extra terms required to fulfill the remaining vector Ward identities do not contribute to $\si_{CVE}$. In general, these latter terms are nonlocal, as the axial Ward identity is not corrected at $T\neq 0$, and is saturated by the conventional $T=0$ vertex \cite{adler}, which is singular as $k^2\rightarrow 0$. The finite temperature vertex is more involved due to extra tensor structures associated with the fluid frame \cite{Itoyama:1982up}, and we have verified that a vertex consistent with all the Ward identities exists, and which reduces to a {\it local} expression on the spatial slice following from $W_{\rm anom}$ for spatial momenta. This is consistent with the spatial vertex for the zero-modes noted above, and used in \cite{Hou:2012xg,Golkar:2012kb}, but the full vertex is non-local in four dimensions.
 
We conclude this Subsection by noting that the AVV vertex structure, visible in the two-loop contributions in four dimensions, is also present in the leading one-loop contribution to $\si_{CVE}$ in a background chemical potential, $\langle J_A^i T^{0j}\rangle \propto c_g \mu_V^2$ computed in \cite{Landsteiner:2011cp}. Varying twice with respect to $\mu_V$, we find $\langle J_A^i J_V^0 J_V^0 T^{0j}\rangle \propto \delta/\delta g_{0j} \langle J_A^i J_V^0 J_V^0 \rangle \propto c_g$. The connection between  $\langle J_A^i J_V^0 J_V^0 \rangle$ and the AVV anomaly coefficient in this case arises indirectly via the constraints imposed by the vector Ward identities that link the longitudinal and transverse components of the AVV vertex. 
 
 \section{Discussion}
\label{S:discuss}

In this work we sought to understand the fate of anomaly-induced transport with dynamical gauge fields. Our results are summarized in Section~\ref{S:introduction}, and are captured in the derivative expansion for the effective action~\eqref{E:Seffn}-\eqref{E:Seff2}, and the two-point functions of currents~\eqref{E:Kubo1}, \eqref{E:Kubo2} that determine the appearance of the anomaly coefficients in hydrostatic response. We find that anomaly-induced transport is subject to perturbative corrections. We explicitly studied one-loop contributions to the spatial effective action, reproducing for example IR-sensitive corrections to the chiral vortical conductivity.

In the remainder of this Section, we comment on the chiral magnetic effect, magnetohydrodynamics, and the application of these results to the Standard Model, necessarily focusing on perturbative regimes in which the corrections to anomalous transport remain small.

\begin{itemize}

\item {\it Axial chemical potential}: Discussions of anomalous transport often involve the axial chemical potential. However, in the presence of dynamical $U(1)_V$ fields, this is a subtle quantity to define in equilibrium. We have made use of the source $\mu_A$ for the axial charge $Q_A = \int d^3 x J_A^0$, but this is not a true chemical potential, since $Q_A$ is not conserved. However, in the presence of the $AVV$ anomaly, one can define a {\it conserved} axial charge $Q_5 = \int d^3 x (J_A^0 - c_g  \ep^{0ijk} V_i \ptl_j V_k)$~\cite{Rubakov:2010qi} (at least when the field strength of the axial source, $(F_A)_{\mu\nu}$ vanishes). The corresponding current is not $U(1)_V$ gauge invariant, but the charge $Q_5$ is invariant, and one can consider a time-independent thermal ensemble defined by $H-\mu_5 Q_5$, with $\mu_5$ now interpreted as a genuine axial chemical potential. We now consider the implications of this source when $U(1)_V$ is dynamical. Indeed, within the spatial effective theory, the presence of this source leads to $\tilde{c}_{VV} = -2c_g \mu_5$. This has the effect of shifting the static pole of the spatial photon away from zero, leading to anti-screening \cite{Redlich:1984md,Niemi:1985ir} (see also the recent discussion in \cite{Khaidukov:2013sja}). If this constant value of $\mu_5$ is realized in terms of a dynamical field, this anti-screening observed on the spatial slice likely implies a dynamical instability. 
One method to obtain an axial chemical potential $\mu_5$ is to turn on a four-dimensional axion coupling $aF_{\mu\nu}\tilde{F}^{\mu\nu}$ with an axion profile linear in time $a\propto t$, so that $\mu_5 \propto \ptl_t a$. In this case, there is indeed an unstable growing mode for the magnetic field, and thus a consistent treatment would involve the full non-equilibrium dynamics.

\item {\it Chiral magnetic effect}: The chiral magnetic effect \cite{Vilenkin:1980fu, Alekseev:1998ds, Fukushima:2008xe, Kharzeev:2009p} can be understood qualitatively as a contribution to the electromagnetic current aligned along an external magnetic field. However, the precise realization depends on the definition of the current and whether the $U(1)_V$ field is dynamical. We consider first the case of a global (i.e. external) $U(1)_V$ magnetic field. In the presence of the $AVV$ anomaly and the source $\mu_A$, the consistent and covariant currents differ by a Bardeen-Zumino polynomial which is precisely of the form required to shift the coefficient of the `CME-like' relation $J_{V,cov}^i \propto \mu_A B^i$. However, this coefficient depends on whether one considers the covariant or consistent current, and indeed it vanishes for the $U(1)_V$-invariant consistent current. The recent literature \cite{Fukushima:2008xe, Kharzeev:2009p,Rubakov:2010qi} has discussed a more precise relation to the $AVV$ anomaly, which emerges when working with the true axial chemical potential $\mu_5$ discussed above. The additional Chern-Simons term in $Q_5$ then ensures that the consistent current takes the chiral magnetic form, $J_V^i = -2c_g\mu_5 B^i$, with the relation to the anomaly fixed by the unique definition of $\mu_5$ as conjugate to the conserved axial charge. 

When the $U(1)_V$ field is dynamical, we are forced to work with the $U(1)_V$-invariant consistent currents and thus the source $\mu_A$ does not lead to the CME \cite{Rubakov:2010qi}. If one allows for additional {\it constant} CPT-odd sources, then a CME-like relation for the current can arise. For example, in \eqref{E:JV33} we have $\langle \JV^i \rangle = \tilde{C}_V T {\cal B}^i$, where 
$\mathcal{B}^i=\epsilon^{ijk}\langle F_{V,jk}\rangle/2$ is the magnetic field, determined by $J_{\rm ext}$ and other external sources. The coefficient $\tilde{C}_V$ is a CPT-odd constant, which needs to be fixed by matching to the UV theory, and is not a priori dictated by anomalies. 
A more precise relation between $\tilde{C}_V$ and the AVV anomaly may emerge (although the anomalous symmetry does not relate the two) when we study a particular theory where we can keep track of the CPT-violating sources, like $\mu_5$~\cite{Rubakov:2010qi,Hou:2011ze}. However, when $U(1)_V$ is dynamical, as discussed above we find $\tilde{c}_{VV} = -2c_g\mu_5$ which results in anti-screening of the magnetic field, and likely a dynamical instability. This suggests that any consistent treatment of the CME in the presence of dynamical gauge fields must go beyond an equilibrium framework (see e.g.~\cite{Warringa:2012bq}).

\item {\it Magnetohydrodynamics}: In this paper our primary focus was the impact of anomalies on hydrostatic response. However, our analysis was motivated by an attempt to understand the constraints that equilibrium matching imposes on hydrodynamics. Thus in Section~\ref{S:dyn2}, we briefly discussed the structure of hydrodynamics with dynamical (but classical) $U(1)_V$ gauge fields, i.e. magnetohydrodynamics. Note that this is not quite textbook MHD, in that we treated the magnetic field as a first-order quantity in the derivative expansion. It would clearly be of interest to develop this framework further, to take anomalies into account at the level of effective field theory, and make contact with existing approaches~\cite{Huang:2011dc}. It may also prove worthwhile to study the form of constitutive relations using the fluid-gravity framework within AdS/CFT. The framework of Section~\ref{S:effectiveS} should also describe holographic examples with dynamical $U(1)_V$ gauge fields, including the alternate quantization of a bulk gauge field in asymptotically AdS$_4$ spacetimes~\cite{Marolf:2006nd} and the standard quantization in AdS$_3$ backgrounds~\cite{Marolf:2006nd,Jensen:2010em}.

\item {\it Quark-gluon plasma}: The $U(N_f)_L \times U(N_f)_R$ chiral currents in QCD decompose into non-singlet $SU(N_f)_A$ and singlet $U(1)_A$ axial currents. The latter $U(1)$ current receives a large gluonic contribution to the anomaly (associated with mass of the $\eta'$ meson), implying that the degrees of freedom associated with this current are not expected to contribute in the hydrodynamic regime. However, the non-singlet currents have purely electromagnetic anomalies, allowing for a quantitative identification of anomaly-induced physics, e.g. the rate for $\pi^0 \rightarrow \gamma\gamma$ decays. In this case, the presence of small perturbative corrections that we have discussed here should not significantly modify the notion of anomalous transport. The extension of the present work to Cartan components of the non-abelian axial global currents is straightforward, and may be of interest for the study of anomalous transport in the quark-gluon plasma.

\item {\it Mixed anomalies}: When the mixed gauge-gravitational anomaly~\eqref{E:4dMatch} is nonzero, it will provide the primary contribution to the chiral vortical response
at sufficiently high temperatures via the  $T^2$ term in~\eqref{E:equalityCon}.  Currents with non-vanishing mixed anomalies are therefore particular interest as they may exhibit an enhanced chiral vortical conductivity as a macroscopic manifestation of this anomaly.  The chiral lepton number current is an example, and early cosmology may provide a high temperature regime in which the mixed gauge-gravitational anomaly could play a role. 
We recall that the conventional electroweak V$_{B+L}$AV anomaly of the vectorial $U(1)_{B+L}$ baryon plus lepton number current in the Standard Model is the only source of baryon and lepton number violation in the Standard Model. The associated sphaleron transitions above the electroweak scale have been utilized in various models of baryogenesis. It would be interesting if the mixed anomaly for the chiral lepton current were also to play a cosmological role.

\item {\it Additional gapless modes}: We have discussed the corrections to anomaly-induced transport that arise when the vector symmetry is gauged. Another way that anomaly-induced response may differ from Section~\ref{S:review2} is through the presence of additional gapless modes which contribute to hydrostatic response. Examples discussed in the literature include a $U(1)$ superfluid, and QED at zero temperature and nonzero chemical potential. In a $U(1)$ superfluid there is an extra massless Goldstone mode $\phi$, and additional gauge-invariant scalars $\epsilon^{ijk} D_i\phi \partial_j \hat{A}_k$ and $\epsilon^{ijk}D_i\phi \partial_j \mathfrak{a}_k$, where $D_i\phi = \ptl_i \phi - \hat{A}_i$, which give rise to magnetic and vortical response beyond the anomalies~\cite{Bhattacharyya:2012xi}. In zero temperature QED there is a sharp Fermi surface, and a recent 
 two-loop calculation~\cite{Gorbar:2013upa} has shown that the Fermi surface leads to a correction to the Chiral Separation Effect, (i.e. the $\mathcal{O}(k)$ part of the two-point function $\langle J_V^i(k) J_A^j(-k)\rangle$), which in the presence of screening is fixed by anomalies~\cite{Jensen:2012jy}. 
  In summary, we can categorize the corrections to anomaly-induced response as due to the dynamics of (e.g. gauge) fields entering the anomaly vertex, and/or  the breakdown of screening and the presence of additional gapless modes.
\end{itemize}

\subsection*{Acknowledgements}
We would like to thank S.~Golkar, C.~Hoyos, M.~Kaminski, D.~Kharzeev, R.~Loganayagam, M.~Pospelov, I.~Shovkovy, D.~Son, L.~Yaffe, and A.~Yarom for useful discussions and/or comments on the manuscript. This work was supported in part by NSERC, Canada.

\bibliography{refs}
\bibliographystyle{utphys}
\end{document}